\documentstyle[11pt,epsfig]{article}
\newcommand \be{\begin{equation}}
\newcommand \ba{\begin{eqnarray}}
\newcommand \ee{\end{equation}}
\newcommand \ea{\end{eqnarray}}
\newcommand{\lp}{\left(}
\newcommand{\rp}{\right)}

\begin{document}

\title{Endogenous versus Exogenous Crashes in Financial Markets}
\thispagestyle{empty}

\author{Anders Johansen$^1$ and Didier Sornette$^{2,3}$\\
$^1$ Ris\o \ National Laboratory, Department of Wind Energy \\
Frederiksborgvej 399, P.O. 49, DK-4000 Roskilde, Denmark \\
$^2$ Institute of Geophysics and
Planetary Physics and Department of Earth and Space Science\\ 
University of California, Los Angeles, California 90095\\
$^3$ Laboratoire de Physique de la Mati\`{e}re Condens\'{e}e\\ CNRS UMR6622 and
Universit\'{e} de Nice-Sophia Antipolis\\ Parc
Valrose, 06108 Nice Cedex 2, France \\
e-mails: jan.anders.johansen@ris\o.dk and sornette@moho.ess.ucla.edu}

\date{\today}
\maketitle

\abstract{
In a series of papers based on analogies with statistical physics 
models, we have proposed that most financial crashes are the climax of 
so-called log-periodic power law signatures (LPPS) associated with speculative 
bubbles \cite{SJ1998,JS1999,JSL1999,JLS2000,SJ2001}. In addition, a large body 
of empirical evidence supporting this proposition have been presented 
\cite{SJB,SJ1998,JLS2000,JS2000,emergent,SJ2001}. Along a
complementary line of research, we have established that, while the vast 
majority of drawdowns occurring on the major financial markets have a 
distribution which is well-described by a stretched exponential, the
largest drawdowns are occurring with a significantly larger rate than 
predicted by extrapolating the bulk of the distribution and should thus be
considered as {\it outliers} \cite{outl1,SJ2001,outl2,crashcom}. Here, these 
two lines of research are merged in a systematic way to offer a classification 
of crashes as either events of an endogenous origin associated with preceding 
speculative bubbles or as events of an exogenous origin associated with the 
markets response to external shocks. 

We first perform an extended analysis of the distribution of drawdowns in
the two leading exchange markets (US dollar against
the Deutsmark and against the Yen), in the major world stock markets, 
in the U.S. and Japanese
bond market and in the gold market, by introducing the concept 
of ``coarse-grained drawdowns,'' which allows for a certain degree of fuzziness
in the definition of cumulative losses and improves on the statistics of our 
previous results. Then, for each identified outlier, we check whether LPPS
are present and take the existence of LPPS as the qualifying signature
for an endogenous crash: this is because a
drawdown outlier is seen as the end of a speculative unsustainable 
accelerating bubble
generated endogenously. In the absence of LPPS, we are able to identify 
what seems to have been
the relevant historical event, {\it i.e.}, a new piece of information
of such magnitude and impact that it is seems reasonable to attribute 
the crash to it, in agreement with the standard view of the efficient market 
hypothesis. Such drawdown outliers are classified as having an exogenous 
origin. 
Globally over all the markets analyzed, we identify 49 outliers, of which 
25 are classified as endogenous, 22 as exogeneous and 2 as associated
with the Japanese anti-bubble. Restricting to the world market indices, 
we find 31 outliers, of which 19 are endogenous, 10 are exogenous and 
2 are associated with the Japanese anti-bubble.
The combination of the two proposed detection techniques, one for drawdown 
outliers
and the second for LPPS, provides a novel and systematic taxonomy of crashes
further subtantiating the importance of LPPS.
}

\thispagestyle{empty}
\pagenumbering{arabic}
\newpage
\setcounter{page}{1}

\section{Introduction} 

The characterization of large stock market moves, and especially large 
negative price drops, is of profound importance for risk management
and portfolio allocation. According to standard economic theory, the complex 
trajectory of stock market prices is the faithful reflection of the continuous 
flow of news that are interpreted and digested by an army of analysts and 
traders \cite{Cutler}. Accordingly, large market losses should result from 
really bad surprises. It is indeed a fact that exogenous shocks exist, as 
epitomized by the recent events of Sept. 11, 2001 and the coup in the 
Soviet Union on Aug. 19, 1991, which move stock market prices and create strong
bursts of volatility. However, it could be argued that precursory fingerprints 
of even these events were known to some, suggesting the possibility that the 
action of these informed agents may have been reflected in part in stock market
prices prior to the advent of the shock. This question is another formulation
of the problem of market efficiency and to what degree is there some residual 
private information that is not fully reflected in prices \cite{Fama}. Thus, a 
key question is whether large losses and gains are indeed slaved to exogenous 
shocks or on the contrary may result from an endogenous origin in the dynamics 
of that particular stock market. The former possibility requires the risk 
manager to closely monitor the world of economics, business, political, social,
environmental ... news for possible instabilities. This approach 
is associated with standard ``fundamental'' analysis. The later endogenous 
scenario requires the investigation of signs of instabilities to be found in 
the market dynamics itself and could rationalize in part so-called 
``technical'' analysis.

As a first step to address this question from a statistical view point, we 
ask whether or not one can distinguish very large losses from the rest of 
the population of smaller losses. Of course, very large losses are naively 
distinct from the rest simply by their sheer size. The issue is 
not to qualify naively a large loss simply by its magnitude, which would 
be a trivial and uninformative definition, but to ask whether there are 
distinctive statistical properties that distinguish big losses from the rest 
of the population of losses. According to the definition of the Engineering 
Statistical Handbook ``An outlier is an observation that lies an abnormal 
distance from other values in a random sample from a population'' 
\cite{handbook}. In a sense, this definition leaves it up to a 
consensus process to decide what will be considered abnormal. However, before 
abnormal observations can be singled out, it is necessary to characterize 
normal observations. Hence, the question we want to address in this paper is 
whether the largest losses seen on the financial markets are merely ``amplified
small losses'' or something entirely different. We do this by following two 
independent but complementary lines of investigation, namely a statistical
analysis of drawdowns in various markets and a case-study analysis of 
speculative bubbles in the same markets. In our previous analysis and 
identification of outliers on the major financial markets \cite{outl1,outl2}, 
drawdowns (drawups) were simply defined as a continuous decrease (increase) 
in the value of the price at the close of each successive trading day. Hence, 
a drawdown (drawup) was terminated by {\it any} increase (decrease) in the 
price no matter how small. In section \ref{defdraws}, 
we will generalise the definition of drawdowns
(drawups) and introduce the concept of ``coarse-grained'' drawdowns (drawups). 
Drawdowns (drawups) of the previous kind analyzed in \cite{outl1,outl2} will be
refereed to as ``pure'' drawdowns.

Analyzing a stock market index or a currency exchange rate  using the 
distribution of drawdowns (or the complementary quantity drawups) rather 
than the more standard distribution of returns has the advantage that 
correlations of order two and higher are in part taken into account
in this one-point statistics. These drawdowns
(drawups) may identify transient bursts of dependences in successive returns. 
As our definition represents a ``worst case scenario'' of loss, see section
\ref{defdraws}, it is different from the one of Grossman \& Zhou and others 
\cite{Grossman,Cvitanic,Checkhlov}, where it is the present loss from the last 
maximum of the price. Furthermore, we will not address the portfolio allocation
problem based on risks quantified by drawdowns, a very important problem which 
is left for a future publication.

By studying several variants of drawdowns (runs of cumulative losses with 
different degrees of fuzziness, see section \ref{defdraws}), we show that the 
largest drawdowns in general 
are outliers to the vast majority ($\geq$ 98\%) of drawdowns and cannot be 
described by an extrapolation of the distribution of small and intermediate 
losses, a property that we refer to as the ``king'' effect \cite{Lahe}.
This analysis extends an increasing amount of evidence showing that 
the distribution of the largest negative market moves belongs to a population 
different from that of the smaller moves 
\cite{outl1,outl2,JS2000,SJ2001,LilloMantegna,Mansilla,crashcom}. 

The second step concerns the origin of this king effect: does it reflect the 
arrival of an anomalously serious piece of news or are very large drawdowns 
the outcome of a self-organized dynamical process of the stock market with its
complex interactions between heterogeneous agents of varying sizes, all
subjected to an incessant bombardment of news, each piece of news
being insufficient by itself to explain the presence or absence
of a shock? While it is a common practice to
associate the large market moves and strong bursts of volatility with
external economic, political or natural events \cite{white}, there is simply 
no convincing evidence supporting it. The first indication that a combination 
of these two processes may be responsible for the creation of (drawdown) 
outliers in the stock market was obtained by \cite{outl1}, who
found that for the DJIA in the last century two outliers were associated with a
profound bull-market (the crashes of Oct. 1929 and Oct. 1987) and one outlier 
with a major historical event, namely the outbreak of WWI. Furthermore, this 
question has previously been addressed quantitatively for volatility shocks 
modeled quite accurately by the 
multifractal random walk model of stochastic volatility \cite{SorMRW}. It was 
shown that endogenous and exogenous shocks give rise to different precursory 
as well as relaxation dynamics \cite{Sorhelm}. In other words, the exogenous 
versus endogenous origins of a large volatility shock leave a sufficiently 
strong distinctive imprint in the price dynamics that one can distinguish two 
classes of signatures, a rather fast relaxation to normal volatility levels 
for exogenous shocks compared with an amplitude-dependent slower relaxation 
for endogenous shocks. Interestingly, most of the volatility shocks were found 
to be endogenous \cite{SorMRW}. Here, we study the problem of distinguishing 
between shocks of exogenous versus endogenous origin at larger time scales. 
We stress that a shock of exogenous origin may very well be a large loss on 
another major stock market.
By analyzing the time series of stock market prices prior to the occurrence 
of each of the qualified outliers, we show that for the large majority of 
cases, a distinctive structure in the price trajectories exists. Specifically, 
the majority of drawdown outliers are preceded by a (super-exponential) power 
law price appreciation decorated by accelerating (log-periodic) oscillations
or log-periodic power law signatures (LPPS)
These structures have previously been found to characterize periods preceding 
financial crashes and can be rationalized with a rational bubble model 
\cite{JSL1999,JS1999,JLS2000,SJ2001}. We also find that a small fraction of 
the drawdown outliers belong to another exogenous class and result from 
uncontroversial strong destabilizing pieces of new information, such as 
declarations of war. The present study thus extends these previous works by 
offering a systematic analysis of drawdown outliers rather than focusing on 
the bubbles and crashes as done before. Specifically, the present study 
distinguishes itself for previous works and from our previous reports in three 
aspects:
\begin{itemize}
\item we first develop and extend an objective measure of ``crashes'' which 
is used to distinguish them from events during normal times. This systematic 
classification recovers all the studied crashes reported in previous 
publications (except the crash of 1937) and adds new events. Previously
published crashes with LPPS will be marked $\star$ in the tables. Bubbles 
identified by the outlier analysis presented here, {\it i.e.}, outliers which 
have prior LPPS bubbles {\it not} identified in previous works will be marked 
by $\dagger$.

\item we then analyze the price time series preceding all these objectively 
defined crashes to test for the presence of LPPS.

\item doing this, we classify two types of crashes, endogenous ones which 
are characterized by a preceding speculative bubble with LPPS that became 
unsustainable and exogenous crashes when the market was subjected to a very 
strong external perturbation.
\end{itemize}
The emphasis of this paper thus lies first in the development of a systematic 
and objective definition of crashes and second in the systematic search for 
LPPS in the time series preceding the crash. This allows us to address previous
criticisms of possible data-picking and to assess the robustness and 
significance of the previously proposed precursors. When an outlier
has been detected, we search for both LPPS and for a 
major historical event which may be causally linked to the date of the outlier.
We find in general that no clear piece of striking news can be associated
with outliers preceded by LPPS while outliers without LPPS have in general
been triggered by a great news surprise.
By combining these two methods aimed at detecting anomalous events seen as 
outliers and LPPS qualifying endogenous speculative bubbles, respectively, we 
provide a new objective 
test of the hypothesis that the largest negative markets moves are special and 
form two distinct populations. On the one hand, we identify exogenous crashes 
which can be attributed to extraordinary important external perturbations in
the form of news impacting the market. On the other hand, most of the crashes 
are found to be endogenous and can be seen as the natural deaths of
self-organized self-reinforcing speculative bubbles giving rise to specific 
precursory signatures, specifically LPPS. For this, we generalize our previous 
analysis of drawdowns to drawdowns coarse-grained in amplitude
($\epsilon$-drawdowns). Another generalization of
drawdowns coarse-grained in time ($\tau$-drawdowns) will be reported elsewhere.
In relation with our previous works, we find that
all of the crashes (except the one in 1937) and 
associated bubbles previously analysed for the presence of LPPS are here 
causally
linked with a drawdown outlier. This gives an additional reason to believe 
our previous results
showing that LPPS is a strong discriminator of bubbles preceding strong 
corrections. It also strengthen the conclusion that LPPS, which are by 
construction 
transient structures, are almost uniquely associated with a speculative phase 
announcing a strong change of regime. 

The outline of the paper is as follows. In the next section, we describe in
detail using four concrete examples what we mean by exogenous and endogenous 
crashes and bubbles with LPPS.  In
section \ref{defdraws}, we discuss how our drawdowns (drawups) are defined
and explain the methodology of 
our analysis. Section \ref{wavelet}
presents an analysis of the distribution of deviations of DJIA
from a 4-year smoothed average obtained by using the Daubechies wavelet 
transform,
which confirms independently the existence of anomalous large
events in the tail of the distribution.
In sections \ref{fx}, \ref{sm}, \ref{bm} and \ref{gold}, we 
present
our results on the FX market, major stock markets, the U.S. and Japanese
bond market and the gold market, respectively.  The last section concludes.
The Appendix \ref{data} lists the various 
markets and time periods we have analysed.

\section{Exogenous and Endogenous Crashes and LPPS Bubbles}

\begin{figure}
\begin{center}
\parbox[l]{8.5cm}{
\epsfig{file=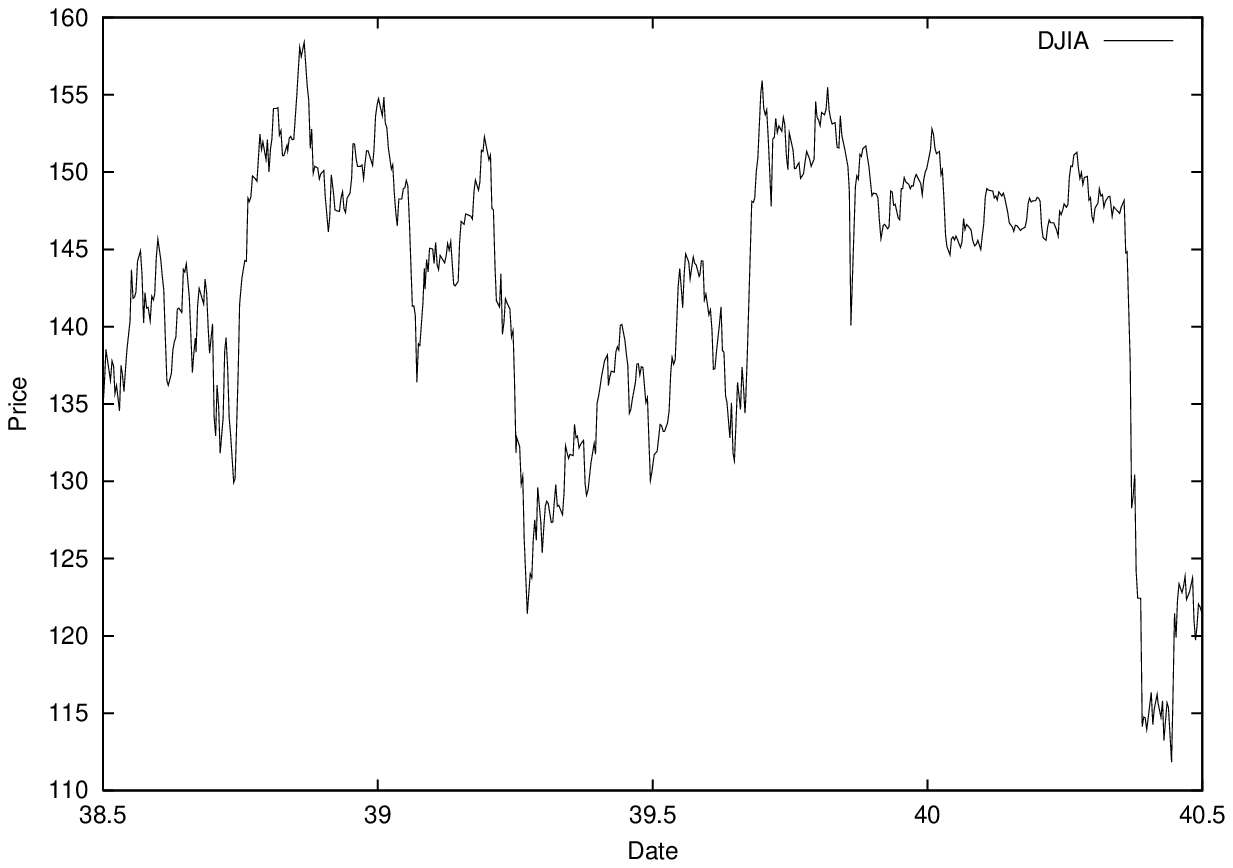,height=8cm,width=8.5cm}
\caption{\label{40shock} The $23.7\%$ $\epsilon$-drawdown in the DJIA in
May 1940 caused by the Nazi invasion of France and Benelux, see table
\protect\ref{outldj}.}}
\hspace{5mm}
\parbox[r]{8.5cm}{
\epsfig{file=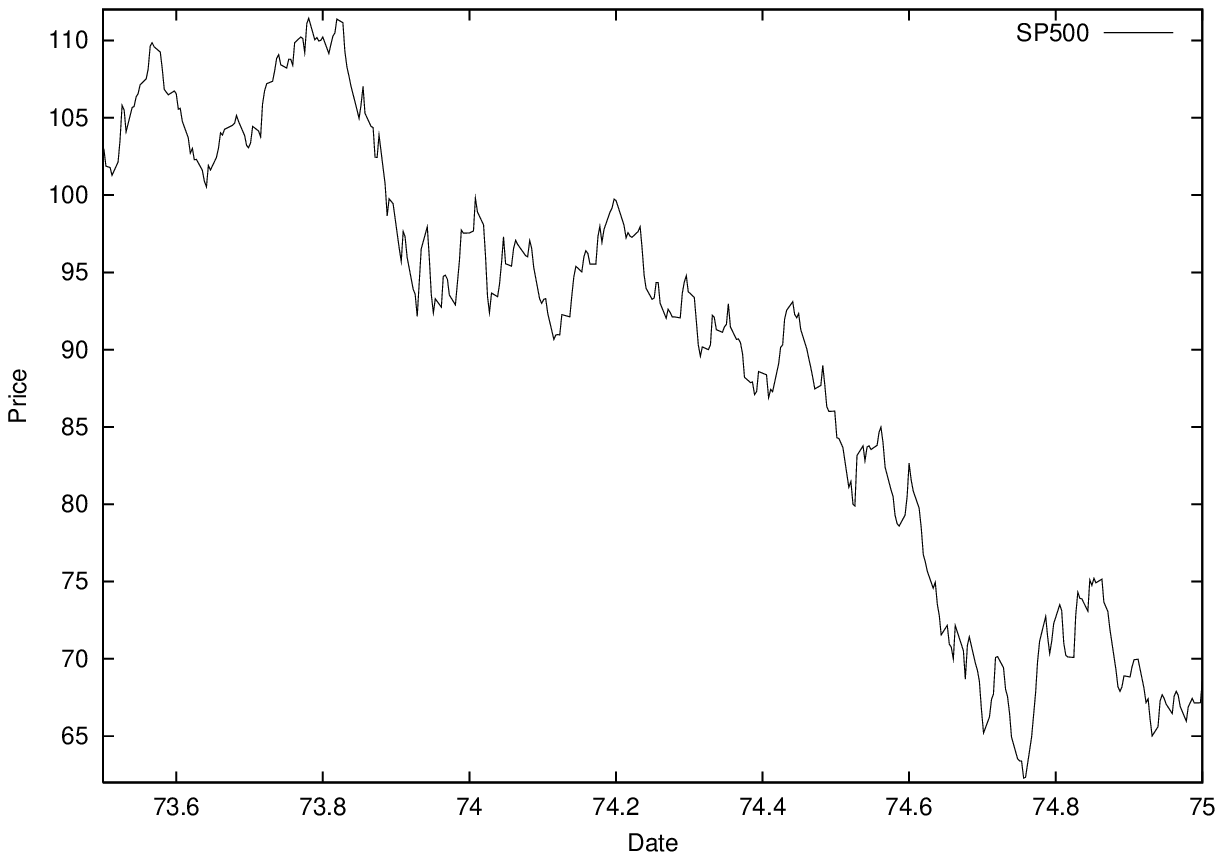=,height=8cm,width=8.5cm}
\caption{\label{74shock} The $11.2\%$ pure drawdown in the SP500 in Sep. 1974
caused by the resignation and subsequent controversial pardoning of 
president R. Nixon, due to the Watergate scandal, see table\protect\ref{outlsp}.
}}
\vspace{1.5cm}
\parbox[l]{8.5cm}{
\epsfig{file=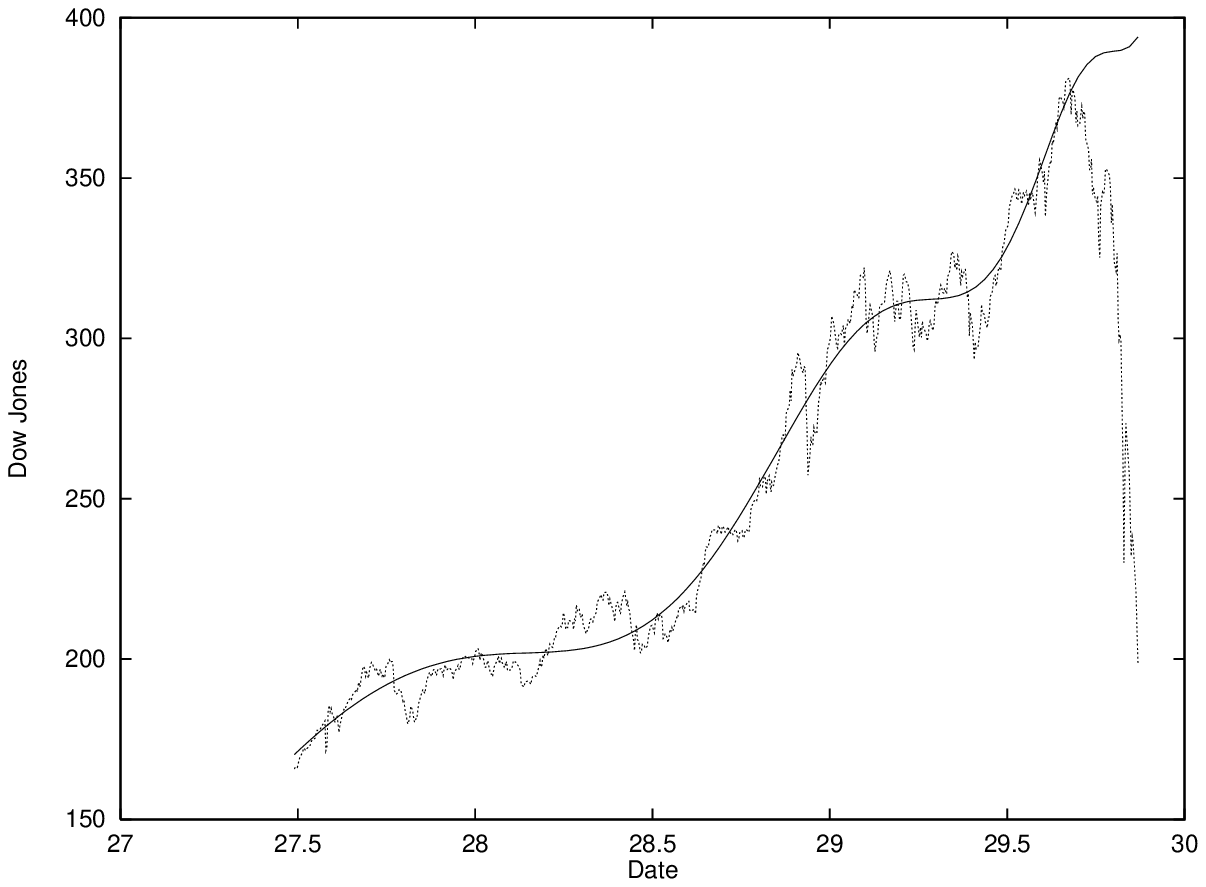,height=8cm,width=8.5cm}
\caption{\label{29bub}. The $23.6\%$ pure drawdown in the DJIA in Oct. 1929.
The fit is eq.~(\ref{lpeq}) where $A\approx 571$, $B\approx  -267$, $C\approx  
-14.3$, $z\approx 0.45$, $t_c\approx 1930.22$, $\phi \approx 4$, $\omega 
\approx 7.9$. } }
\hspace{5mm}
\parbox[r]{8.5cm}{
\epsfig{file=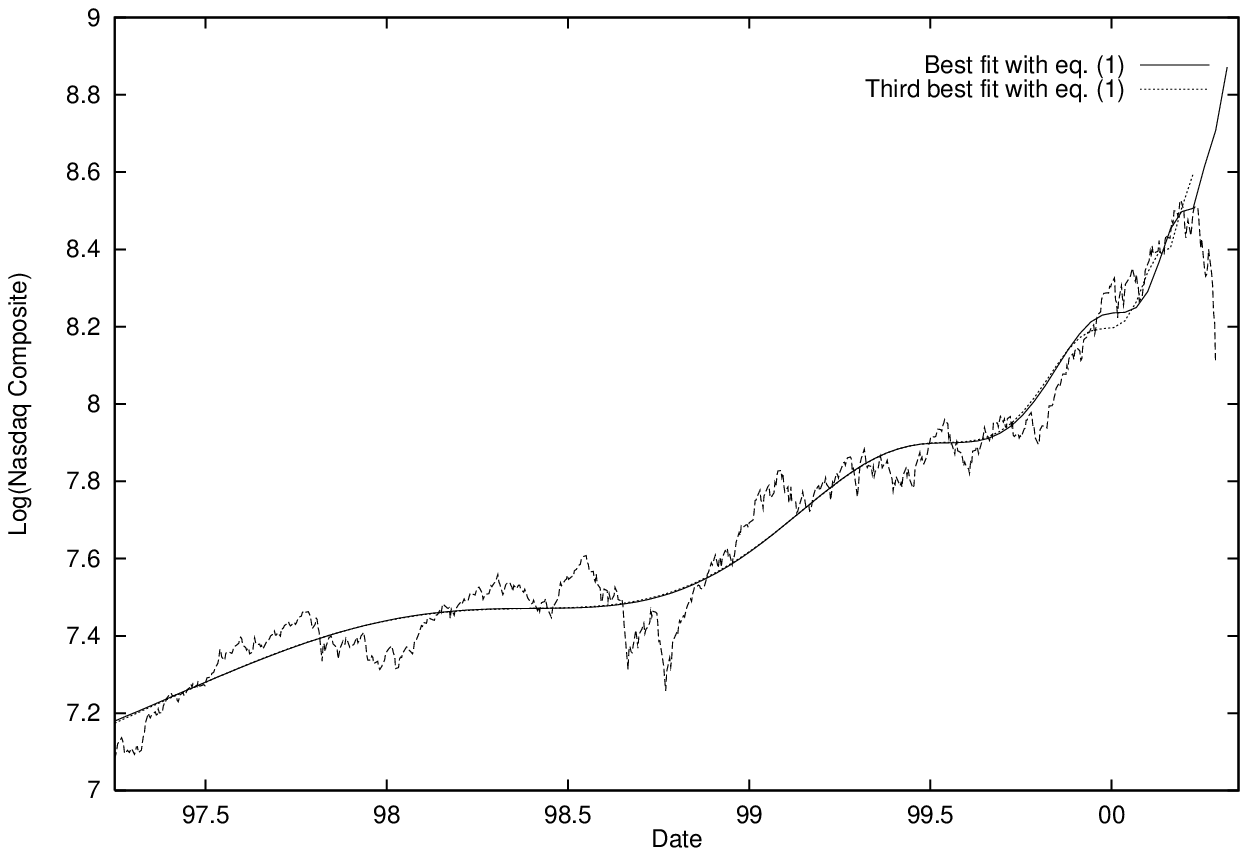,height=8cm,width=8.5cm}
\caption{\label{00bub} The $25.3\%$ pure drawdown in the Nasdaq in April 2000. 
The two fits are eq. (\ref{lpeq}) where $A\approx 9.5;8.8$, $ B\approx 
-1.7;-1.1$, $C\approx 0.06;0.06$, $z\approx 0.27;0.39$, $ t_c \approx 2000.339;
2000.247$, $\phi\approx -0.14;-0.814$, $\omega\approx 7.0;6.5$.}}
\end{center}
\end{figure}

Two good examples of exogenous crashes, {\it i.e.}, large declines in the price
caused by external shocks, are the Nazi invasion of France and Belgium, 
Luxembourg and the Netherlands (Benelux) on May 10th 1940 as well as the 
resignation and the following controversial pardoning of president R. 
Nixon on August 8th and September 8th 1974, see the last entry in tables 
\ref{outldj} and \ref{outlsp} for the parameters of the related drawdowns and 
figures \ref{40shock} and \ref{74shock} for the time series prior to these two 
exogenous crashes. In the two figures, the index show no signs of
a bull-market, even less of an unsustainable bubble, and hence it seems
reasonable to attribute the two large 
declines to these two historical events.

Two good examples of the endogenous class are the arch-typical crash of 
October 1929 and the ``dot.com'' crash of April 2000, see table \ref{outldj} 
and \ref{outlnas} for the parameters of the related drawdowns (to be analyzed 
in detail below). These two crashes can be seen as nothing but efficient 
deflations \footnote{resembling the impact of the famous ``invisible hand'' 
of Adam Smith, delayed over the long time scales of the speculative bubbles.} 
of extended stock market bubbles in both cases funneled by heavy investment in 
so-called  ``new economy'' companies\footnote{That the crash of Oct. 1929 
caused the Great Depression is a part of financial folklore, but nevertheless 
untrue. For instance, using a regime switching framework, \protect\cite{Coe} 
finds that a prolonged period of crisis began not with the 1929 stock market 
crash but with the first banking panic of October 1930.}. Figures 
\ref{29bub} and \ref{00bub} show the bubbles preceding the these
two endogenous crashes with their LPPS. The solid lines are fits 
with a first order LPPS eq.~(\ref{lpeq}), see figure captions for the fit 
parameters. 

\begin{figure}
\begin{center}
\parbox[l]{8.5cm}{
\epsfig{file=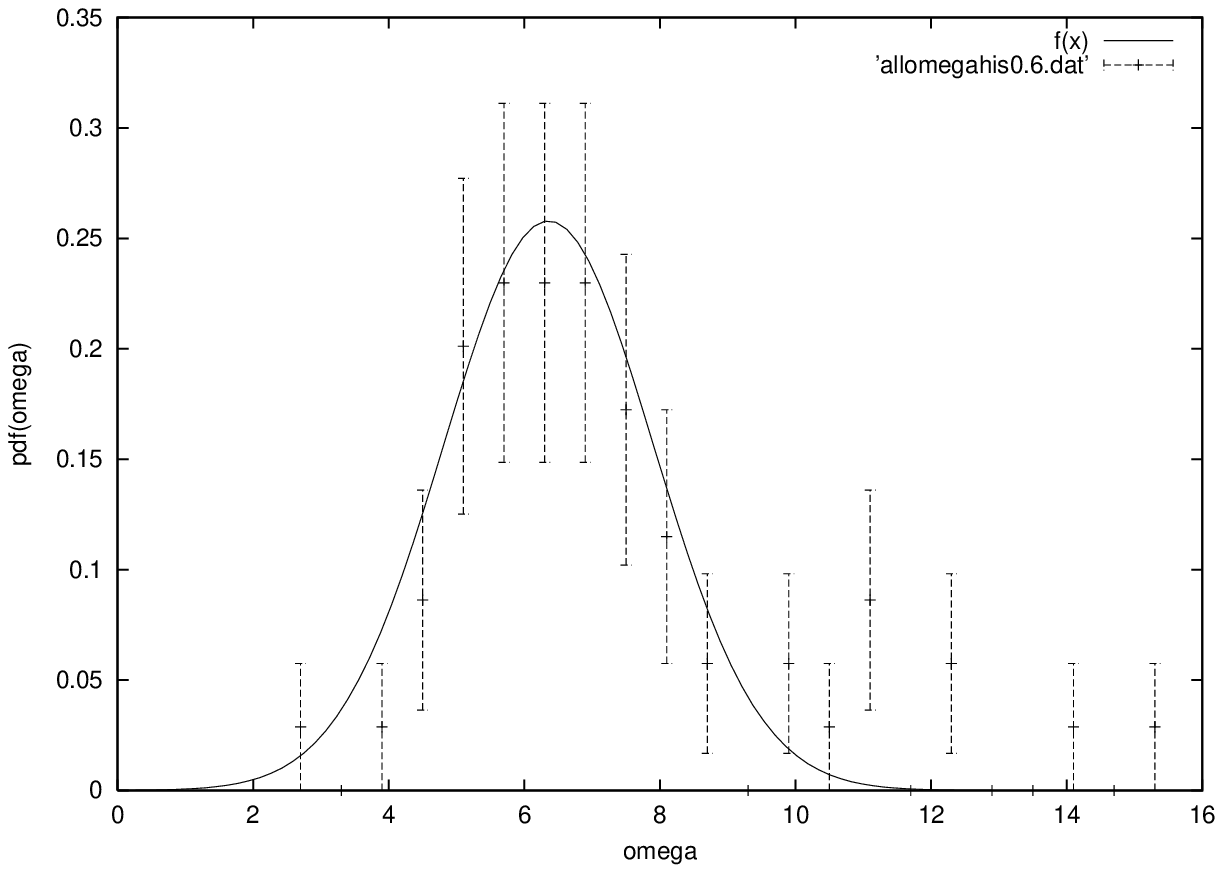,height=8cm,width=8.5cm}
\caption{\label{omegadistrib} Empirical distribution of the log-periodic
angular frequency $\omega$ in eq.~(\protect\ref{lpeq}) for over thirty case studies. 
The fit with a Gaussian distribution gives $\omega \approx 6.36 \pm 1.55$. 
The smaller peak centered on $11-12$ suggests the existence of a second
discernable harmonics at $2 \omega \approx 12$.}}
\hspace{5mm}
\parbox[r]{8.5cm}{
\epsfig{file=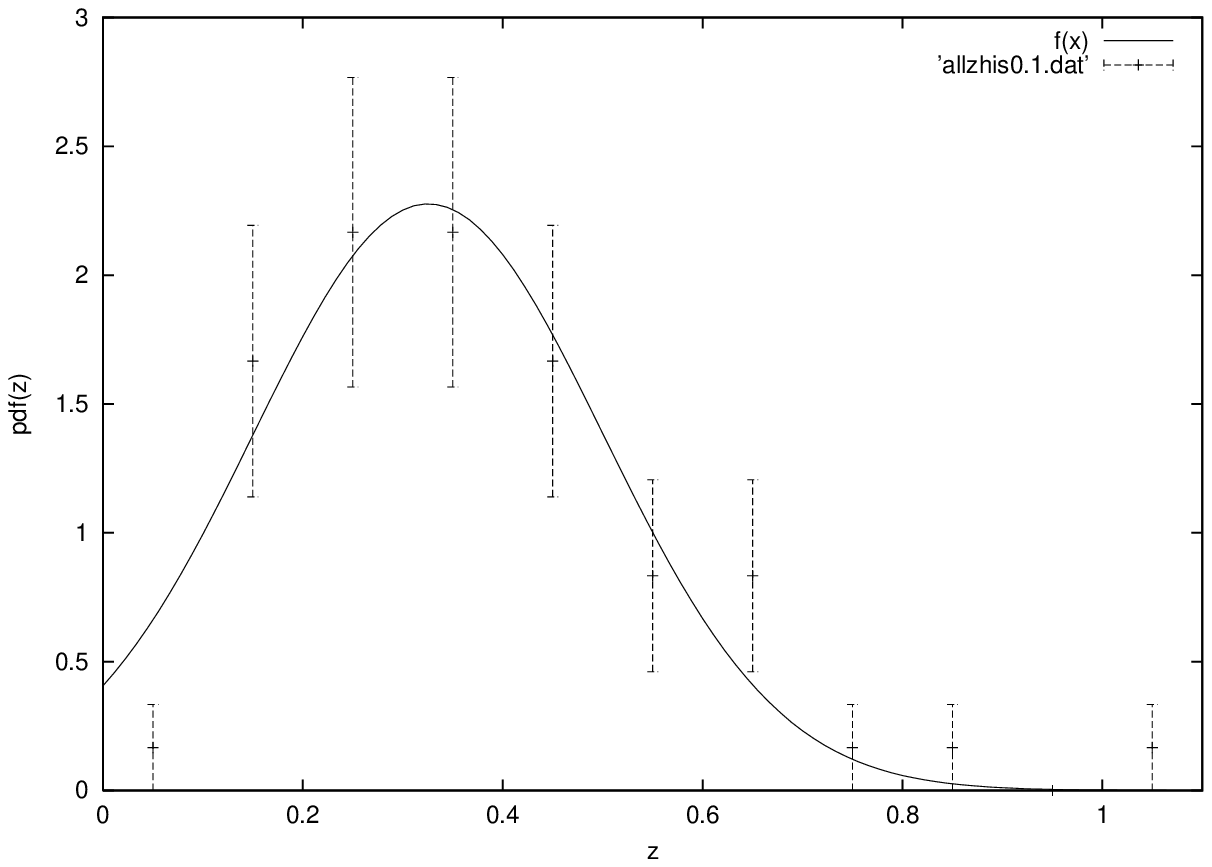,height=8cm,width=8.5cm}
\caption{\label{zdistrib} Empirical distribution of the exponent $z$ of
the power law in eq.~(\protect\ref{lpeq}) for over thirty case studies. The fit with 
a Gaussian distribution gives $\beta \approx 0.33 \pm 0.18$.
}}
\vspace{1.5cm}
\parbox[l]{8.5cm}{
\epsfig{file=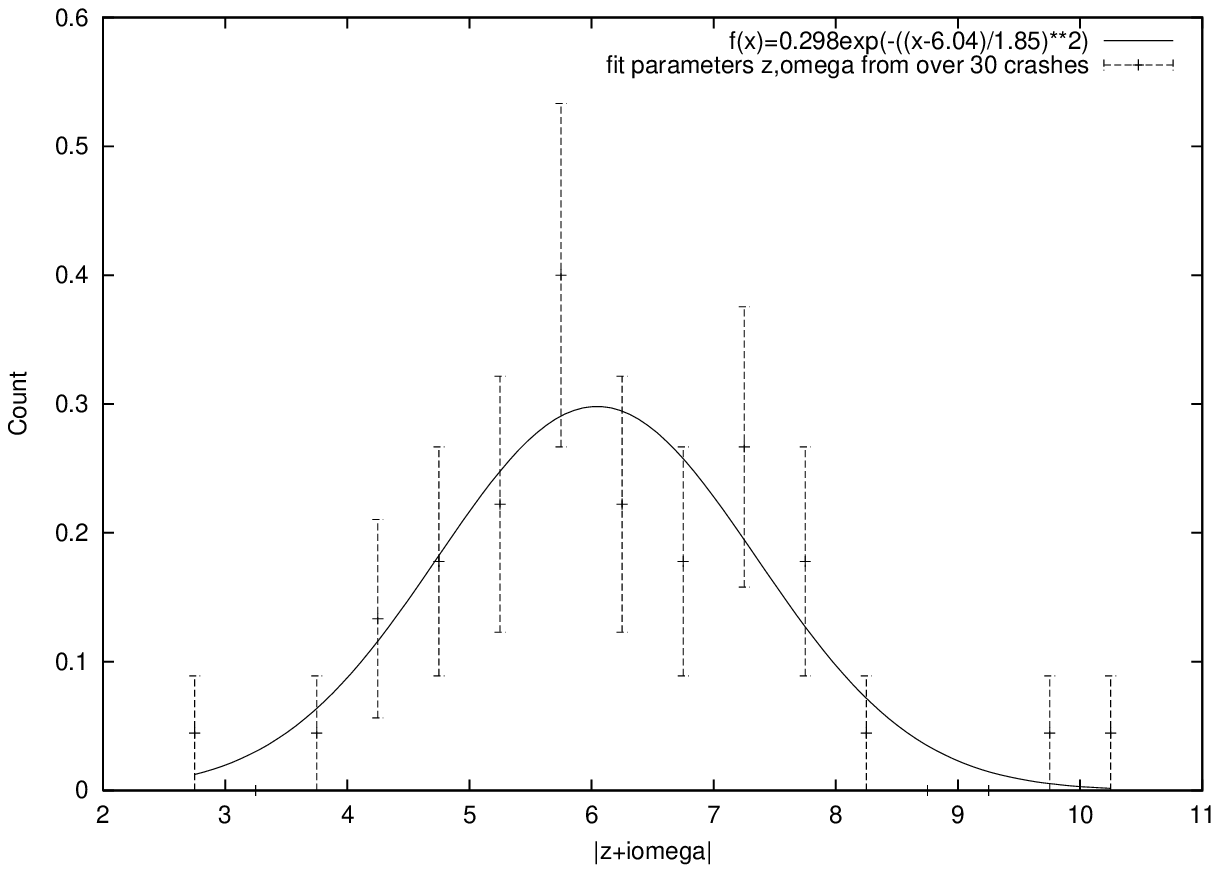,height=8cm,width=8.5cm}
\caption{\label{zomegadistrib} Empirical distribution of the norm $|z+i\omega|$,
of the complex exponent $z+i\omega$, where $z$ and $\omega$ are defined in
eq.~(\protect\ref{lpeq}). The fit with 
a Gaussian distribution gives $|z+i\omega| \approx 6.04 \pm 1.85$.
}}
\hspace{5mm}
\parbox[r]{8.5cm}{
\epsfig{file=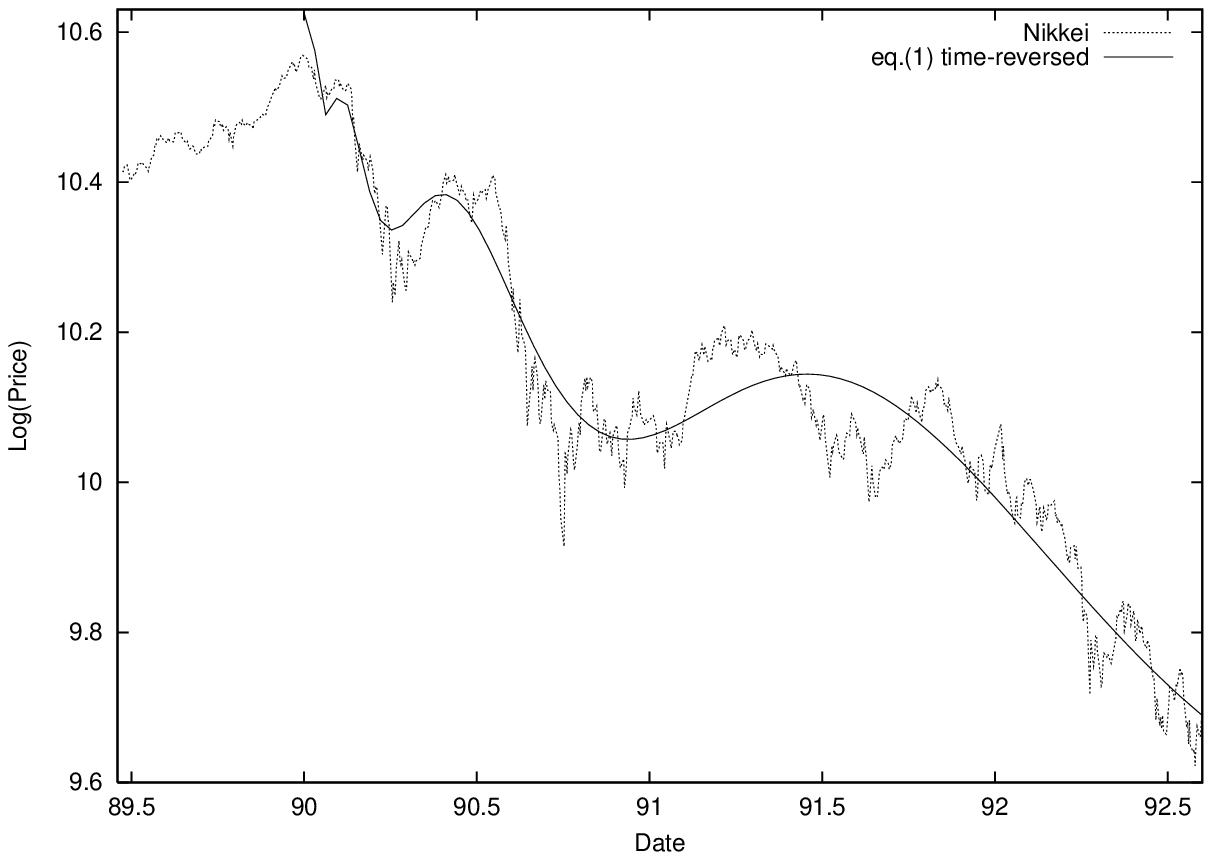,height=8cm,width=8.5cm}
\caption{\label{antinik}. The anti-bubble in the Nikkei starting in 1990. The
fit is eq. (\ref{lpeq}) with argument $t-t_c$, {\it i.e.}, time reversed.
The parameters of the fit are $A\approx 10.7$, $ B\approx -0.54$, $C\approx 
-0.11$, $z\approx 0.47$, $ t_c \approx 1989.99$, $\phi\approx -0.86$, 
$\omega\approx 4.9$.
}}
\end{center}
\end{figure}

In a series of papers, the authors have documented that most speculative 
bubbles on the major stock markets as well as emergent markets present the
following characteristics \cite{JSL1999,JS1999,JLS2000,SJ2001}: (i) a 
super-exponential overall acceleration of the price quantified by a power law 
in the time to the end of the bubble and (ii) oscillations with a frequency 
accelerating approximately in geometrical proportional to the distance to the 
(most probable) time $t_c$ of the end of the bubble. These accelerating 
oscillations have been called log-periodic oscillations because they are 
quantified by the following formula
\be \label{lpeq}
I\lp t\rp = A + B\lp t_c -t\rp^z + C\lp t_c -t\rp^z
\cos \lp \omega \ln \lp t_c -t\rp -\phi \rp ~,                       
\ee
Expression (\ref{lpeq}) has been proposed as the general signature
of cooperative speculative behavior 
coexisting with rational behavior in a general mathematical theory of 
rational bubbles \cite{JSL1999,JLS2000}. This formula 
(\ref{lpeq}) exemplifies the two characteristics of a bubble preceding an 
endogenous crash mentioned above: 
\begin{itemize}
\item[(i)] a faster-than exponential growth of the price 
captured by the power law $\lp t_c -t\rp^z$ with $0<z<1$
\item[(ii)] an accelerating oscillation decorating the accelerating price,
described by the term $\cos \lp \omega \ln \lp t_c -t\rp -\phi \rp$
which leads to local periods converging to zero according to a geometrical 
progression with scale factor $\lambda \equiv \exp \lp 2\pi/\omega \rp$.
\end{itemize}
We stress that, of the 7 parameters present in eq. (\ref{lpeq}), the 3 linear
parameters $A$, $B$ and $C$ are ``slaved'' in the fitting algorithm and 
calculated from the values obtained from the remaining 4 parameters 
\cite{thesis,JLS2000}. Furthermore, $A$, $B$, $C$ and $\phi$ are simply units
and carry no ``structural'' information and $t_c$ is by definition 
event-specific. Hence, only the two variables $z$ and $\omega$ quantifying
the over-all acceleration in the price increase and the oscillations, 
respectively, carry any ``structural'' information about the market dynamics.
This is expressed by the remarkable observation that the values for 
$z$ and $\omega$ are found to be remarkably consistent for a large variety of 
speculative bubbles on different markets \cite{bali}. Specifically, we have 
previously found that 
\be \label{zomegaval}
\omega \approx 6.36 \pm 1.56 \hspace{10mm} z \approx 0.33 \pm 0.18
\ee
for over thirty crashes on the major financial markets, as shown in figure
\ref{omegadistrib} for the distribution of $\omega$, figure \ref{zdistrib} for 
the distribution of $z$ and figure \ref{zomegadistrib} for the distribution of 
the complex exponent $z+i\omega$. Note the existence of a major peak at 
$\omega \approx 6.36 \pm 1.55$ in figure \ref{omegadistrib} and of a secondary 
peak centered on approximately twice this value, suggesting the relevance of a 
second harmonics of the main log-periodic angular log-frequency $\omega$. The 
reason for this is that in general, the best fit in terms of the r.m.s. is also
the best solution in terms of estimating the {\it structual} variables $z$ and 
$\omega$ as well as the most probable time $t_c$ of the end of the bubble, 
see \cite{SJ2001} for a more detailed discussion. However, for a few cases the 
two best fits are included in the statistics which explains the presence of 
the ``second harmonics'' around $\omega \approx 11.5$.   

The consideration of the complex exponent $z+i\omega$ in figure  
\ref{zomegadistrib} comes from the fact that the third term
$C\lp t_c -t\rp^z
\cos \lp \omega \ln \lp t_c -t\rp -\phi \rp$
inequation (\ref{lpeq}) is nothing but the real part of 
the power law $C'\lp t_c -t\rp^{z+\i\omega}$ where
$C'=C e^{-i\phi}$, written with the complex exponent $z+i\omega$. 
Thus, the log-periodic formula (\ref{lpeq})
can be viewed simply as the generalization to the field
of complex numbers of standard power laws \cite{Sordsi}.

Such a narrow range
of variation of a scaling index such as $\omega$ or $z+i\omega$ is in
the physical sciences often attributed to a universal behavior reflecting a 
robust feature of the underlying mechanism of cooperative behavior between 
``units'' leading to the observed critical pattern \cite{Fisher}.
We thus take the qualification of this pattern encoded in the mathematical 
formula (\ref{lpeq}) as the definition of an endogenous bubble ending in a 
drawdown outlier or ``crash.'' Reciprocally, the absence of such a 
pattern is taken as indicating that the shock was not the culmination of an 
endogenous speculative process and may have resulted from an exogenous source.

Here, we combine the search for log-periodic power law signatures
(LPPS) embodied by eq.~(\ref{lpeq}) with the outlier concept described below
(which amounts to
find significant deviations from
eq.~(\ref{stretched})), which were previously persued independently.
Before persuing further, we need to clarify what financial quantity
should be analyzed, in other words, what is the proxy $I\lp t\rp$ in 
eq.~(\ref{lpeq} 
to analyse. Specifically, for the bubble preceding the crash of
Oct. 1929 (figure \ref{29bub}), we have chosen the market price index itself 
and, for the the bubble 
preceding the Nasdaq crash of April 2000, we have chosen the logarithm of the 
market price
(figure \ref{00bub}). In fact, based on the rational bubble model of 
\cite{JSL1999,JLS2000}, it can be shown that, if the magnitude of the crash is 
proportional to the price increase only associated with the
contribution of the bubble, then the 
correct proxy is the price itself; on the other hand, if the magnitude of the 
crash is
proportional to the price, then the correct proxy is the logarithm of the
price \cite{JS1999}. 

We stress that the physical framework underlying eq.(\ref{lpeq}) is invariant 
under an interchange of $t_c$ and $t$. This leads
to the interesting property that eq.~(\ref{lpeq}) with $t-t_c$ as its 
argument describes so-called ``anti-bubbles'', see figure \ref{antinik} for
a specific example. Several cases of such anti-bubbles have been documented in 
many markets \cite{antibub1,antibub2,JSL1999,emergent}, most noteworthy that 
of the Japanese stock market since its collapse in 1990, see see figure 
\ref{antinik}. In fact, a very successful prediction
based on the extension of the mathematical theory for bubbles to anti-bubbles
was published in Jan. 1999 predicting the value of the Nikkei in Jan. 2000
with an accuracy of $\approx 1\%$ \cite{antibub1,antibub2}. More recently,
a similar anti-bubble has been detected 
in the behavior of the US S\&P500 index from 1996 to 
August 2002 \cite{predsz}.

Last, it should be noted that eq.(\ref{lpeq}) is in fact a special solution of
a first order ``Landau-type'' expansion where a general periodic function has 
been chosen as a cosine for simplicity. Including higher order expansions is
in principal straightforward (see \cite{SJ1997,thesis,antibub1} for works 
including second and third order terms), however the technical difficulties 
especially with respect to controlling the fitting algorithm are considerable.
These extensions are not considered in the present work.

\section{Methodology for coarse-grained drawdowns and drawups} \label{defdraws}

\subsection{First evidence of outliers at large time scales using wavelets} 
\label{wavelet}

\begin{figure}
\begin{center}
\parbox[l]{8.5cm}{
\epsfig{file=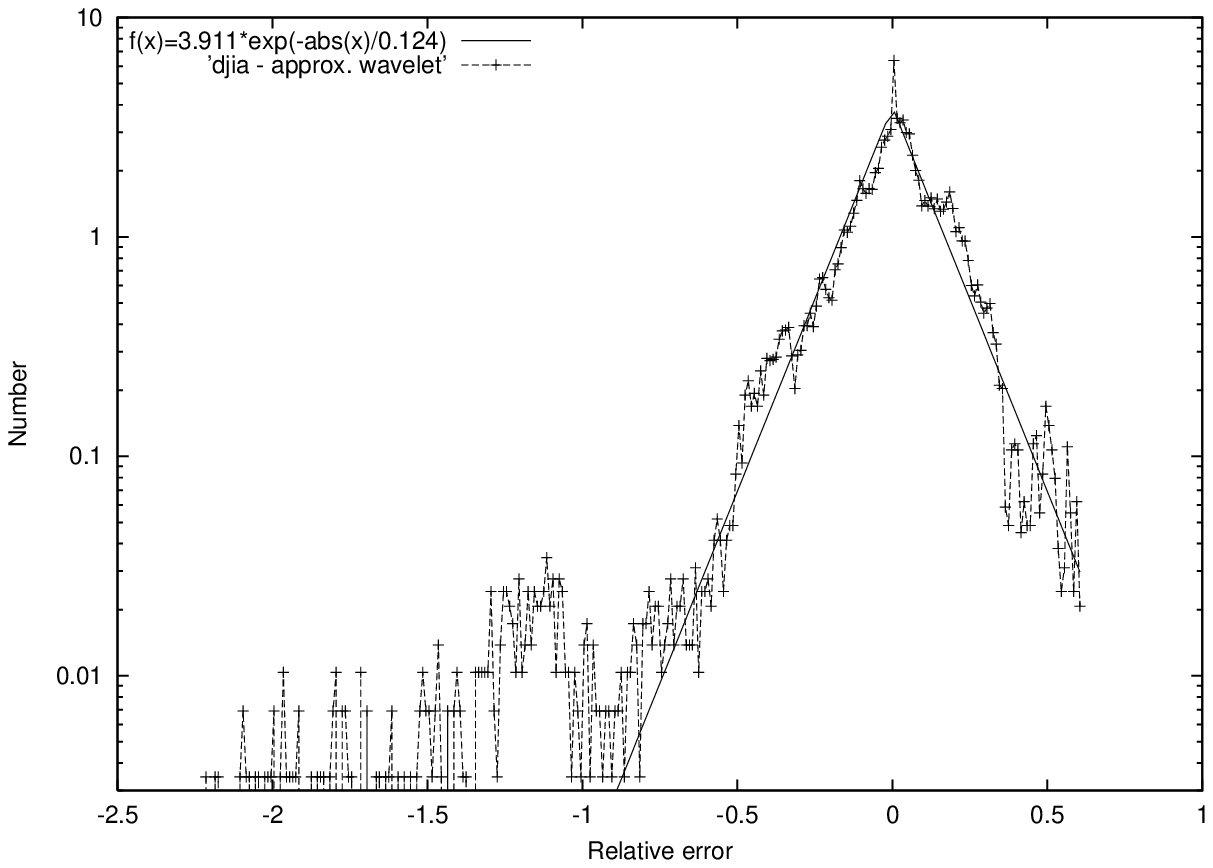,height=8cm,width=8.5cm}
\caption{Histogram of the differences between the DJIA and the approximating 
wavelet (window of 1024 days) shown in figure \protect\ref{djiawave}. 
\protect\label{wave} } }
\hspace{5mm}
\parbox[r]{8.5cm}{
\epsfig{file=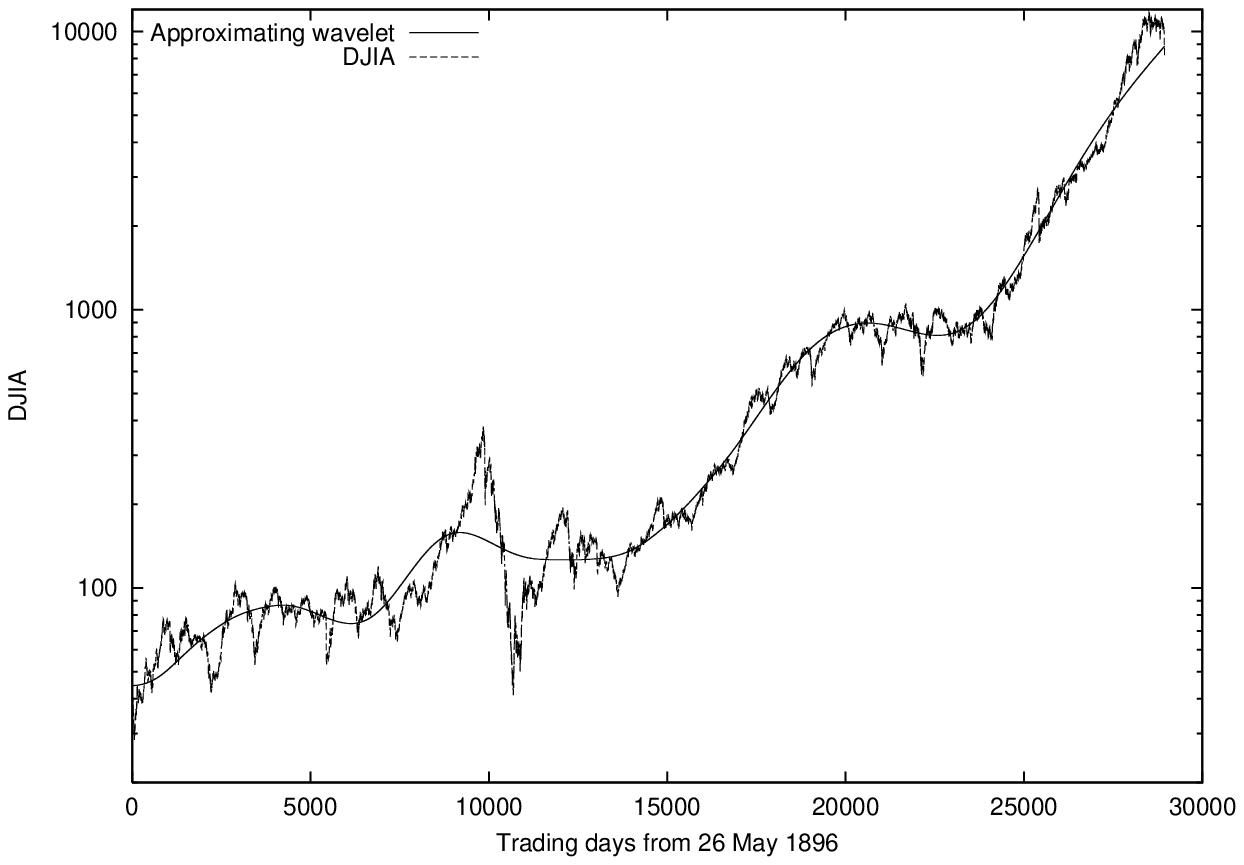,height=8cm,width=8.5cm}
\caption{The history of DJIA and and the approximating wavelet 
(window of 1024 days). \protect\label{djiawave} } }
\end{center}
\end{figure}

Before presenting the more detailed analysis of $\epsilon$-drawdowns,
it is useful to give a broader view. For this, a smoothed version of the DJIA 
index in the last century using a wavelet filter has been constructed
\cite{Simonsen-EPJB-2002}.  Specifically, a  Daubechies'
wavelet of order 24 at a scale
corresponding to a time window of $1024$ trading days has been used to 
model long-term (meaning larger than the time window) trends in the DJIA 
\cite{Book:NR-1992}. This is achieved by first transforming the data to the 
wavelet-domain, setting all wavelet coefficient corresponding to scales larger 
than $1024$ trading days to zero and transforming back to the time domain 
again. Then, this approximating wavelet is subtracted from the original DJIA 
price time series to obtain a residual time series. 
Figure \ref{wave} shows the histogram of the residues, that is, of the 
differences between the original DJIA index and the wavelet-filtered version
at the scale of approximately 4 years. 

The distribution of residuals is made of two parts. The bulk part of
small and intermediate size residuals is well-fitted by an exponential
distribution, qualified as the straight lines in this log-linear 
representation. The largest negative residuals deviate significantly from the 
extrapolation of the
exponential distribution by shaping a secondary peak. This qualifies the
existence of a second regime of ``outliers'' for the distribution of 
large negative drops. Interestingly, there are no outliers for the positive
residuals. Thus, this wavelet analysis substantiates the asymmetry previously 
established for drawdowns and drawups \cite{outl2} as well as the outlier 
concept for drawdowns, however, for surprisingly long time scales. 

This analysis is complementary to the drawdown analysis presented below
which focuses on short time scales, {\it i.e.}, days. We stress that figure 
\ref{wave} has been obtained without the need for any parametric adjustment, 
except for the choice of the scale in the wavelet. This analysis provides 
an extension of the outlier concept at larger time scales than the daily scale
used in the $\epsilon$-drawdown analysis presented below and hence quantifies
the existence of outliers on time scales much longer than those belonging to
crashes and runs. In figure \ref{djiawave}, we see that all the negative
residue outliers belongs to the Great Depression in the 1930ties, which 
not suprisingly is the only outlier on time scales longer than $\approx 4$ 
years in the last century. A more thorough outlier analysis of the relevant 
time scales is left for a future publication \cite{JSS}.

\subsection{Definitions of coarse-grained drawdowns and drawups}

In our previous analysis and identification of outliers in the major financial
 markets \cite{outl1,outl2}, drawdowns (drawups) were simply defined as a 
continuous decrease (increase) in the value of the price at the close of each
successive trading day (daily close). Hence, a drawdown (drawup) was terminated
by {\it any} increase (decrease) in the price no matter how small. However, 
this definition of drawdowns (drawups) is sensitive to noise, {\it i.e.}, to 
random uncorrelated as well as correlated fluctuations in the price. 
Simulations adding noise to the time series analysed have already showed that 
the distributions of drawdowns are robust to i.d.d. noise of ``reasonable'' 
magnitude \cite{outl2} (we will comment further on this below). We now 
investigate this question in more detail by filtering the data before the 
distribution of drawdowns (drawups) is calculated. 

There are two straightforward ways for 
defining such coarse-grained drawdowns (drawups). First, we may ignore 
increases (decreases) below a certain fixed magnitude (absolute or relative to 
the price). This is called price coarse-grained drawdowns or 
``$\epsilon$-drawdowns'', where $\epsilon$ is the minimum (relative or 
absolute) positive (negative) fluctuation which will terminate the drawdown 
(drawup). If $\epsilon=0$, we call the drawdown (drawup) ``pure''. For 
a non-zero $\epsilon$, we let 
the drawdown (drawup) continue if the amplitude of the fluctuation between 
consecutive daily closes is smaller than $\epsilon$. In the present paper, the 
emphasis will be on these $\epsilon$-drawdowns, where the $\epsilon$ used is 
the {\it relative} threshold or the minimum relative positive (negative) 
fluctuation which will terminate the drawdown (drawup). Second, we may ignore 
increases (decreases) that occur over a time horizon smaller than some horizon 
regardless of magnitude. This is called temporally coarse-grained drawdowns or 
``$\tau$-drawdowns'', where $\tau$ is the horizon. For $\tau=0$ this is 
equivalent to pure drawdowns. Specifically, when a movement in the reverse 
direction is identified, the drawdown (drawup) is nevertheless continued {\it 
if } the index within a horizon of $\tau$ days contains a continuation of the 
downward (upward) trend. This defines a $\tau$-drawdown and will not be 
discussed further in this paper. A thorough analysis using $\tau$-drawdowns 
will be left for a future publication \cite{tauhori}.

\subsection{Distribution of reference for drawdowns and drawup and 
methodology for qualifying an outlier}

Our previous analysis \cite{outl1,outl2} of the cumulative distributions of 
pure drawdowns in the U.S. stock market indexes (DJIA, SP500 and NASDAQ) as 
well in other western and emergent market indexes as well as individual stocks
and in the DM/US\$ exchange rate have shown that they are well-parameterized by
a minor generalisation of an exponential function called the stretched 
exponential
\be \label{stretched}
N(x) = Ae^{-bx^z}~,
\ee
where it is found empirically that $z\approx 0.8-1.1$. The case $z=1$ recovers 
an exponential distribution which is justified in the case where the returns 
are sequentially uncorrelated and the marginal distributions are exponential 
or super-exponentially distributed \footnote{that is, the convergence of the 
left and right tails of the cumulative distribution to $0$ and $1$ respectively
is no slower than exponential.}. Taking the logarithm of expression
(\ref{stretched}) is convenient and efficient for the numerical fits to 
the data and will be used in our subsequente analysis.

The specific
functional form of eq. (\ref{stretched}) has not particular importance
as long as it fits well the bulk of the distribution of drawdowns
and our goal is not to defend this parameterization more than any other one.
However, we need a good model of the bulk of the distribution of
drawdown in order to be able to ask the question about the existence
or abscence of outliers. We find expression (\ref{stretched})
convenient for this purpose as it is the 
simplest extension of the most natural null-hypothesis of an exponential 
function. However, an objective quantitative definition of an outlier based on 
is eq. (\ref{stretched} not straightforward. Hence, we will use the intuitively
very appealing, though not quite objective, definition that an outlier is a 
drawdown that does not fit in the continuation of the statistics obtained from 
the bulk of the distribution containing {\it at least} $95\%$ of the 
distribution. Thus, the detection of an outlier will boil down to the detection
of a significant deviation from the fit with (\ref{stretched}) in the tail of 
the distribution. We shall see that, in general, a ``break'' and/or ``gap'' in 
the distribution can easily be identified around a drawdown amplitude of 
$10-15\%$.  The label ``outlier'' is then reserved for drawdowns with a 
magnitude well above this ``break'' or ``gap''.

\subsection{Price coarse-graining algorithm}

The algorithm used for calculating price coarse-grained drawdowns (drawups) is 
defined as follows. We first identify a local maximum (minimum) in the price 
and then look for a continuation of the downward (upward) trend ignoring 
relative movements in the reverse direction smaller than $\epsilon$. 
Specifically, when a movement in the reverse direction is identified, the 
$\epsilon$-drawdown ($\epsilon$-drawup) is nevertheless continued {\it if } 
the relative magnitude of this reverse move is less than $\epsilon$. Those 
very few drawdowns (drawups) initiated by this algorithm which end up as 
drawups (drawdowns), {\it i.e.}, as the complementary quantity, are discarded. 

The amplitude $\epsilon$ of the coarse-graining is expressed in units of the 
volatility $\sigma$ defined by 
\ba \label{voleq}
\sigma^2=N^{-1}\Sigma_{i=1}^N [r_{i+1}-{\rm E}[r]]^2.
\ea 
where
$r_{i+1} \equiv \log p\lp t_{i+1}\rp - \log p\lp t_i \rp$ is the return
from time $i$ to $i+1$ and ${\rm E}[r]$ is its historical average.

As the large drawdowns contribute to $\sigma$, we can expect that $\epsilon$ 
should be taken smaller than $\sigma$ in order not to destroy the possible 
bursts of dependence we are looking for. We find below that the choices 
$\epsilon = \sigma/4$ to $\sigma/2$ give reasonable and robust results, 
improving on the definition of drawdowns with $\epsilon =0$. Since we know 
little about the nature of the noise present in financial data, we adopt the 
procedure of using systematically $\epsilon = 0, \sigma/4, \sigma/2$ and 
$\sigma$, which allows us to test for the robustness of the results. For larger
$\epsilon \geq \sigma$, the larger $\epsilon$ is, the closer to a Gaussian
distribution is the distribution of $\epsilon$-drawdowns, and thus the exponent
$z$ of (\ref{stretched}) obtained from a fit to the distribution of 
$\epsilon$-drawdowns increases above $1$. This is expected since, in the limit 
$\epsilon$ very large, the $\epsilon$-drawdowns are nothing but the returns 
over long stretches of price time series, which by the central limit theorem 
converge in distribution to the Gaussian law in the limit of long time 
intervals.

\section{$\epsilon$-drawdown outliers of foreign exchange markets} \label{fx}

\subsection{Historical introduction}

The foreign exchange market (FX) is by far the largest market in terms of 
volume. Of all possible pairs of currencies, we analyze the two pairs that 
account for more than $80\%$ of the total trading volume, namely the US\$ 
against the Deutch Mark (DM) and against the Yen acknowledging the leading role
of the US\$, see below. 

A short account of the history of the FX market after World War II is 
appropriate here. Under the Bretton Woods Treaty (New Hampshire, 1944), the 
price of gold was set at \$35 per ounce. Only the dollar had a direct gold 
parity. The gold content of the other currencies was established only 
indirectly, by means of a fixed parity with the dollar. The fixed parities were
intended to be permanent: a pound sterling was to be worth $2.80$ against the 
dollar, a dollar was to be 625 lire, 360 yen, and so on. Fluctuations were to 
be confined to a plus or minus $1\%$ band. These fixed parities provided the 
inestimable benefit of price predictability: they meant that international 
traders would know that dollar bills of exchange used in international 
import-export transactions could be expected to vary no more than plus or minus
$1\%$ over their three month or six month lifetimes. The United States was 
expected to buy and sell gold in settlement of international transactions. If 
the United States ran a payments deficit with the rest of the world, then the 
rest of the world might ask for settlement in gold at the rate of \$35 per 
ounce. Due to international pressure, on December 17, 1971, the dollar was 
formally and officially devalued with respect to gold by $8.57\%$, while 
Germany upvalued the mark by $13.57\%$ and Japan the yen by $16.9\%$. Sweden 
and Italy each devalued by $1\%$ with respect to gold. New bands of fluctuation
of $\pm 2.25\%$ were applied to all countries. However,  within less than 
fourteen months and despite a second devaluation of the dollar, this agreement 
was swept away by international money flows and, by 1973, the Bretton Wood 
system collapsed completely. Hence, the turmoil seen on the FX markets in the 
first half of 1973 is most likely related to the collapse of the Bretton Wood 
system and will be described and classified as ``BW shocks''.

\subsection{The DM and the Yen against the US\$}

\begin{figure}
\begin{center}
\parbox[l]{8.5cm}{
\epsfig{file=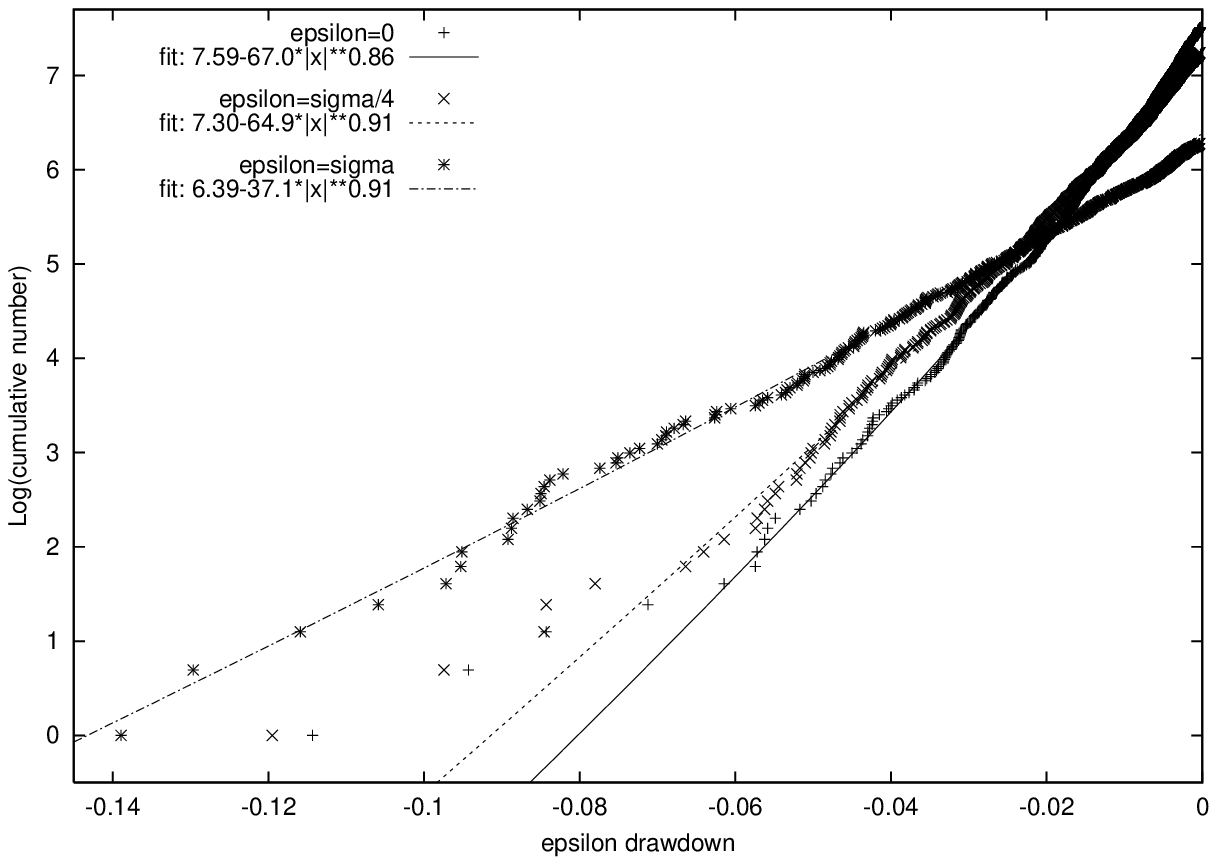,height=8cm,width=8.5cm}
\caption{\label{usdmepsdd} Logarithm of the cumulative distribution of 
$\epsilon$-drawdowns in the DM/US\$ exchange rate using an $\epsilon$ of $0$, 
$\sigma /4$ and $\sigma$,where $\sigma=0.0065$ has been obtained from the data.
}}
\hspace{5mm}
\parbox[r]{8.5cm}{
\epsfig{file=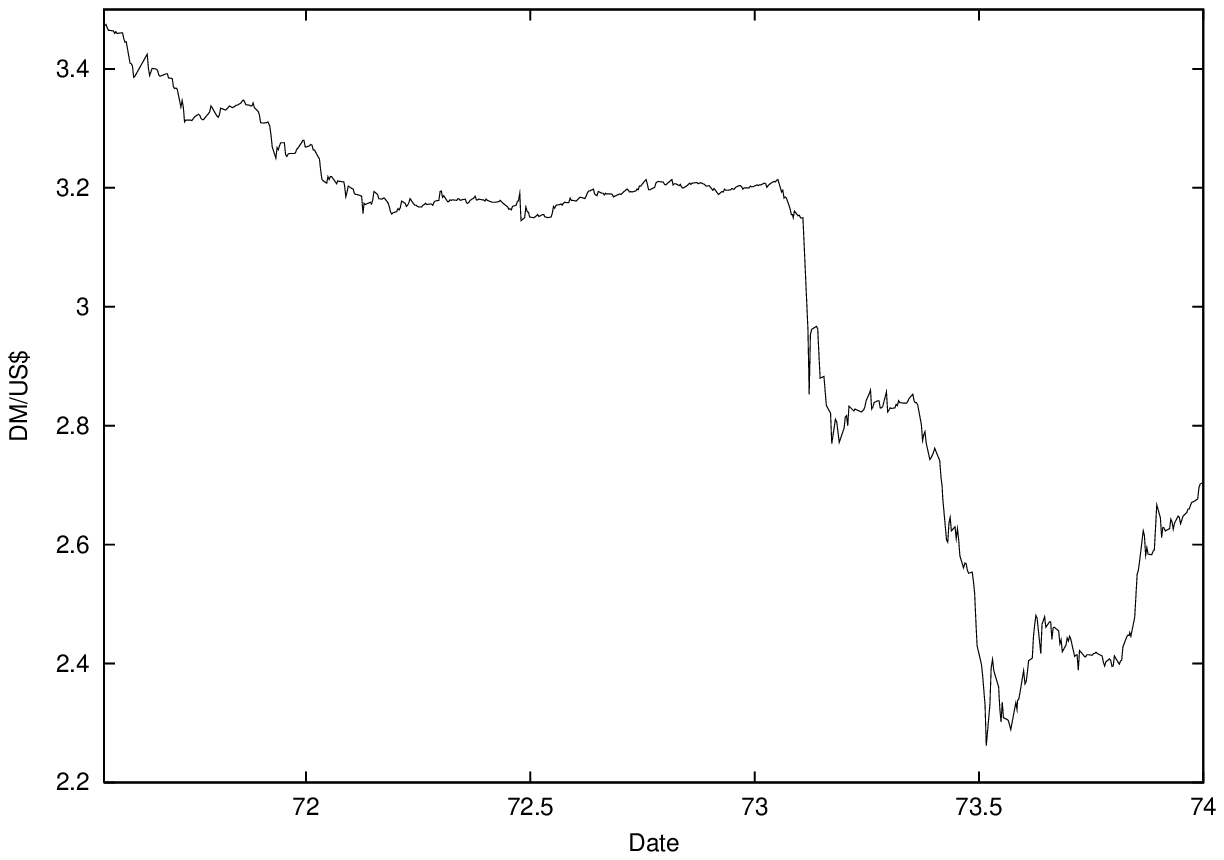=,height=8cm,width=8.5cm}
\caption{\label{usdm73}DM/US\$. The exchange rate between DM and US\$ in the 
period 1971.5 to 1974. The two BW crashes of the US\$ in 1973.1 and 1973.5 are 
easily identified by eye.}}
\vspace{1.5cm}
\parbox[l]{8.5cm}{
\epsfig{file=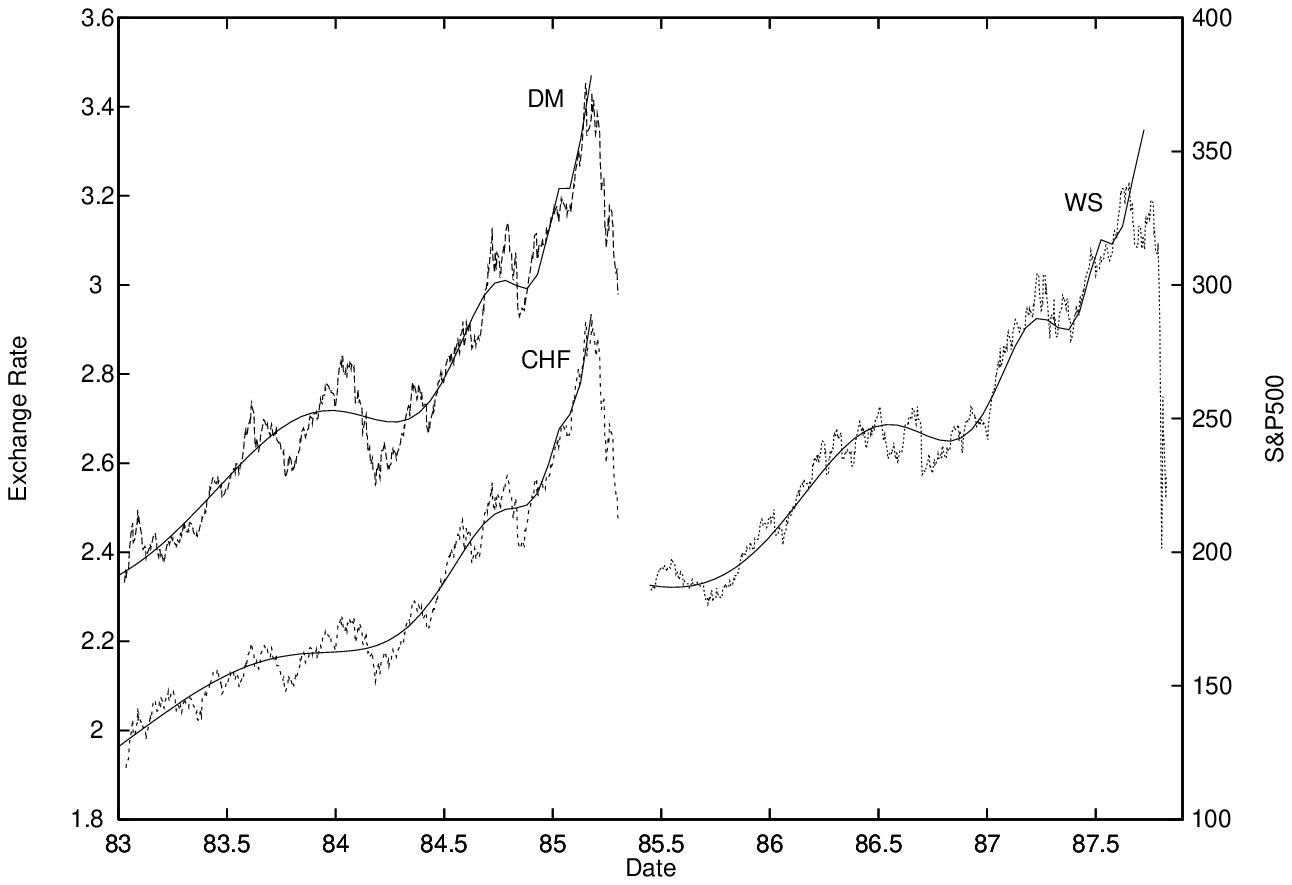,height=8cm,width=8.5cm}
\caption{\label{usdm85}. The US\$ bubble ending in 1985. For the DM the fit 
is eq. (\ref{lpeq}) where  $A \approx  3.88  $, $ B\approx  -1.2 $, $C \approx
0.08 $, $z \approx 0.28 $, $ t_c \approx   85.20  $, $ \phi \approx   -1.2 $, 
$\omega \approx 6.0$. For illustration purposes, we also show the fit for the 
CHF/US\$ exchange rate and that of the SP500 prior to the Oct. 1987 crash.
}}
\hspace{5mm}
\parbox[r]{8.5cm}{
\epsfig{file=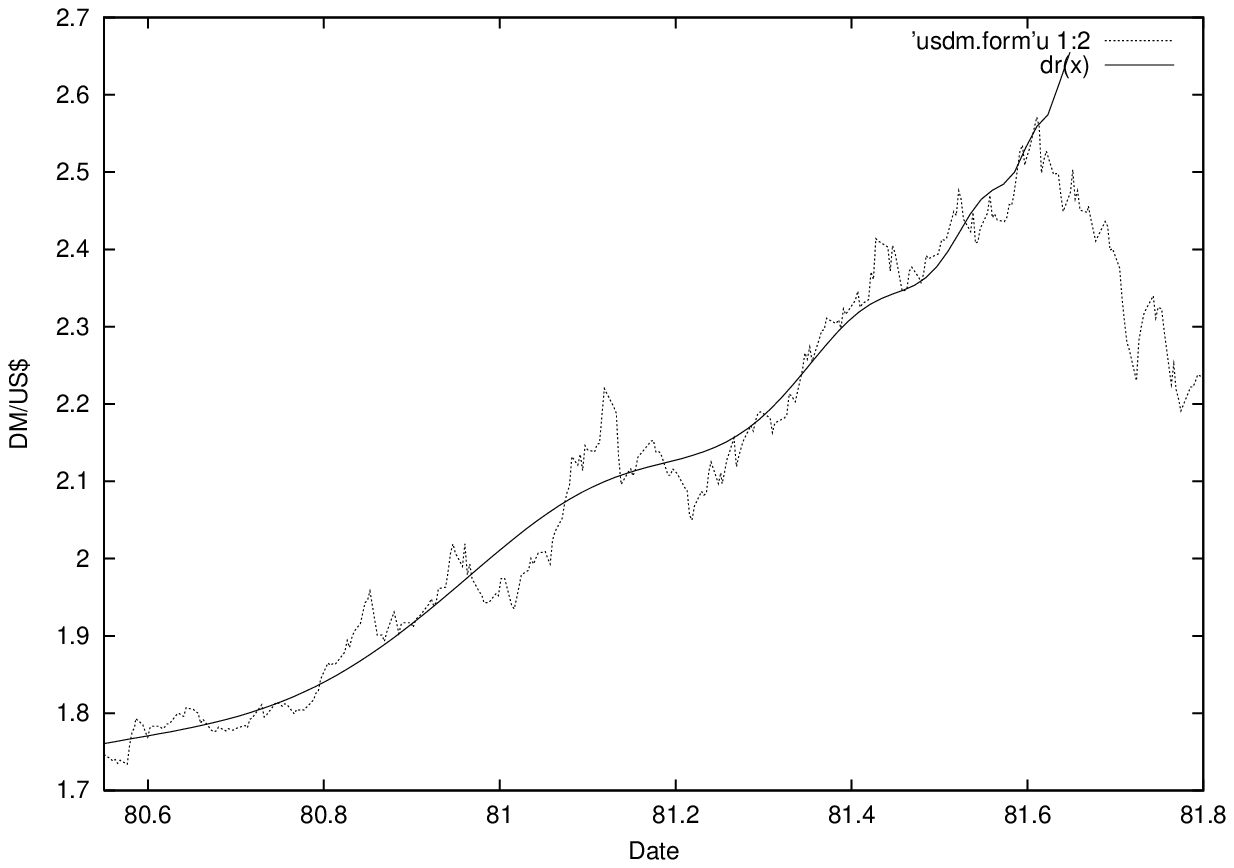,height=8cm,width=8.5cm}
\caption{\label{usdm81}The DM/US\$ bubble ending in 1981. The fit is eq. 
(\ref{lpeq}) where  $A \approx  2.70  $, $ B\approx  -0.90 $, $C \approx   
0.04 $, $z \approx 0.58 $, $ t_c \approx   81.65  $, $ \phi \approx   -1.2 $, 
$ \omega \approx   7.5$ 
}}
\end{center}
\end{figure}

\begin{figure}
\begin{center}
\parbox[l]{8.5cm}{
\epsfig{file=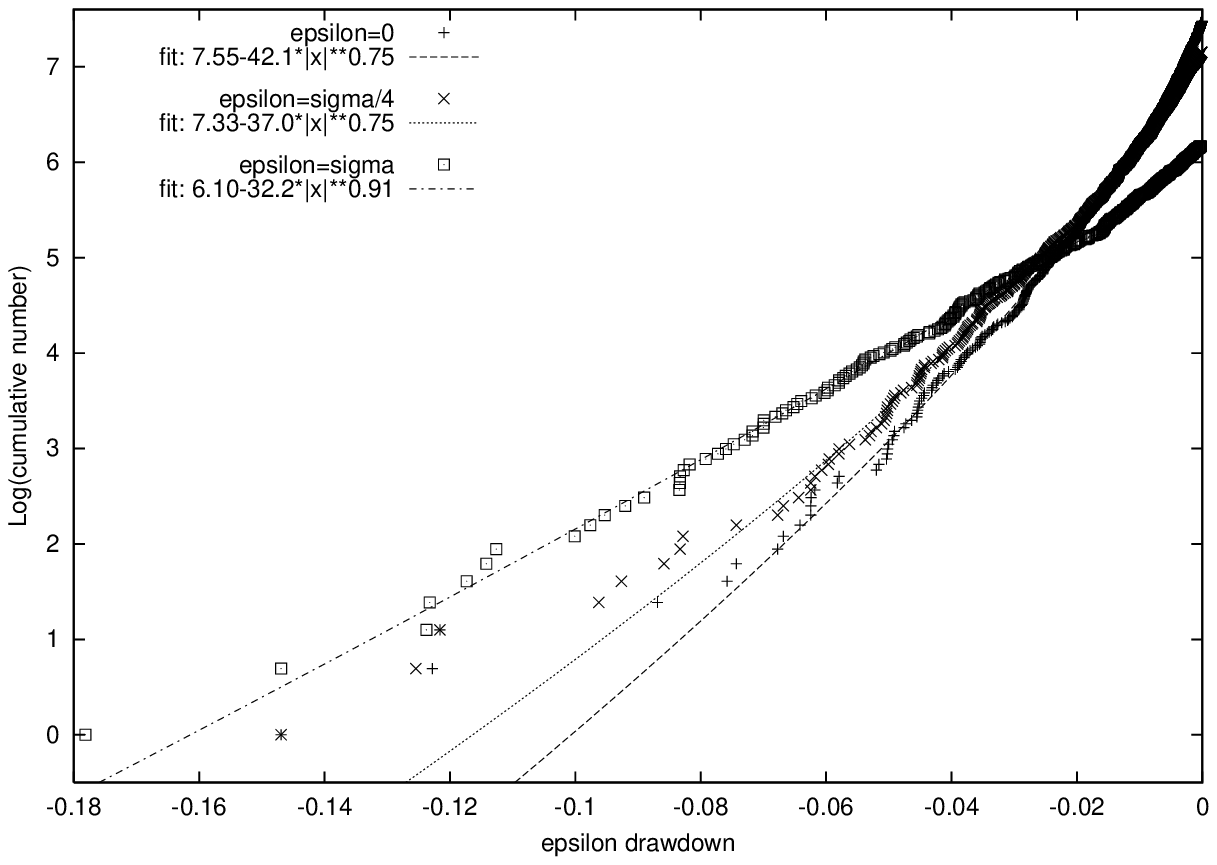,height=8cm,width=8.5cm}
\caption{\label{usyenepsdd}Yen/US\$. Logarithm of the cumulative distribution 
of $\epsilon$-drawdowns in the Yen/US\$ exchange rate using an $\epsilon$ of 
$0$, $\sigma /4$ and $\sigma$,where $\sigma=0.0065$ has been obtained from the 
data.
}}
\hspace{5mm}
\parbox[r]{8.5cm}{
\epsfig{file=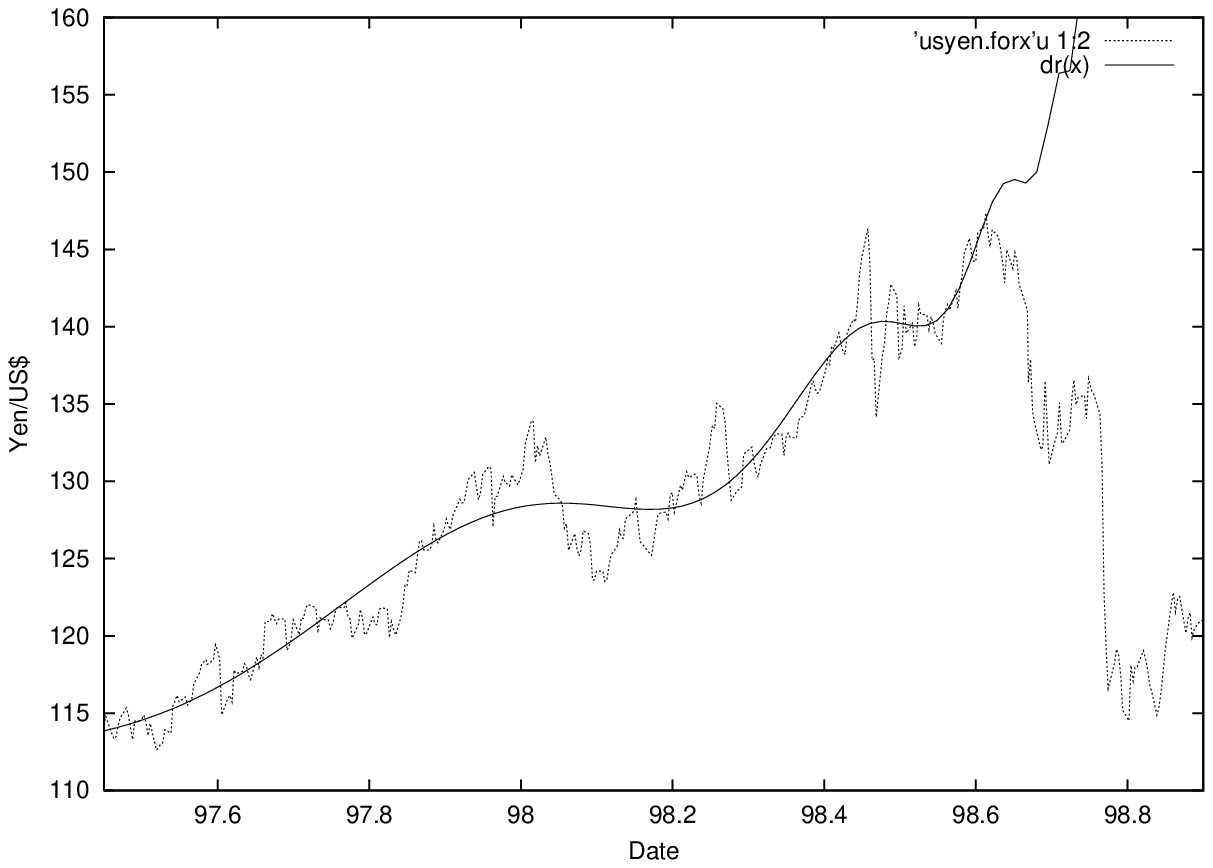,height=8cm,width=8.5cm}
\caption{\label{usyen98}.The Yen/US\$ bubble ending in 1998. The fit is eq. 
(\ref{lpeq}) where  $A \approx  182  $, $ B\approx  -61 $, $C \approx   2.9 $, 
$z \approx   0.27 $, $ t_c \approx   98.76  $, $ \phi \approx   1.7 $, 
$ \omega \approx   6.8$ 
}}
\vspace{1.5cm}
\parbox[l]{8.5cm}{
\epsfig{file=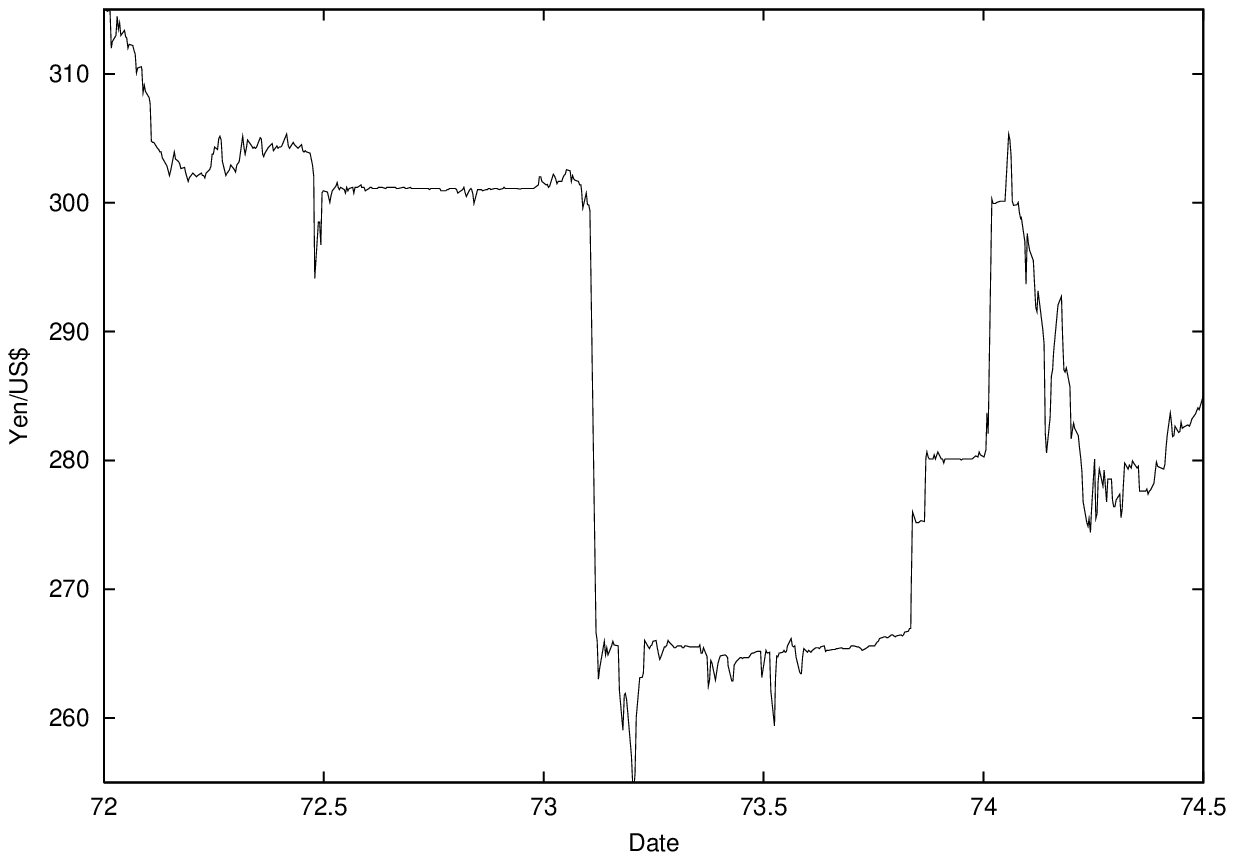,height=8cm,width=8.5cm}
\caption{\label{usyen73} The exchange rate between Yen and US\$ in the period 
1972 to 1974.5. The BW crash of the US\$ in 1973.1 is easily identified by eye.
}}
\hspace{5mm}
\parbox[r]{8.5cm}{
\epsfig{file=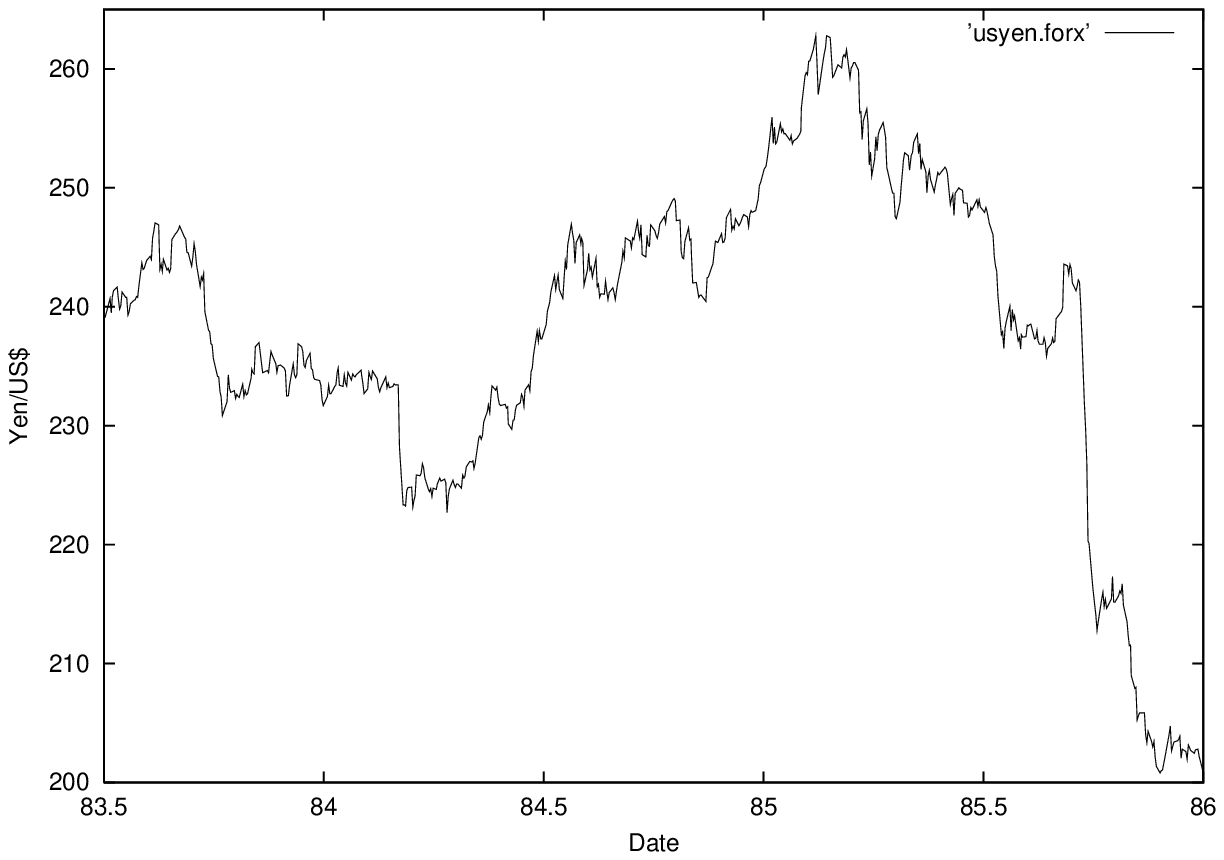=,height=8cm,width=8.5cm}
\caption{\label{usyen85} The Yen/US\$ exchange rate in the period 1983.5 to 
1986. The crash of the US\$ in 1985.7.1 is easily identified by eye. 
}}
\end{center}
\end{figure}

The two data sets analysed here spans from January, 4, 1971 to 
May, 19, 1999. The last date ensures the absence of any possible distortion 
due to the conversion of the DM to the Euro, which is not analyzed due to its 
still short existence. Figures \ref{usdmepsdd} and \ref{usyenepsdd} present 
the cumulative distribution of $\epsilon$-drawdowns of the DM/US\$ and of the 
Yen/US\$ respectively, with $\epsilon$ taking the values $0$, $\sigma /4$ and 
$\sigma$, where $\sigma$ is the volatility, defined as the standard deviation 
of the daily log-return eq. (\ref{voleq}). For $\epsilon=0$ and $\sigma/4$, 
respectively four and three $\epsilon$-drawdowns lie above a naive visual 
extrapolation of the rest of the bulk of the cumulative distribution. The break
in the cumulative distribution lies approximately around an $\epsilon$-drawdown
of $6.5\%$. In contrast, for $\epsilon = \sigma$, all $\epsilon$-drawdowns seem
to belong to the same distribution as to be expected.

To make this more 
quantitative, we fit the three cumulative distributions over the whole range of
$\epsilon$-drawdowns excluding those five (for $\epsilon=0$) and four (for 
$\epsilon=\sigma/4$) presumed outliers to the stretched exponential function 
(\ref{stretched}). The fits as well as the parameters of the fits are shown in 
figures \ref{usdmepsdd} and \ref{usyenepsdd}. These fits of the cumulative 
distributions of $\epsilon$-drawdowns for $\epsilon=0$ and $\sigma /4$ confirm 
the visual impression that there are four outliers, the fifth one being only 
present for $\epsilon=0$ is less clear-cut. The dates, sizes and durations (in 
number of trading days) of these outliers are given in Tables \ref{outlusdm} 
and \ref{outlusyen}.

\begin{table}[]
\begin{center}
\begin{tabular}{|c|c|c|c|c|c|c|c|c|c|}\hline
$\epsilon$& Date     & Size          & Duration & Class   & $\epsilon$ & Date            & Size      & Duration & Class   \\ \hline
$0$       & 1973.485 & $11.4\%$      & 8 days   & BW Shock& $\sigma/4$ & 1973.469        & $11.9\%$  & 12 days  & BW Shock\\\hline
$0$       & 1973.108 & $9.4\%$       & 2 days   & BW Shock  & $\sigma/4$ & 1973.089        & $9.7\%$   & 7 days   & BW Shock\\ \hline
$0$       & 1985.716$\star$& $8.4\%$& 6 days   & Bubble  & $\sigma/4$ & 1985.716$\star$& $8.4\%$   & 6 days   & Bubble  \\ \hline
$0$       & 1981.697$\dagger$&$7.1\%$&6 days& Bubble  & $\sigma/4$ & 1981.688$\dagger$ &$8.4\%$   & 9 days   & Bubble  \\ \hline
$0$       & 1995.169 & $6.1\%$       & 4 days   & No Outlier   & $\sigma/4$ & 1981.103        & $7.8\%$   &13 days   & No Outlier       \\ \hline
\end{tabular}
\caption{\label{outlusdm} List and properties of outliers identified in the 
distribution of $\epsilon$-drawdowns for $\epsilon=0$ (left part)
and $\epsilon=\sigma/4$ (right part) of the DM/US\$ exchange data. The outliers
are ranked by decreasing amplitudes. The term ``shock'' refers to an outlier 
which has been triggered by an event exogenous to the market. BW Shock refers 
to the the collapse of the Bretton Wood system. The term ``bubble'' embodies 
the idea that the corresponding outlier corresponds to a crash ending a 
speculative endogenous LPPS bubble. The event with a $\star$ has already been 
studied in previous publications. Bubbles identified by the outlier analysis 
presented here, {\it i.e.}, outliers which have prior log-periodic bubbles not
identified in previous works will be marked by $\dagger$.}
\end{center}
\end{table} 

Having identified four or five outliers, the next question is whether we can 
classify them and pinpoint their origin. For this, the time series 
preceding and including the four outliers of the US\$ in DM are in figure 
\ref{usdm73} for the two events of 1973.5 and 1973.1, in figure \ref{usdm85} 
for the event of 1985.7 and in figure \ref{usdm81} for the event of 1981.7. In 
addition, the US\$ expressed in Swiss Franc and the S\&P500 index are also 
shown in figure \ref{usdm85}, to stress the similarity between the three time 
series. The three time series of figure \ref{usdm85} and the time series of 
figure  \ref{usdm81} before the $\epsilon$-drawdown outliers are all 
preceding by a pattern which has been found previously to be characteristic of 
speculative LPPS bubbles \cite{JSL1999,JS1999,JLS2000,SJ2001}. The fits with 
eq. (\ref{lpeq}) of the three time series of figure \ref{usdm85} and the time 
series of figure  \ref{usdm81} preceding the $\epsilon$-drawdown outliers are 
shown as smooth continuous lines and the parameters obtained from the fits are 
given in the figure captions. 

According to the fits presented in figures \ref{usdm85} and \ref{usdm81}, the 
two FX events of the DM/US\$ of 1985.7 and 1981.7 qualify as the end of a 
speculative unsustainable bubble that became unstable and burst into an 
endogenous crash. The 1985.7 event is particularly noteworthy as the dollar 
crashed against most other currencies after reaching record-heights under the 
Reagan administration. In contrast, the two events of 1973.1 and 1973.5 shown 
in figure \ref{usdm73} have occurred without any of the qualifying structures 
described previously for the other events quantified by expression 
(\ref{lpeq}). Recalling the historical account of the FX market given above, it
is reasonable to attribute these two events to the collapse of the Bretton Wood
system. Last, the two events ranked fifth for $\epsilon=0$ and $\epsilon=\sigma
/4$, respectively do not classify as outliers but as ``normal'' events since 
they belong to the bulk of the distribution and have been included for 
illustration purposes only. For $\epsilon=\sigma/4$, we have the 1995.2 event 
which was caused by the crises in Mexico and subsequent devaluation of the
Peso\footnote{The US\$ was vulnerable to this crisis because of the NAFTA 
agreement between USA, Canada and Mexico, a fact well-illustrated by the 
bail-out by the Clinton administration that followed the Mexican crises.}. As 
for the event of 1981.1, the authors have not found any qualifying historical 
event.

\begin{table}[]
\begin{center}
\begin{tabular}{|c|c|c|c|c|c|c|c|c|c|}\hline
$\epsilon$ & Date            & Size      & Duration & Class        &  $\epsilon$ & Date     & Size       & Duration& Class   \\\hline
$0$        & 1998.749$\star$& $14.7\%$  & 7 days   & Bubble       & $\sigma/4$ & 1998.749$\star$ & $14.7\%$& 7 days &Bubble\\\hline
$0$        & 1973.103        & $12.3\%$  & 5 days   & BW Shock     & $\sigma/4$ & 1973.097 & $12.5\%$   & 7  days & BW Shock \\\hline
$0$        & 1985.716        & $12.2\%$  & 11 days  & Shock &$\sigma/4$ & 1985.716 & $12.2\%$ & 11 days & Shock \\\hline
$0$        & 1990.756        & $8.7\%$   &12 days   &  Shock  & $\sigma/4$ & 1990.740 & $9.6\%$ & 15 days & Shock  \\\hline
\end{tabular}
\caption{\label{outlusyen} Same as table \ref{outlusdm} but for the Yen/US\$ 
exchange data. The event with a $\star$ has already been studied in previous 
publications.} 
\end{center}
\end{table} 

Similar results are found for the Yen/US\$ exchange data, for which for 
$\epsilon$-drawdown outliers are identified and listed in table 
\ref{outlusyen}. The event ranked fourth is borderline 
but has again been included for illustration purposes. Figure \ref{usyen98} 
shows that the largest outlier qualifies as an endogenous crash terminating 
a speculative bubble phase.
It is difficult not to associate this bubble with the prolonged bubbles in 
western stock markets fueled by heavy investments in the emergent market of 
Russia and Eastern Europe, which were followed by a general crash triggered by 
the default of the ruble and by severe economic problems in Russia. This 1998 
crash ending a speculative bubble have previously been studied in 
\cite{JSL1999}
and we refer to that paper for further details. The evolution of the Yen/US\$ 
exchange rate preceding its second largest $\epsilon$-drawdown outlier is 
shown 
in figure \ref{usyen73} and does not exhibit the qualifying structure 
described 
by expression (\ref{lpeq}). Recalling the historical account of the FX market 
given above, it is reasonable to attribute this event to the collapse of the 
Bretton Wood system. Figure \ref{usyen85} shows the Yen/US\$ exchange rate 
preceding the 1985.7 outlier. Here, the case is ambiguous, since the date of 
the largest drawdown is approximately 6 month after the date of the end of 
the 
rising trend identified as the date of the maximum. As with the DM, we can 
observe a signature of an increasing trend starting in early 1984 and ending 
in 
early 1985, however rather smoothly as if a speculative bubble has been 
aborted 
in its main course before its full ripening. The crash does not occur until 
1985.7, as we also saw in table \ref{outlusdm}, which suggests that it was 
triggered by the crash of the US\$ against the other main currency, namely the 
DM. This event is thus intermediate between a genuine endogenous bubble 
destabilized into a crash and an exogenous event. We refer to it as ``Shock'' 
in table \ref{outlusyen} as no LPPS can be established. The same scenario can 
be used to explain the borderline event dated 1990.765. It is causally linked 
with two large drawdowns in the Nikkei dated 1990.622 and 1990.723, see table 
\ref{outlnik}, and one in the Nasdaq dated 1990.622, see table \ref{outlnas}. 
Hence, it is quite likely an exogenous event caused by the collapse of the 
Japanese stock market and should be classified as a ``Shock'' in table 
\ref{outlusyen}.

In sum, we have identified three clear-cut cases of endogenous speculative 
bubbles leading to an $\epsilon$-drawdown outlier (the 1981.7 and 1985.7 
events on the DM/US\$ exchange rate and the 1998.7 event on the Yen/US\$ 
exchange rate). The three events in 1973.1 and 1973.5 can be associated with 
the breakdown of the Bretton Wood system (two for the DM/US\$ exchange rate
and one for the Yen/US\$ exchange rate) and thus qualify as exogenous. The 
1985.7 Yen/US\$ event can be associated with the collapse of the US\$ against 
the other major currency DM in 1985.7. The 1990.8 borderline Yen/US\$ event 
likewise can be associated with the collapse of the Japanese real-estate market
and the subsequent events on the Japanese stock market in 1990.6 and 1990.7 and
on NASDAQ in 1990.6. Hence, the seven clear outliers found on the FX market 
(DM and Yen against US\$) contains three LPPS bubbles and four shocks of which 
three could be linked with the collapse of the Bretton-Woods system in 1973 and
one (borderline) with the collapse of the Japanese real-estate and two outliers
on the Nikkei stock market as well as one on the NASDAQ. Last, the 1985.7 
Yen/US\$ event can be linked with the collapse of the US\$ against the DM and 
the CHF. This means that the majority (namely four) of outliers found here on 
the FX market can be classified as exogenous, whereas three have prior LPPS
bubbles and should be classified as endogenous. As we shall see in the next 
section, this situation is reversed for the stock markets, where the majority
of outliers are of endogenous origin.

\section{$\epsilon$-drawdown outliers of major stock markets} \label{sm}

In the previous section on the FX markets, we found that the selection of 
$\epsilon$-drawdown outliers is robust with respect to different degrees of 
filtering quantified by the amplitude of $\epsilon$ in the range 
$\epsilon=0-\sigma/2$. As we shall see below, the situation is not quite as 
simple for the major stock markets, where not only the rank of say the largest 
five events is permutated by changing $\epsilon$ but the duration and/or size 
of the $\epsilon$-drawdowns may be drastically modified. Two good illustrations
are found in table \ref{outldj}: 1) the crash caused by the outbreak of 
World-War I (WWI) is prolonged from 4 days to 64 days using a non-zero 
$\epsilon$, however not adding significantly to its size; 2) a $11.4\%$ drop 
over 2 days caused by the Nazi invasion of France, Belgium, Luxembourg and the 
Netherlands on May, 10th 1940 is amplified to a $23.7\%$ drawdown over 44 days.
This suggests the need for more sophisticated measures than pure size to 
quantify drawdowns. An alternative and perhaps more appropriate ordering could 
be obtained by defining a combined measure of drawdown size and ``drawdown 
velocity'' equal to drawdown size divided by crash duration. This will not be 
pursued further here.

\subsection{The U.S. markets} \label{usstock}

\begin{figure}
\begin{center}
\parbox[l]{8.5cm}{
\epsfig{file=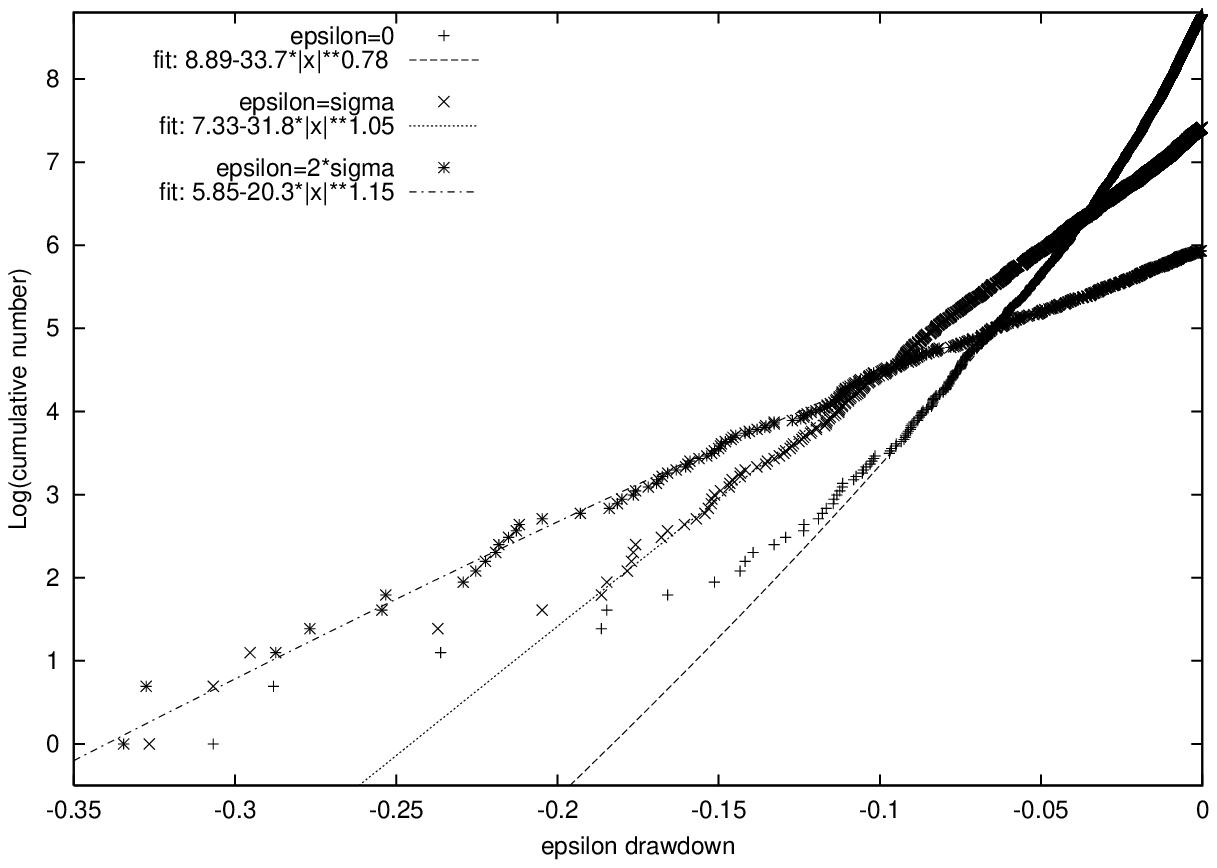,height=8cm,width=8.5cm}
\caption{\label{djepsdd} Logarithm of the cumulative distribution of 
$\epsilon$-drawdowns in the DJIA using an $\epsilon$ of $0$, $\sigma $ and 
$2\sigma $, where $\sigma=0.011$ has been obtained from the data.}}
\hspace{5mm}
\parbox[r]{8.5cm}{
\epsfig{file=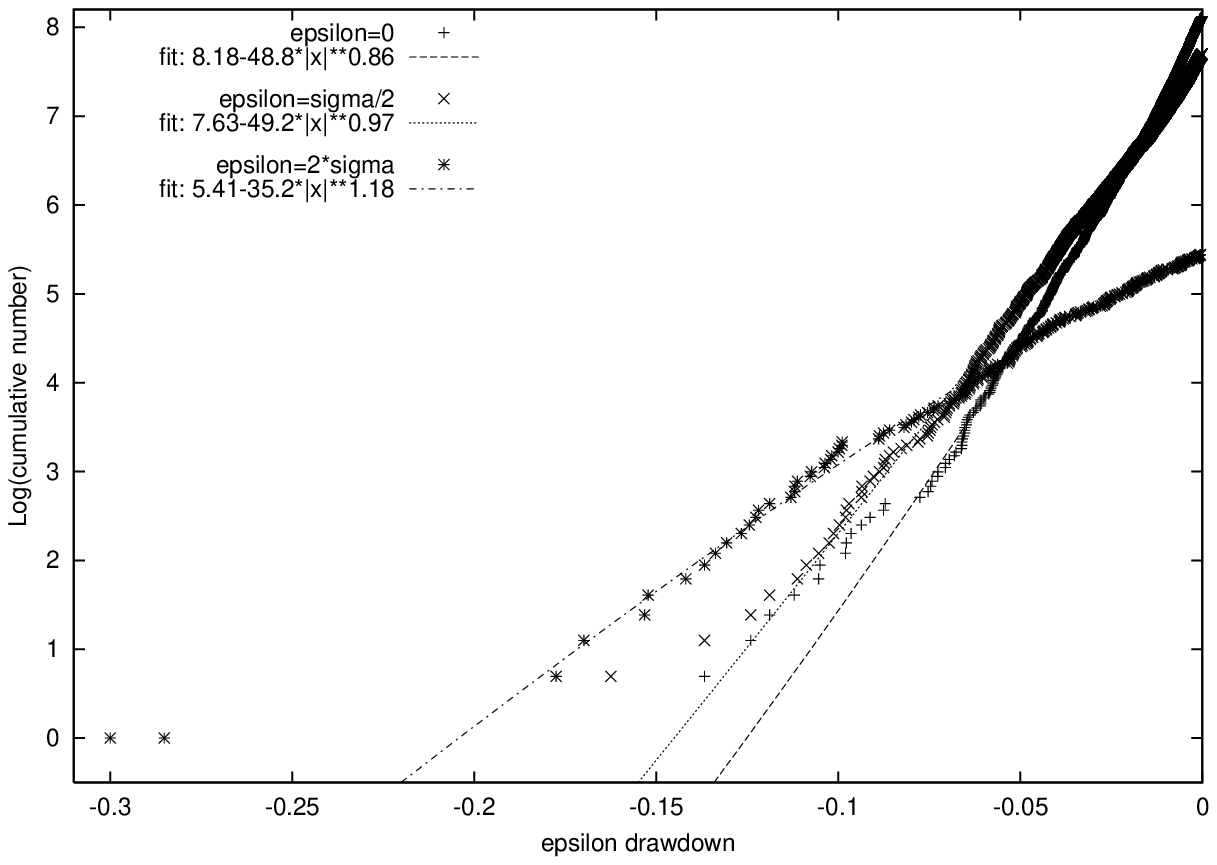,height=8cm,width=8.5cm}
\caption{\label{spepsdd} Logarithm of the cumulative distribution of 
$\epsilon$-drawdowns in the SP500 using an $\epsilon$ of $0$, $\sigma /2$ and 
$2\sigma $, where $\sigma=0.009$ has been obtained from the data.}}
\vspace{1.5cm}
\parbox[l]{8.5cm}{
\epsfig{file=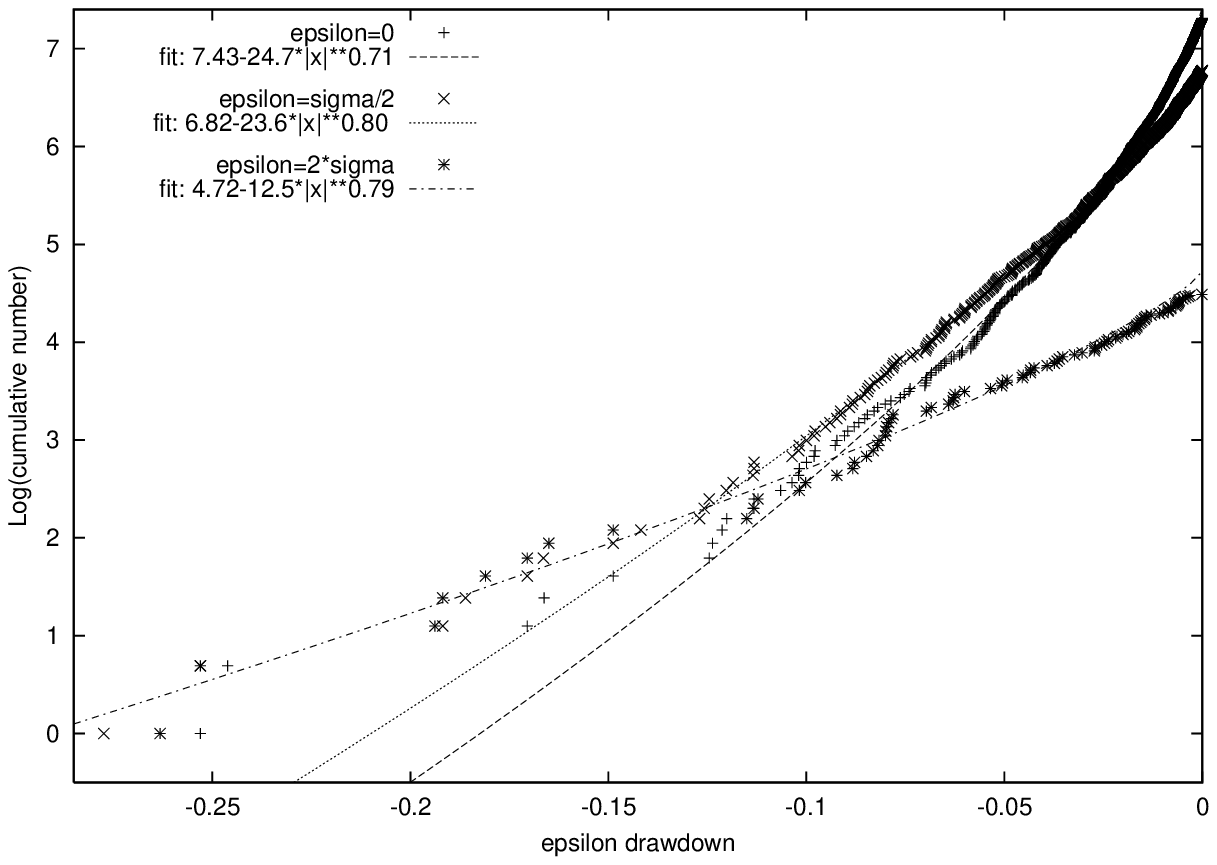,height=8cm,width=8.5cm}
\caption{\label{nasepsdd}  Logarithm of the cumulative distribution of 
$\epsilon$-drawdowns in the NASDAQ using an $\epsilon$ of $0$, $\sigma /2$ and 
$2\sigma $, where $\sigma=0.010$ has been obtained from the data.}}
\hspace{5mm}
\parbox[r]{8.5cm}{
\epsfig{file=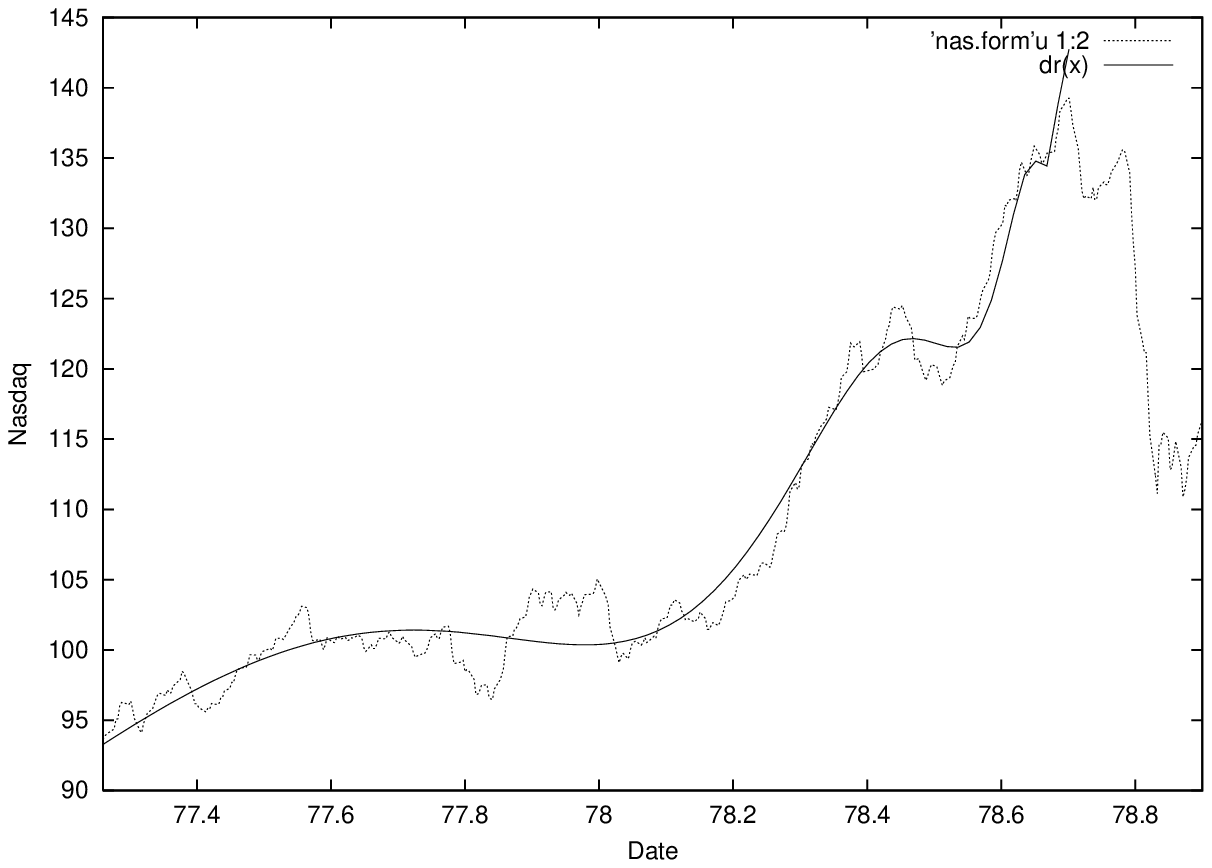,height=8cm,width=8.5cm}
\caption{\label{nas1978} The NASDAQ bubble ending in 1978. The fit is eq. 
(\ref{lpeq}) where  $A \approx  154  $, $ B\approx  -57 $, $C \approx   5.8 $, 
$z \approx 0.35 $, $ t_c \approx   78.71  $, $ \phi \approx   0.8 $, $ \omega 
\approx  4.5$.
}}
\end{center}
\end{figure}

Figures \ref{djepsdd}, \ref{spepsdd} and \ref{nasepsdd} show the 
$\epsilon$-drawdown distributions for the DJIA, SP500 and NASDAQ index, 
respectively, obtained by the same analysis as for the FX market. The time 
series extend from 01/01 1900 to 17/07 2000 for the DJIA, from 29/11 1940 to 
17/07 2000 for SP500 and from 05/02 1971 to 17/07 2000 for the NASDAQ.

Tables \ref{outldj}, \ref{outlsp} and \ref{outlnas} show the four, five and six
largest events for the DJIA, SP500 and NASDAQ respectively. The three largest
$\epsilon$-drawdown outliers both for $\epsilon=0$ and $\epsilon=\sigma$ in 
table \ref{outldj} are associated with three well-known crashes of the past 
century. The fourth ranked outlier for $\epsilon=0$ belongs to the Great 
Depression while the fourth ranked outlier for $\epsilon=\sigma$ is a drawdown 
related to WWII, specifically the Nazi invasion of France, Belgium, Luxembourg 
and the Netherlands on May 10th 1940.

The 1987.8 and 1929.8 outliers are nothing but the famous crashes of Oct. 1929 
and 1987, already analyzed in depth in \cite{SJB,SJ1997}. The price
time series preceding these two crashes have been fitted with (\ref{lpeq}) over
more than two years of data and over almost 8 years with a simple extension 
based on a second-order expansion of the mathematical theory of critical 
crashes
\cite{SJ1997}. These two crashes thus qualify as endogenous events following 
self-organized speculative LPPS bubbles.

The 1914.5, 1933.5 and 1940.3 events are characterized by preceding time series
that can absolutely not be represented by formula (\ref{lpeq}). According to 
our classification, they are exogenous. Indeed, they are associated with 
external shocks, respectively the outbreak of WWI, the political maneuvering of
president F. D. Roosevelt \footnote{Roosevelt New Deal policy included the 
passing of the Securities Acts of 1933 and 1934 as well as going of the Gold 
standard in 1933.} upsetting the financial markets and the Nazi invasion of 
France, Belgium, Luxembourg and the Netherlands on May 10th 1940.

\begin{table}[]
\begin{center}
\begin{tabular}{|c|c|c|c|c|c|c|c|c|c|}\hline
$\epsilon$ & Date     & Size       & Duration &  Class &$\epsilon$ & Date     & Size     & Duration  &  Class\\ \hline
$0$        & 1987.786$\star$ & $30.7\%$   & 4 days  & Bubble & $\sigma$ & 1914.374 & $32.7\%$   & 64 days & Shock  \\ \hline
$0$        & 1914.579 & $28.8\%$   & 2 days    & Shock & $\sigma$ & 1987.786$\star$ & $30.7\%$   & 4 days  & Bubble   \\ \hline
$0$        & 1929.818$\star$ & $23.6\%$   & 3 days   &  Bubble & $\sigma$ & 1929.810$\star$ & $29.5\%$   & 6 days & Bubble\\ \hline
$0$        & 1933.549 & $18.6\%$   & 4 days   & Depression & $\sigma$ & 1940.261 & $23.7\%$   & 44days  & Shock  \\ \hline
\end{tabular}
\caption{\label{outldj} Same as table \ref{outlusdm} but for the DJIA stock 
market index. The events with a $\star$ have already been studied in previous 
publications.}  
\end{center}
\end{table} 

Table \ref{outlsp} gives the list of the five largest $\epsilon$-drawdowns in 
the SP500 index since 1940. All outliers except the 1974.7 event have price 
time series preceding them which can be fitted well with formula (\ref{lpeq}) 
and thus qualify as endogenous crashes. All these events are marked with a 
$\star$ to remind us that they have already been identified and analyzed in 
previous studies. The 1987.7 and 1987.8 events are two big drops associated 
with the crash of October 1987. The 1998.6 event is the crash associated with 
the ruble crisis and Russian default which was already discussed in our above 
analysis of the FX and has been analyzed specifically in \cite{JS1999}. The 
1962.3 event is the (rather slow) crash ending the "tronic" boom of the early 
1960s and was discovered in a blind search \cite{thesis,JLS2000}. The event 
1946.6 is also preceded by a log-periodic power law structure described by 
formula (\ref{lpeq}) and has been analyzed in \cite{SJ2001}. \cite{SJ2001}
has in addition identified a strong accelerating log-periodic signal in the 
price time series ending in 1937 in a slow crash, which is not detected by our
selection of $\epsilon$-drawdown outliers.

The  1974.7 event is a drawdown of $11.2\%$ which does not qualify as 
endogenous as it can not be fitted by formula (\ref{lpeq}). Actually, it is 
associated with a well-known external shock, namely the political turmoil 
caused by the resignation and the pardoning of president R. Nixon on August 
8th and September 8th 1974. 

\begin{table}[]
\begin{center}
\begin{tabular}{|c|c|c|c|c|c|c|c|c|c|}\hline
$\epsilon$ & Date             & Size       & Duration & Class & $\epsilon$  & Date             & Size       & Duration& Class\\\hline
$0$        & 1987.784$\star$ & $28.5\%$   & 4 days & Bubble  &  $\sigma/2$ & 1987.784$\star$ & $28.5\%$   & 4 days  & Bubble\\\hline
$0$        & 1962.370$\star$ & $13.7\%$   & 9 days & Bubble  &  $\sigma/2$ & 1946.636$\star$ & $16.2\%$   & 9 days  & Bubble\\\hline
$0$        & 1998.649$\star$ & $12.4\%$   & 4 days & Bubble  &  $\sigma/2$ & 1962.370$\star$ & $13.7\%$   & 9 days  & Bubble\\\hline
$0$        & 1987.805$\star$ & $11.8\%$   & 3 days & Bubble  &  $\sigma/2$ & 1998.649$\star$ & $12.4\%$   & 4 days  & Bubble\\\hline
$0$        & 1974.721         & $11.2\%$   & 9 days & Shock   &  $\sigma/2$ & 1987.805$\star$ & $11.9\%$   & 3 days  & Bubble\\\hline
\end{tabular}
\caption{\label{outlsp} Same as table \ref{outlusdm} but for the SP500 stock 
market index. The events with a $\star$ have already been studied in previous 
publications.}  
\end{center}
\end{table}

Table \ref{outlnas} gives the characteristics of the six largest 
$\epsilon$-drawdowns found in the NASDAQ time series. As with the SP500 index, 
all events but one (the 1990.6 event) have preceding price time series 
inflating as in a speculative bubble that can be very well fitted by formula 
(\ref{lpeq}) and qualify as endogenous. The crashes of 1987 and 1998 were also 
identified by the analysis of the DJIA and SP500 and are hence classified as 
``Bubbles''. The other crashes listed and preceded by a log-periodic power 
law bubble are the crash of April 2000  (of which we are still suffering from
its aftermath \cite{predsz}) 
which was analyzed in depth in \cite{JS2000} and the crash of Oct. 1978 
(unpublished until now) and shown in figure \ref{nas1978}. The value of 
$\omega=4.5$ found in the log-periodic fit of the speculative bubble preceding 
it is rather small compared with previous values in the range given by 
our previous results braketed
in (\ref{zomegaval}) but still within two standard deviations from the mean.

\begin{figure}
\begin{center}
\parbox[l]{8.5cm}{
\epsfig{file=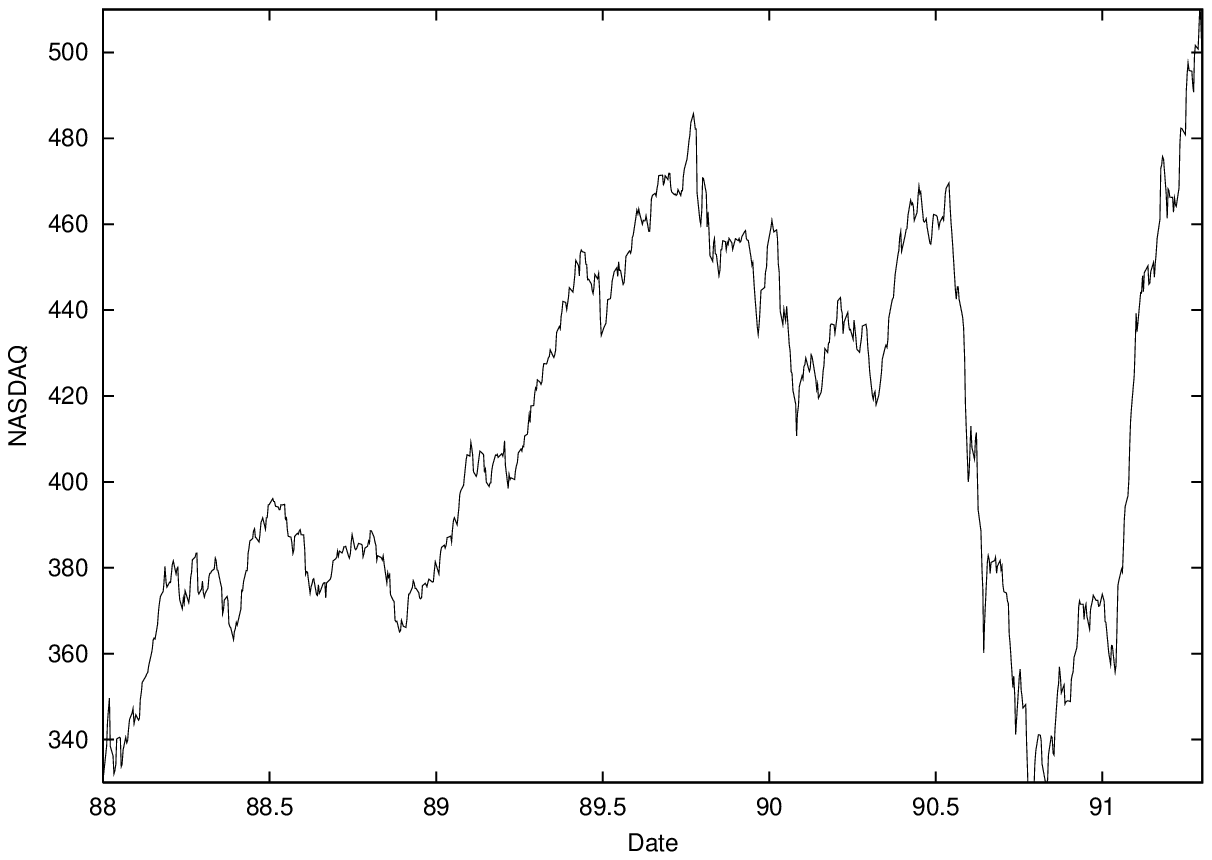,height=8cm,width=8.5cm}
\caption{\label{nas90turmoil} NASDAQ in the period from 1988 to 1991.5. The 
crash in 1990.6 is clearly visible.
}}
\hspace{5mm}
\parbox[r]{8.5cm}{
\vspace{1.5cm}
\epsfig{file=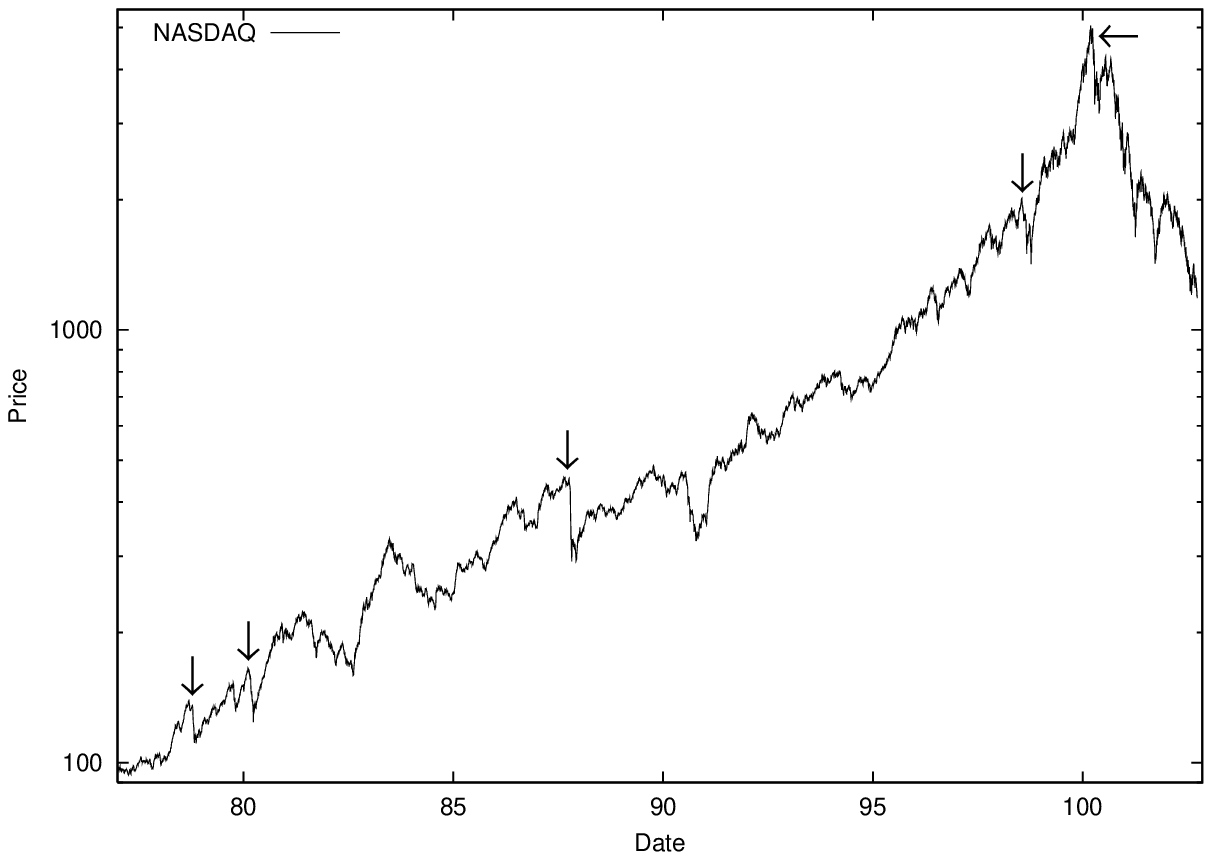,height=8cm,width=8.5cm}
\caption{\label{figNADsum} NASDAQ time series since 1976 with arrows
pointing the endogenous crashes detected by the log-periodic formula 
(\ref{lpeq}) fitted to the time series preceding them. All except the one in 
1980 also show up as $\epsilon$-drawdown outliers. }}
\vspace{1.5cm}
\parbox[l]{8.5cm}{
\epsfig{file=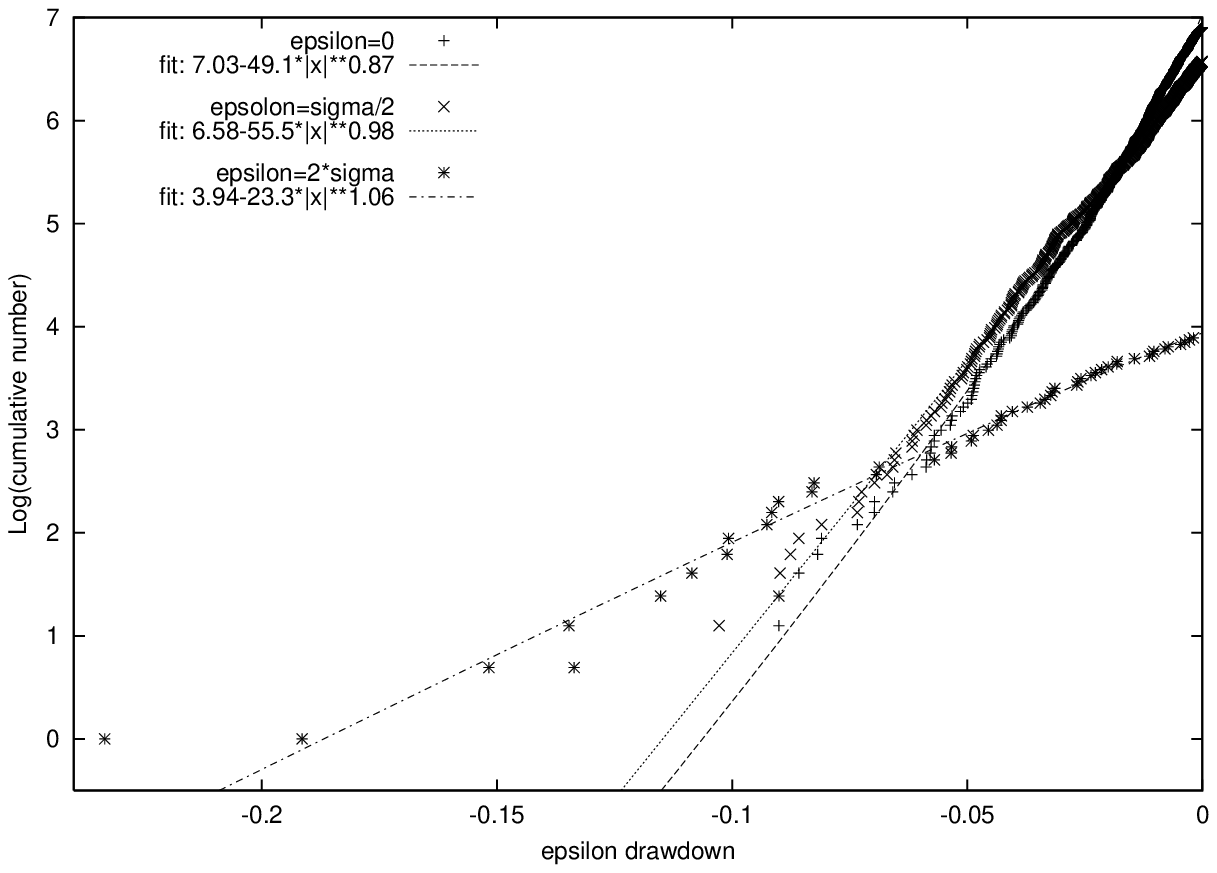,height=8cm,width=8.5cm}
\caption{\label{ftseepsdd} Logarithm of the cumulative distribution of 
$\epsilon$-drawdowns in the FTSE using an $\epsilon$ of $0$, $\sigma /2$ and 
$2\sigma $,where $\sigma=0.010$ has been obtained from the data.
}}
\hspace{5mm}
\parbox[r]{8.5cm}{
\epsfig{file=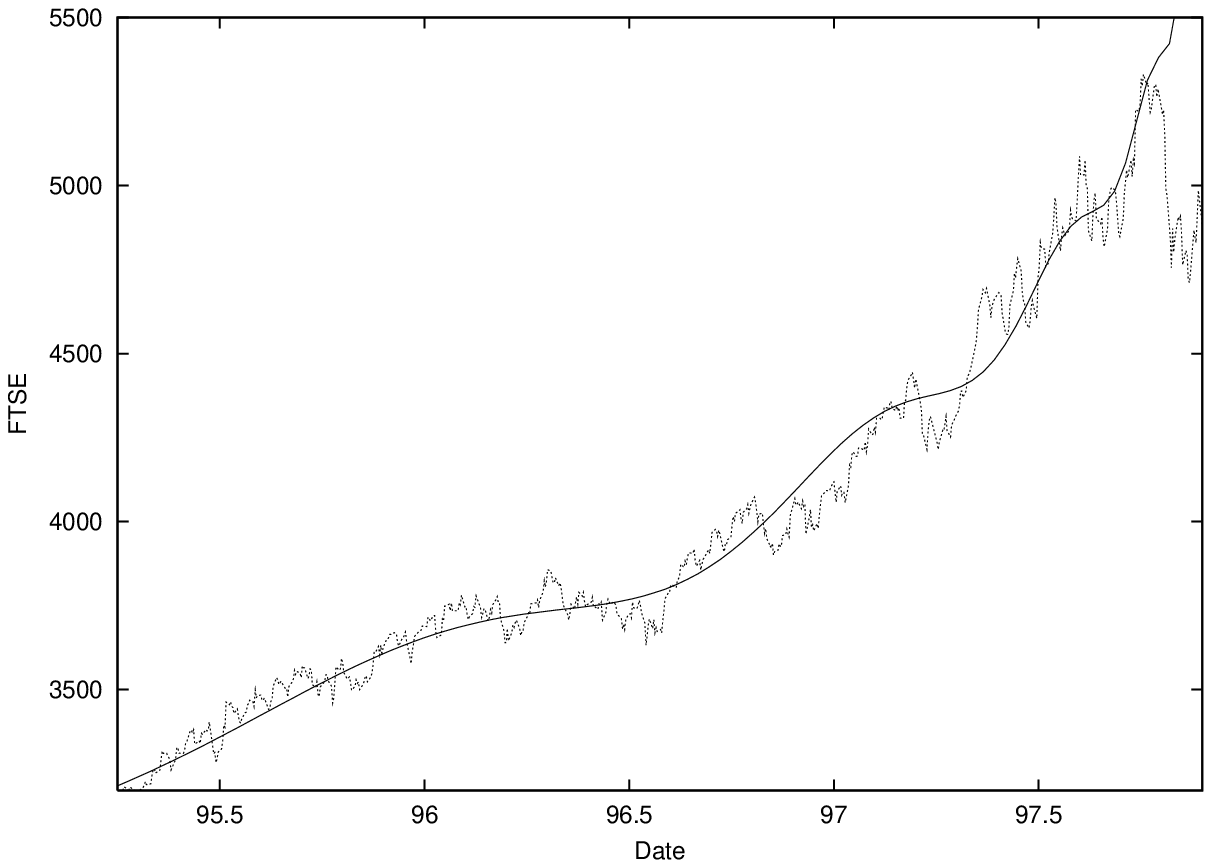,height=8cm,width=8.5cm}
\caption{\label{ftse97} The FTSE bubble ending in 1997. The fit is eq. 
(\ref{lpeq}) where  $A \approx  8212  $, $ B\approx  -4108 $, $C \approx   
74 $, $z \approx 0.18 $, $ t_c \approx   97.93  $, $ \phi \approx   -1.2 $, 
$\omega \approx   7.6$.
}}
\end{center}
\end{figure}

The 1990.6 event was not preceded by any LPPS bubble as can be seen from 
figure \ref{nas90turmoil}. A comparison with table \ref{outlnik} shows that 
the Nasdaq crash of 1990 coincides with the largest $\epsilon$-drawdown of the 
Japanese Nikkei index for $\epsilon=\sigma/2$ as already mentioned in section
\ref{fx}. As documented previously \cite{antibub1}, this $\epsilon$-drawdown
is part of the decay of almost $50\%$ during the year 1990 of the Japanese 
stock market after it reached its all-time peak on Dec. 29th 1989. Hence, it
is quite likely an reaction to the outlier event on the Japanese market and
thus exogenous and should be classified as a shock.

Fig.~\ref{figNADsum} shows the NASDAQ time series since 1976 with arrows
pointing the endogenous crashes detected by the log-periodic formula 
(\ref{lpeq}) fitted to the time series preceding them. All except the one in 
1980 also show up as $\epsilon$-drawdown outliers. The crash in 1980 is rank 
10 and 14 for $\epsilon= 0$ and $\epsilon=\sigma/2$, respectively, and its 
amplitude is 11.3

\begin{table}[]
\begin{center}
\begin{tabular}{|c|c|c|c|c|c|c|c|c|c|}\hline
$\epsilon$ & Date     & Size            & Duration & Class & $\epsilon$ & Date             & Size       & Duration& Class  \\ \hline
$0$        & 2000.268$\star$ & $25.3\%$& 5 days   & Bubble& $\sigma/2$ & 1987.762$\dagger$ & $27.7\%$   & 11 days & Bubble \\ \hline
$0$        & 1987.784 $\dagger$ & $24.6\%$& 5 days   & Bubble& $\sigma/2$ & 2000.268$\star$ & $25.3\%$   & 5 days  & Bubble \\\hline
$0$        & 1987.805 $\dagger$ & $17.0\%$        & 5 days   & Bubble& $\sigma/2$ & 1998.630$\dagger$ & $19.2\%$   & 9 days  & Bubble \\\hline
$0$        & 1998.649$\dagger$ & $16.6\%$& 4 days   & Bubble& $\sigma/2$ & 1998.724$\dagger$ & $18.6\%$   & 9 days  & Bubble \\\hline
$0$        & 2000.374$\star$ & $14.9\%$& 5 days   & Bubble& $\sigma/2$ & 1987.806$\dagger$ & $17\%$     & 5 days  & Bubble \\\hline
$0$        & 1990.622 & $12.5\%$        & 6 days   & Shock & $\sigma/2$ & 1978.753$\dagger$& $16.6\%$   & 21 days & Bubble \\\hline
\end{tabular}
\caption{\label{outlnas} Same as table \ref{outlusdm} but for the NASDAQ stock 
market index. The events with a $\star$ have already been studied in previous 
publications. The bubbles identified by the outlier analysis presented here, 
{\it i.e.}, the outliers which have prior LPPS not identified in previous works
are marked by $\dagger$.}  
\end{center}
\end{table}

The cumulative evidence from analyzing the DJIA, SP500 and NASDAQ indexes is 
that all LPPS bubbles on the US-markets ending in a crash previously published 
by the authors \cite{SJ2001,JS1999,JSL1999,JLS2000,JS2000} are recovered as the
largest outliers of the distribution of $\epsilon$-drawdowns. The sole 
exception is the 1937.2 crash previously published in \cite{SJ2001}, which 
qualifies only as an $\approx 6\%$ pure drawdown and an $\approx 8\%$ 
$\epsilon$-drawdown using $\epsilon=\sigma$. Conversely, all of the identified 
$\epsilon$-drawdown outliers can be linked to either the crash of a LPPS bubble
or to a historical event of major proportion playing the role of an external 
shock. To the first class belongs the events in 1929, 1946, 1962, 1978, 1987, 
1998 and 2000 which all had prior LPPS previously published 
\cite{SJ2001,JS1999,JSL1999,JLS2000,JS2000}. To the second class belong the 
events of 1914 (WWI), 1933 (New Deal etc.), 1940 (Nazi invasion of France and 
Benelux), 1974 (resignation and pardoning of R. Nixon) and 1990 (burst of 
Japanese real estate bubble and the anti-bubble that followed with sharp 
declines) which were all due to external shocks represented by historical 
events.

\subsection{London stock exchange}

\begin{figure}
\begin{center}
\parbox[l]{8.5cm}{
\epsfig{file=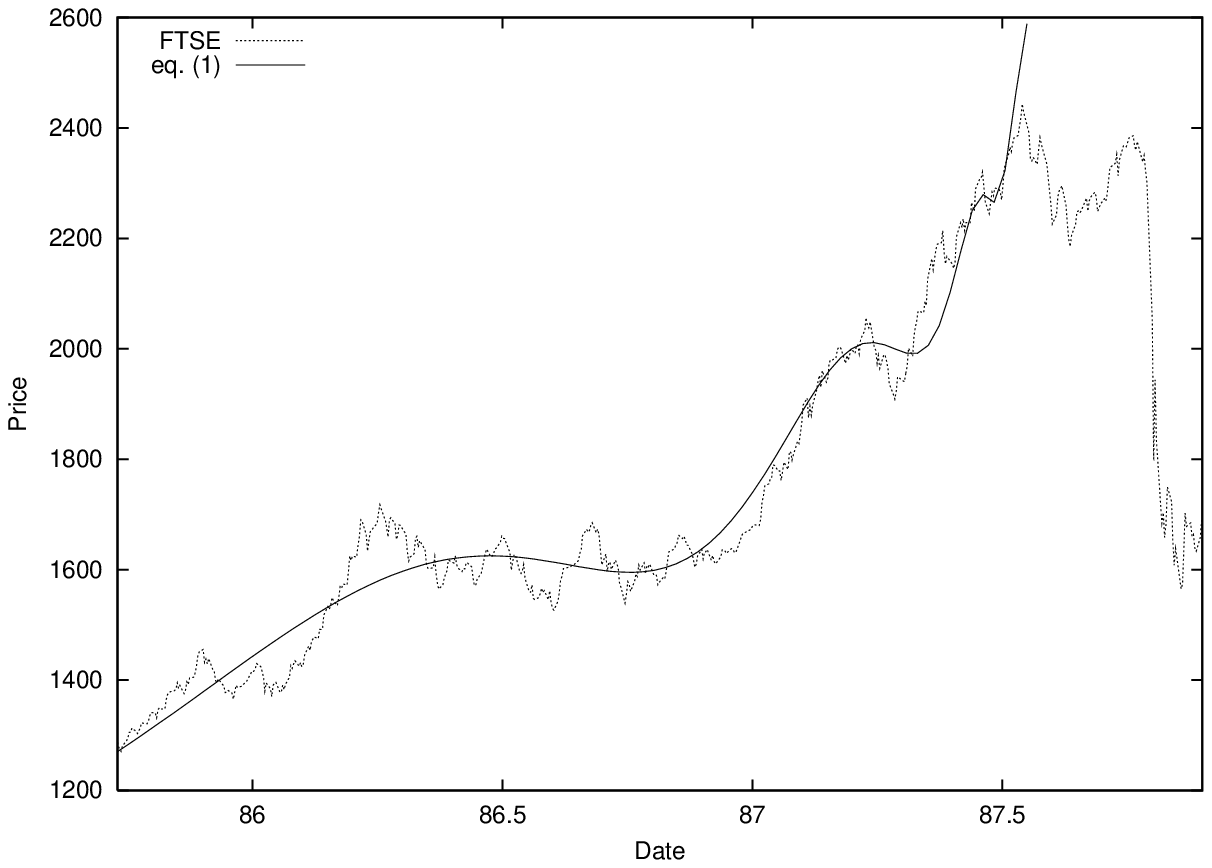,height=8cm,width=8.5cm}
\caption{\label{ftse87} The FTSE bubble ending in 1987. The fit is eq. 
(\ref{lpeq}) where  $A \approx  2884  $, $ B\approx  -1309 $, $C \approx 106 $,
$z \approx 0.30 $, $ t_c \approx   87.55  $, $ \phi \approx   1.1 $, $ \omega 
\approx  5.1$. } }
\hspace{5mm}
\parbox[r]{8.5cm}{
\epsfig{file=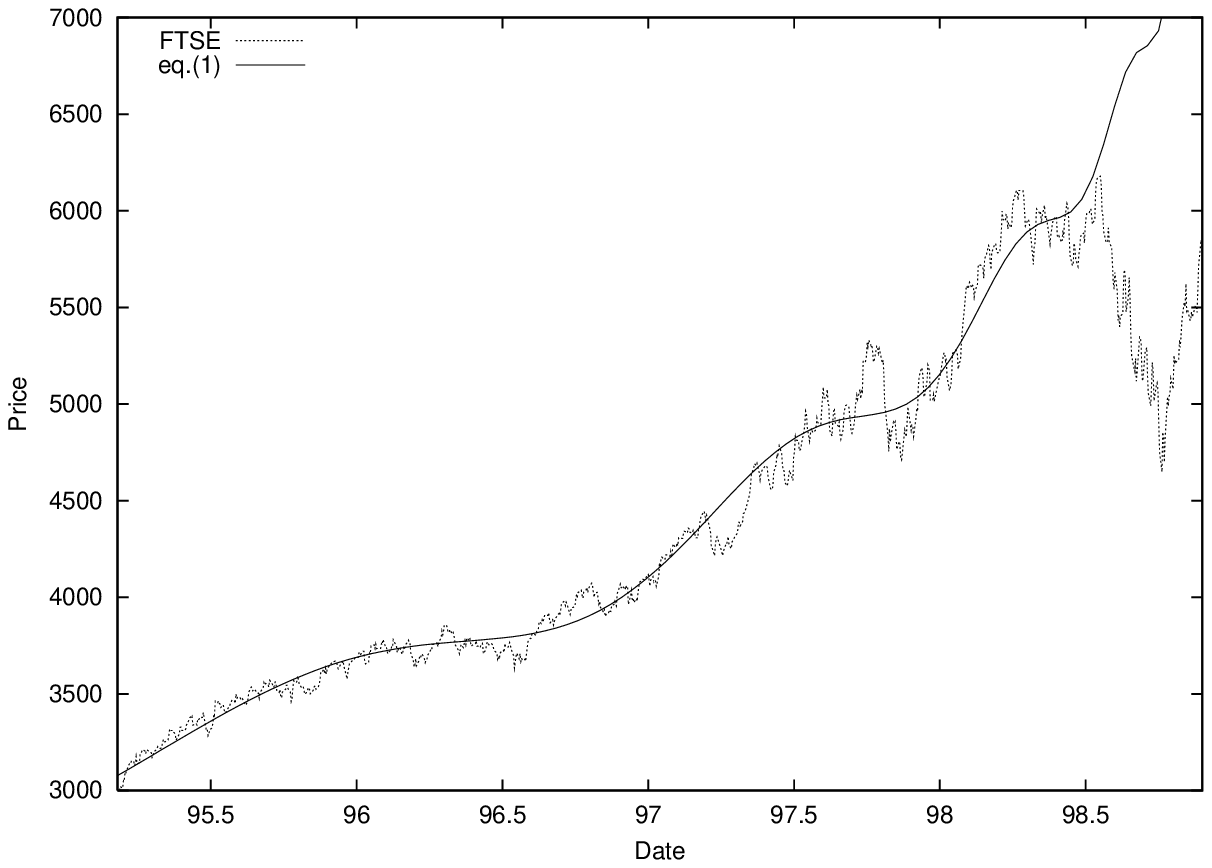,height=8cm,width=8.5cm}
\caption{\label{ftse98} The FTSE bubble ending in 1998. The fit is eq. 
(\ref{lpeq}) where  $A \approx  13122$, $ B\approx  -7847 $, $C \approx  136 $,
$z \approx 0.18 $, $ t_c \approx   98.99  $, $ \phi \approx   3.6 $, $ \omega 
\approx  8.6$. Note that the value for $\omega$ is rather high compared with
the distribution shown in figure \protect\ref{omegadistrib}. This is due to the
outlier shown in figure \protect\ref{ftse97} since the fit picks up this 
``dip'' in the time series.}}
\end{center}
\end{figure}

Figure \ref{ftseepsdd} shows the distribution of $\epsilon$-drawdowns for the 
FTSE index of the London stock exchange. The time series extends from 02/03 
1984 to 13/07 2000. Table \ref{outlftse} shows that the outliers identified for
$\epsilon=0$ and $\epsilon=\sigma/2$ are identical except for a reshuffling of 
the ranks 3 and 4. All these four drawdown outliers are preceded by a LPPS
bubble well-parametrised by (\ref{lpeq}) as shown in figures  \ref{ftse97}, \ref{ftse87}
and \ref{ftse98} for the fits and in the captions for the fit parameters.
Except for the 1998 outlier, the obtained values of the log-frequencies and of the
exponents are in good agreement with (\ref{zomegaval}) and with figures 
\ref{omegadistrib} and \ref{zdistrib}. For the 1998 outlier, 
the rather larger value $\omega \approx 8.6$ is probably due to the
fact that the fit picks up the ``dip'' associated with the
presence of the 1997 outlier.

\begin{table}[]
\begin{center}
\begin{tabular}{|c|c|c|c|c|c|c|c|c|c|}\hline
$\epsilon$ & Date             & Size       & Duration & Class & $\epsilon$ & Date             & Size     & Duration  & Class\\\hline
$0$        & 1987.784$\dagger$         & $23.3\%$   & 4 days   & Bubble& $\sigma/2$ & 1987.784$\dagger$         & $23.3\%$   & 4  days & Bubble\\\hline
$0$        & 1987.805$\dagger$         & $13.4\%$   & 3 days   & Bubble& $\sigma/2$ & 1987.805$\dagger$         & $13.4\%$   & 3 days  & Bubble\\\hline
$0$        & 1998.745$\dagger$         & $9.0 \%$   & 4 days   & Bubble& $\sigma/2$ & 1997.784$\dagger$& $10.3\%$   & 11 days & Bubble\\\hline
$0$        & 1997.805$\dagger$& $9.0 \%$   & 5 days   & Bubble& $\sigma/2$ & 1998.745$\dagger$         & $ 9.0\%$   & 4 days  & Bubble\\\hline
\end{tabular}
\caption{\label{outlftse} Same as table \ref{outlusdm} but for the FTSE stock 
market index. Bubbles identified by the outlier analysis presented here, 
{\it i.e.}, outliers which have prior LPPS not identified in 
previous works are marked by $\dagger$.}
\end{center}
\end{table}

\subsection{Frankfurt stock exchange (DAX)}

\begin{figure}
\begin{center}
\parbox[l]{8.5cm}{
\epsfig{file=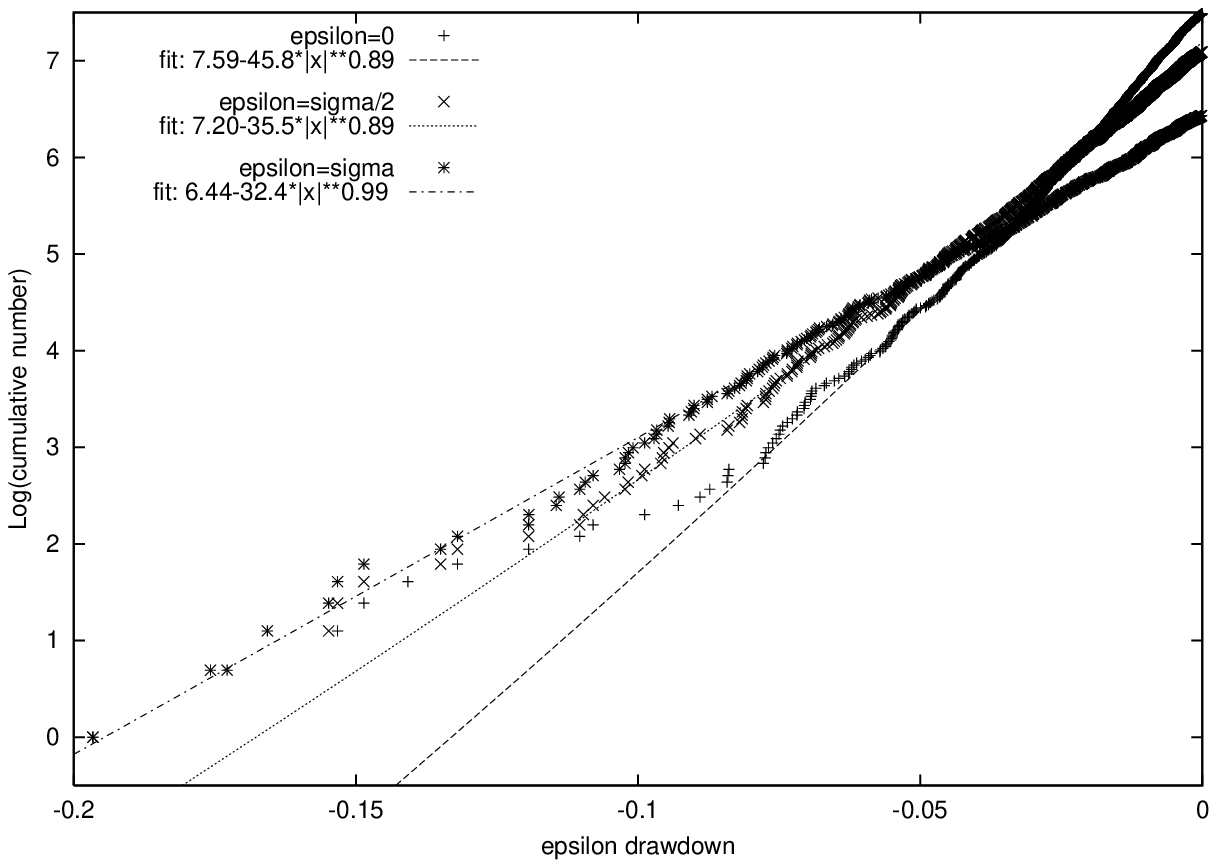,height=8cm,width=8.5cm}
\caption{\label{daxepsdd} Logarithm of the cumulative distribution of 
$\epsilon$-drawdowns
in the DAX using an $\epsilon$ of $0$, $\sigma /2$ and $\sigma $,where 
$\sigma=0.011$ 
has been obtained from the data.}}
\hspace{5mm}
\parbox[r]{8.5cm}{
\epsfig{file=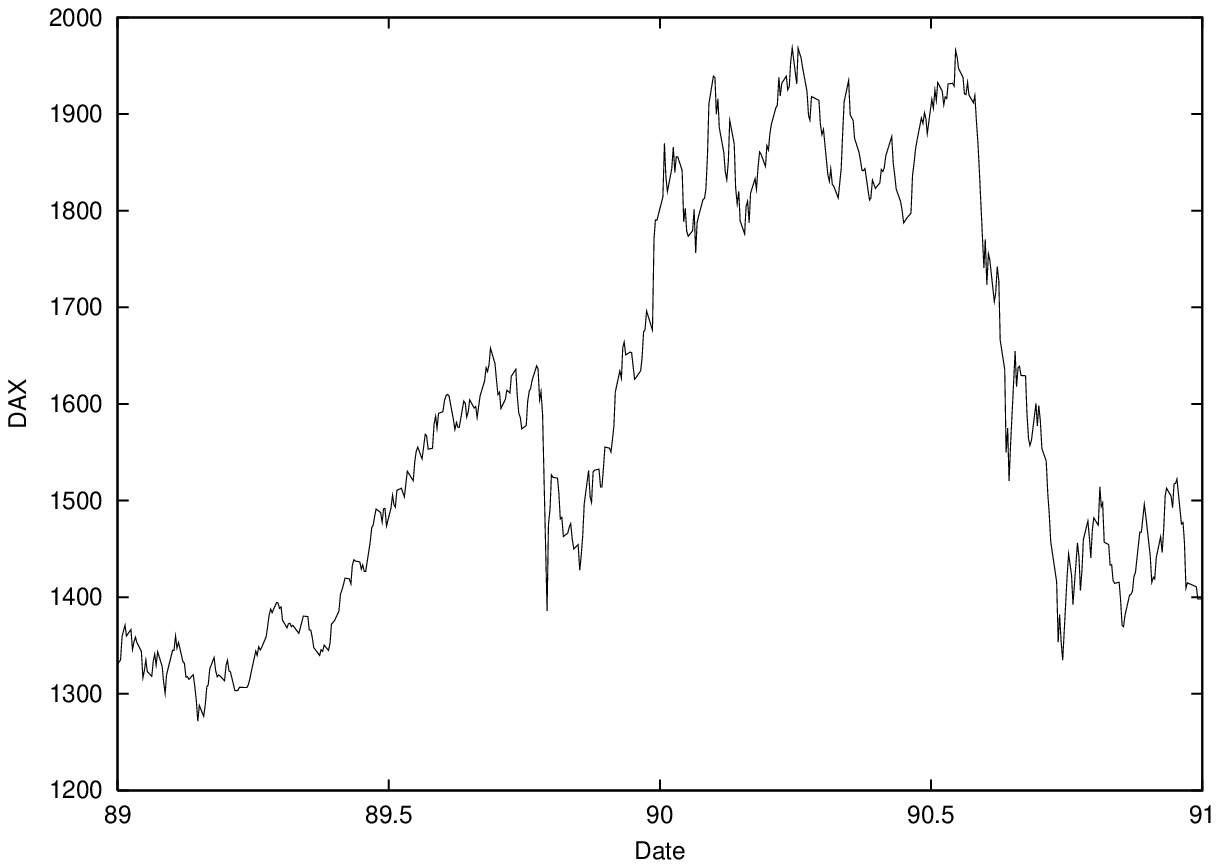,height=8cm,width=8.5cm}
\caption{\label{dax1989}. DAX in the period from 1989 to 1991. The crashes 
in 1989.8 and 1990.7 are clearly visible}}
\vspace{1.5cm}
\parbox[l]{8.5cm}{
\epsfig{file=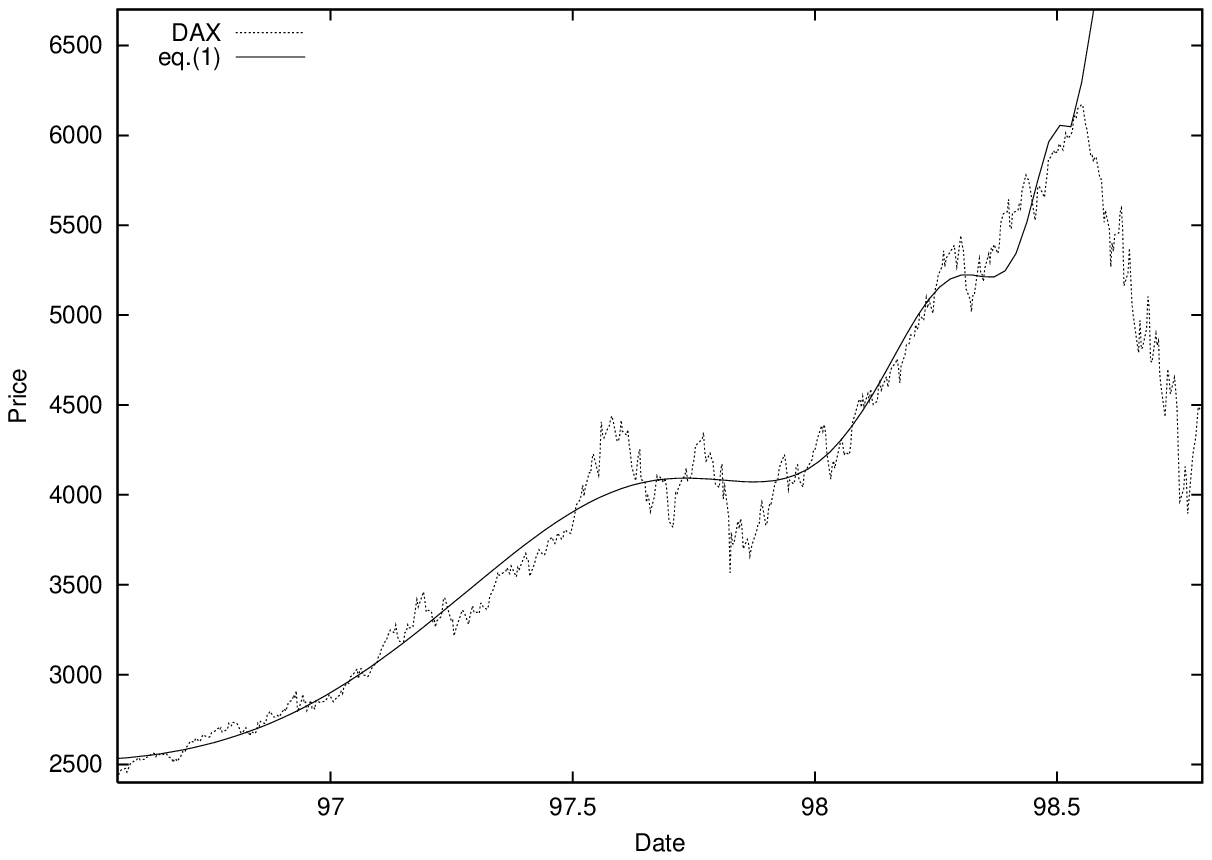,height=8cm,width=8.5cm}
\caption{\label{dax1998} The DAX bubble ending in 1998. The fit is eq. 
(\ref{lpeq}) where  $A \approx  8343$, $ B\approx  -4553 $, $C \approx  257 $,
$z \approx 0.28 $, $ t_c \approx   98.61  $, $ \phi \approx   0.2 $, $ \omega 
\approx  5.7$. } }
\hspace{5mm}
\parbox[r]{8.5cm}{
\epsfig{file=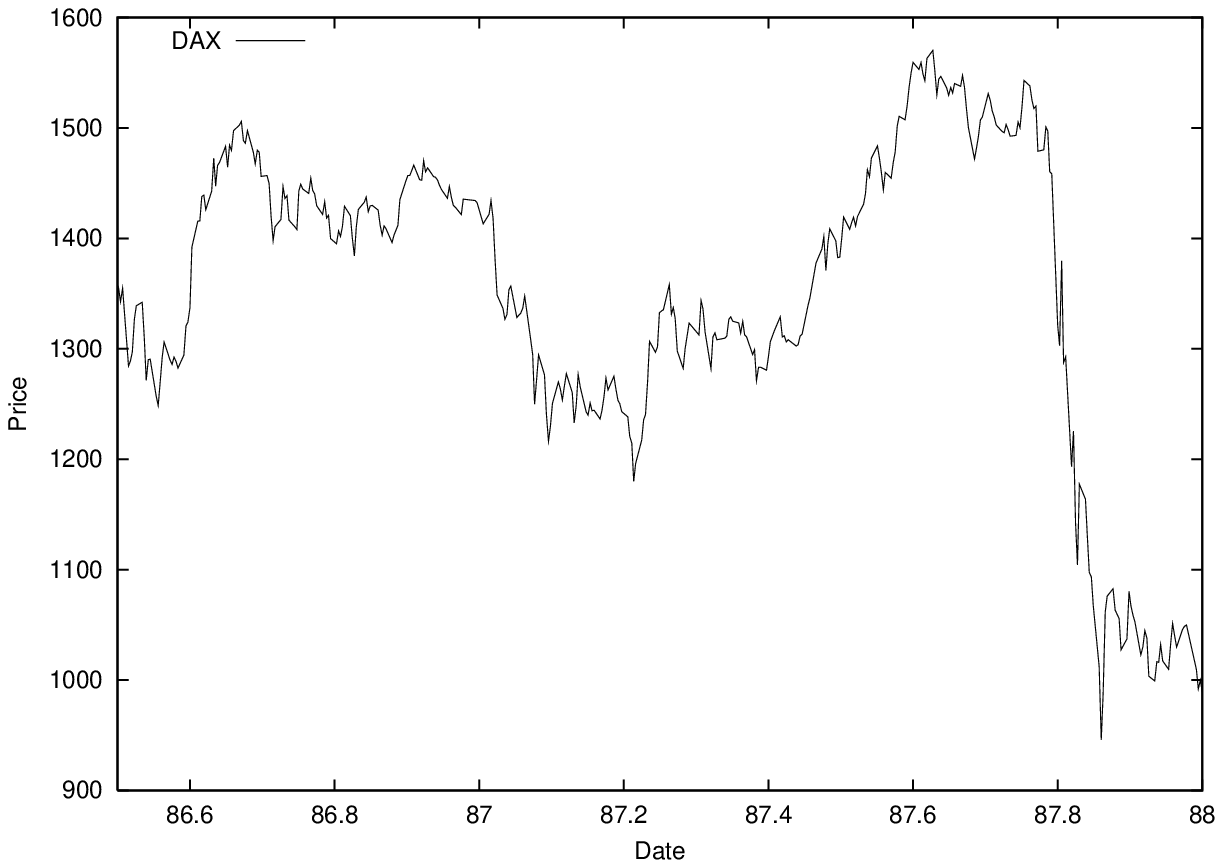,height=8cm,width=8.5cm}
\caption{\label{dax1987}. DAX in the period from 1986.5 to 1988. The crash of 
Oct. 1987 is clearly visible}}
\end{center}
\end{figure}

The time series for the Frankfurt stock exchange (DAX) goes from 02/01 1970 to 
13/07 2000. Beside the well-known crashes of 1987 and 1998 whose preceding time
series for the stock markets visited so far are well-fitted by the log-periodic
power law formula (\ref{lpeq}), the $\epsilon$-drawdown outliers allow us to 
identify another familiar date, namely the 1990.7 event already discussed for 
the NASDAQ. As with the NASDAQ event of the same date, no prior bubble can be 
qualified by a fit with (\ref{lpeq}), as shown in figure \ref{dax1989} and as 
with the NASDAQ, we conclude that the origin of this large drawdown is 
exogenous and most likely due to the burst of the Japanese real estate bubble 
and the ``anti-bubble'' that followed with sharp declines 
\cite{antibub1,antibub2}. In figure \ref{dax1987}, we see the DAX prior to
the 1987 outlier and we must conclude that no LPPS is present and that the
crash was of exogenous origin, specifically caused by the collapse of the 
U.S. markets.  In figure \ref{dax1998}, we see the DAX prior to the 1998 
outlier and we must conclude that this outlier is of endogenous origin with
LPPS. Furthermore, the values obtained for $\omega$ and $z$ are in good 
agreement with (\ref{zomegaval}).

The analysis of the $\epsilon$-drawdown outliers identifies two new 
dates, 1970.3 and 1989.8. Unfortunately, the data at our disposal is not 
sufficient to decide whether the 1970.3 event was caused by a LPPS bubble or 
not. However, this slow crash coincides with the so-called ``liquidity crisis''
of May 1970\footnote{Between 1965 and 1968, government spending in the US
increased from \$118 billion to \$179 billion of which half was related to 
defense. As a result, the inflation rate increased to nearly 6\% by 1970.
By May 1970, this excessive demand for money led to a liquidity crisis in which
interest rates rose to new heights. The largest railways in America, Penn 
Central Railway, went bankrupt as a consequence. Confidence was restored when 
Congress passed the Securities Investor Protection Act of 1970.}.
With respect to the 1989.8 crash, we see from figure \ref{dax1989} that 
this event was definitely not caused by a LPPS bubble. Historically, it is 
tempting to relate this event to the unification of Germany. In fact, the exact
date of the crash is 16th October 1989 which is also the date where the Central
Committee of the SED took control and forced general secretary Honecker to 
resign from his office as head of state and party leader. That the financial 
system of then West-Germany reacted in panic on these events is no great 
surprise. This means that with the exception of the slow crash or ``liquidity 
crisis''of May 1970, all large drops in the DAX in the period 02/01 1970 to 
13/07 2000 was either caused by an historically important exogenous shock or 
an endogenous speculative bubble.

\begin{table}[]
\begin{center}
\begin{tabular}{|c|c|c||c|c|c|c|c|c|c|}\hline
$\epsilon$ & Date     & Size       & Duration & Class &  $\epsilon$ & Date     & Size     & Duration & Class \\ \hline
$0$        & 1987.830 & $19.7\%$   & 7 days   & Shock& $\sigma/2$ & 1987.830 & $19.7\%$   & 7  days & Shock\\ \hline
$0$        & 1970.345 & $17.3\%$   & 13 days  & ?     & $\sigma/2$ & 1970.345 & $17.3\%$   & 13 days & ?     \\ \hline
$0$        & 1990.699 & $15.3\%$   & 9 days   & Shock & $\sigma/2$ & 1989.773 & $15.5\%$   & 5  days & Shock \\ \hline
$0$        & 1998.743$\dagger$ & $14.9\%$   & 4 days   & Bubble& $\sigma/2$ & 1990.699 & $15.3\%$   & 9 days  & Shock \\ \hline
$0$        & 1989.781 & $14.1\%$   & 2 days   & Shock & $\sigma/2$ & 1998.743$\dagger$ & $14.9\%$   & 4 days  & Bubble\\ \hline
\end{tabular}
\caption{\label{outldax} Same as table \ref{outlusdm} but for the DAX stock 
market index. Bubbles identified by the outlier analysis presented here, 
{\it i.e.}, outliers which have prior LPPS not identified in 
previous works are marked by $\dagger$}
\end{center}
\end{table}

\subsection{Tokyo stock exchange}

\begin{figure}
\begin{center}
\parbox[l]{8.5cm}{
\epsfig{file=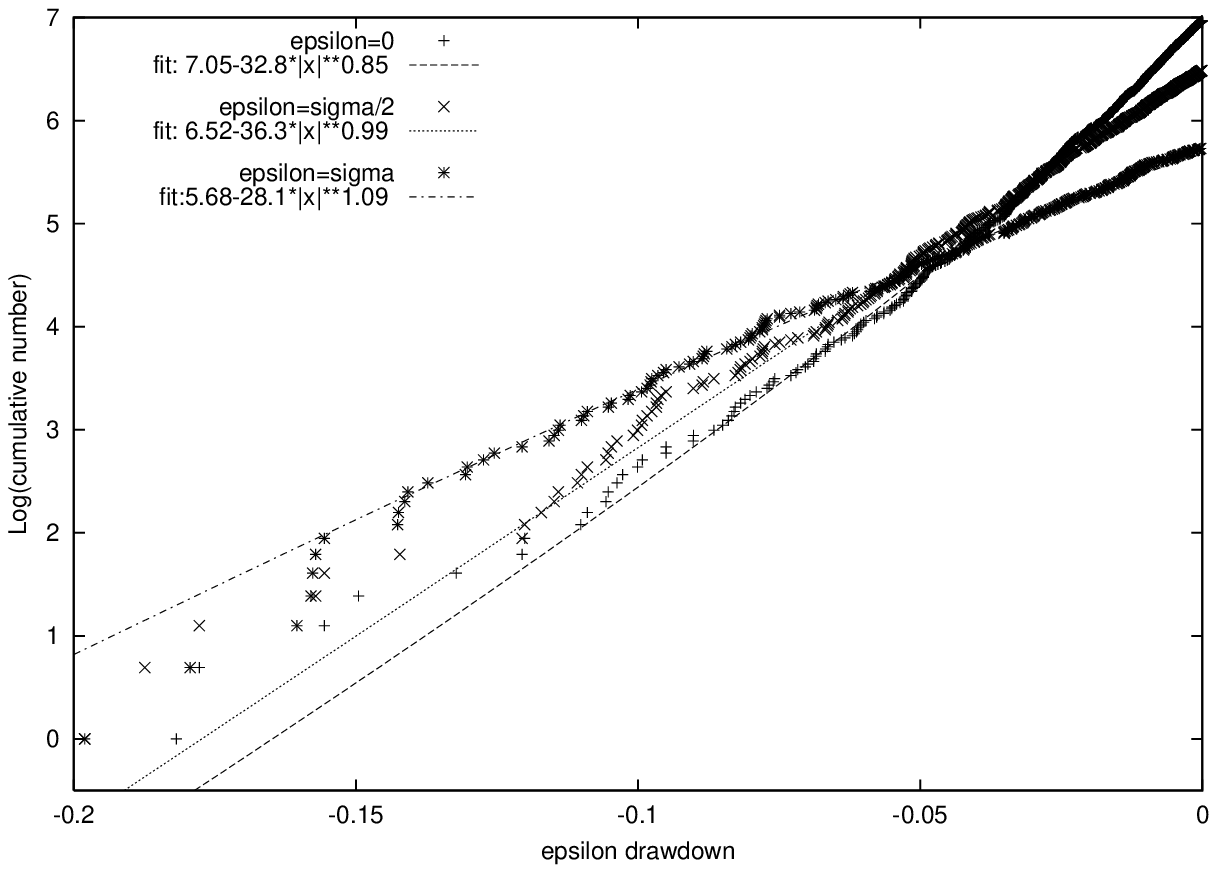,height=8cm,width=8.5cm}
\caption{\label{nikepsdd} Logarithm of the cumulative distribution of 
$\epsilon$-drawdowns in the Nikkei using an $\epsilon$ of $0$, $\sigma /2$ 
and $\sigma $,where $\sigma=0.015$ has been obtained from the data.}}
\hspace{5mm}
\parbox[r]{8.5cm}{
\epsfig{file=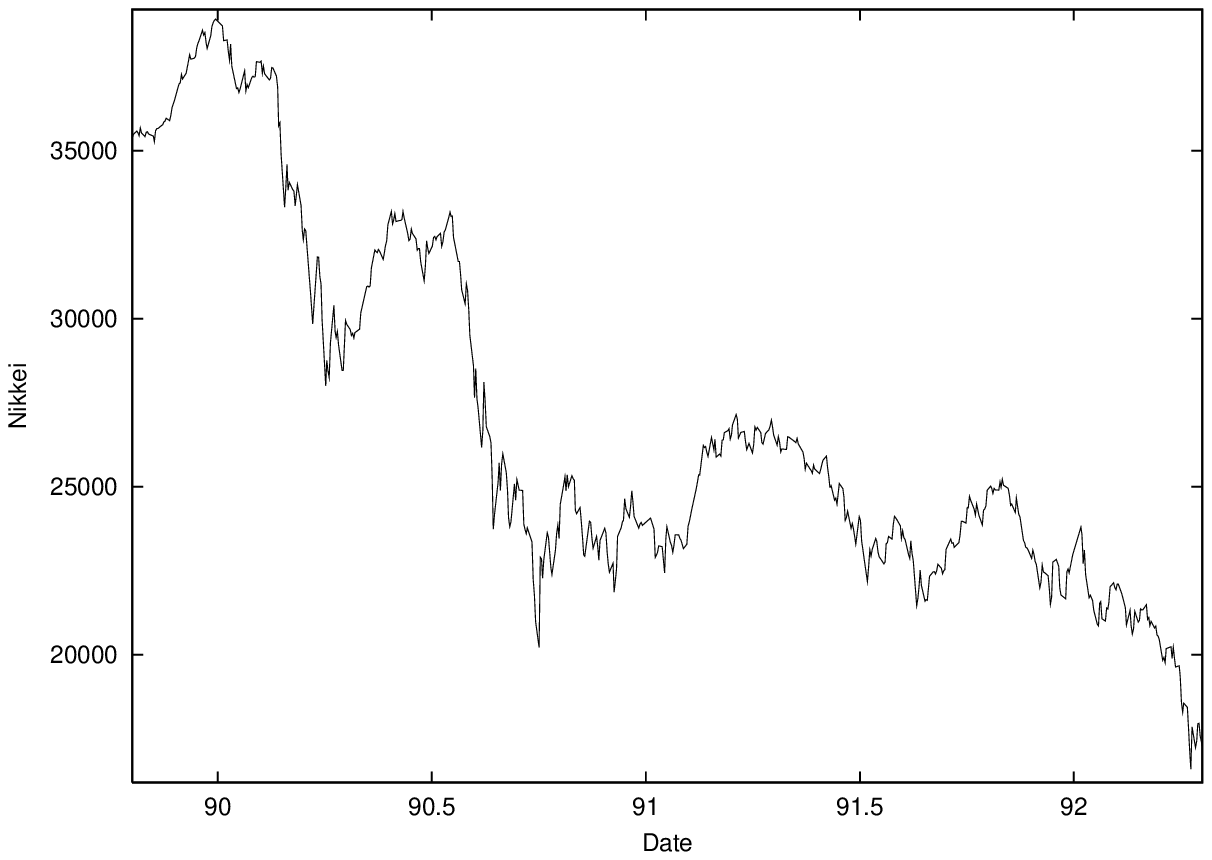,height=8cm,width=8.5cm}
\caption{\label{nik1990} Nikkei in the period from 1989.8 to 1992.3. The 
crashes of Aug. and Sept. 1990 are clearly visible.}}
\vspace{1.5cm}
\parbox[l]{8.5cm}{
\epsfig{file=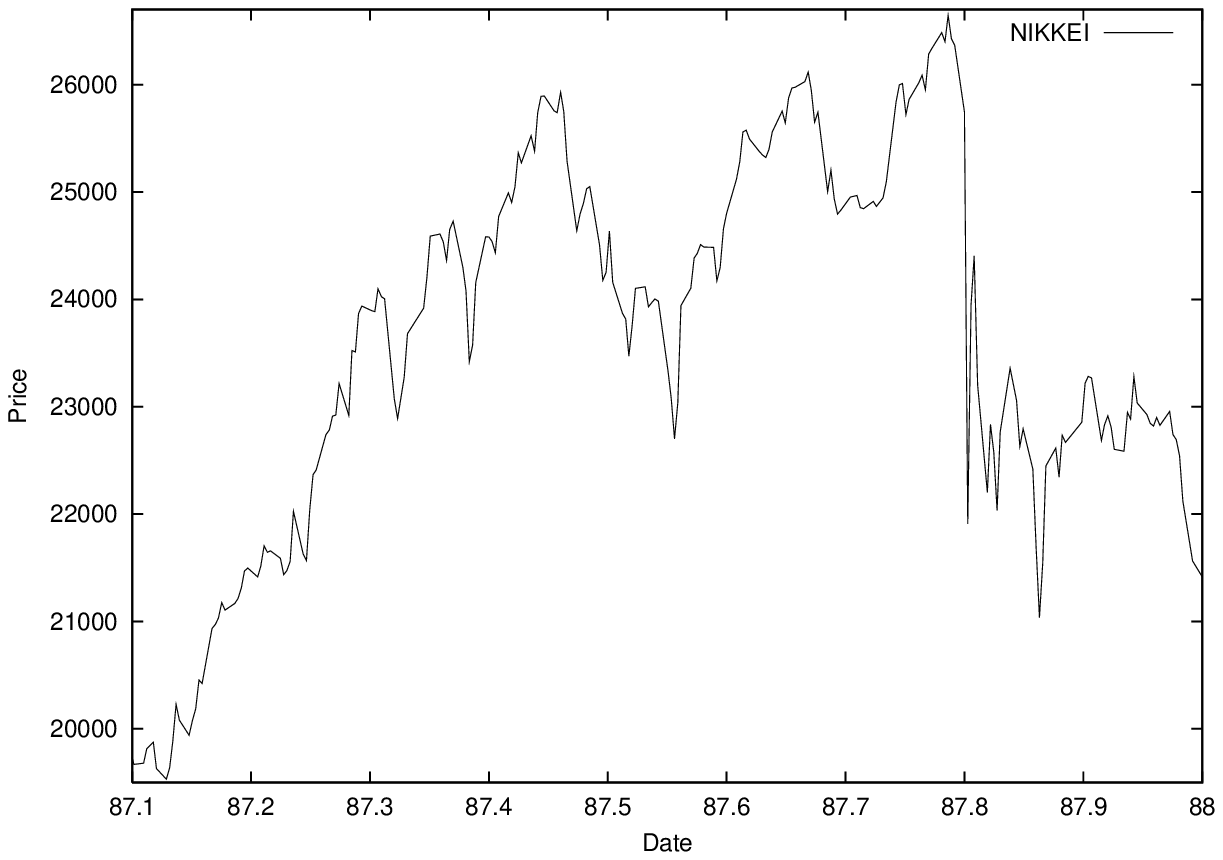,height=8cm,width=8.5cm}
\caption{\label{nik1987} Nikkei in the period from 1987.1 to 1988. The 
crashes of Oct. 1987 is clearly visible.}}
\hspace{5mm}
\parbox[r]{8.5cm}{
}
\end{center}
\end{figure}

The analyzed time series goes from 05/01 1973 to 10/03 2000. Among the 3
events listed in table \ref{outlnik}, we find the well-known date of the 
crash of 1987. In figure 
\ref{nik1987}, we see that, as with the outlier on the DAX of the same time, no
LPPS is present. Hence, we must conclude that the crash was of exogenous 
origin, specifically caused by the collapse of the U.S. markets.

The remaining two events are Aug. and Sept. 1990. Figure \ref{nik1990} shows 
the Nikkei index from 1989.9 to 1992.3. The Nikkei index reached its all-time
 high on the last trading day of 1989 (29th of Dec.) and has since then 
followed a downward trend punctuated by decelerating oscillations with large 
amplitudes.
 In two prior publications \cite{antibub1,antibub2}, we have shown that the 
behavior of the Nikkei index since the beginning of 1990 could be understood
as the symmetric to a speculative bubble, which we termed an ``anti-bubble'', 
reflecting also imitation and herding processes leading to positive feedbacks 
similarly to what occurs during speculative bubbles. The difference is that 
the positive feedbacks reinforce the speculative bearish phase rather than a 
bullish phase. The degree of symmetry, after the critical time $t_c$ 
corresponding to the all-time high, is characterized by a power law decrease 
of the price (or of 
the logarithm of the price) during the anti-bubble as a function of time 
$t>t_c$ and by decelerating/expanding log-periodic oscillations.
Another good
example is found for the gold future prices after 1980, after its
all-time high. The Russian market prior to and after its
speculative peak in 1997 also constitutes a remarkable example
where both bubble and anti-bubble structures appear simultaneously
for the same $t_c$. This is however a rather rare occurrence,
probably because accelerating markets with log-periodicity often
end-up in a crash, a market rupture that thus breaks down the
symmetry ($t_c-t$ for $t<t_c$ into  $t-t_c$ for $t>t_c$). 
A remarkable similarity in the behavior of the US S\&P500 index from 1996 to 
August 2002 and of the Japanese Nikkei index from 1985 to 1992 (11 years shift)
has also been found recently, with particular emphasis on the structure of the 
bearish phases \cite{predsz}.

The 1990.6 and 1990.7 events (Aug. and Sept. 1990) occurred during the 
descending phase of the Nikkei decelerating log-periodic oscillation as shown 
in figure \ref{antinik}. We thus propose to see them as a consequence of the 
anti-bubble regime in its first stage where the power law decay and the 
log-periodic oscillations combine to create large drops in the descending 
phases of the oscillations. We thus capture this concept by the term 
``Anti-Bubble'' in table \ref{outlnik}.

\begin{table}[]
\begin{center}
\begin{tabular}{|c|c|c|c|c|c|c|c|c|c|}\hline
$\epsilon$ & Date            & Size       & Duration &     Class & $\epsilon$ & Date     & Size     & Duration&  Class     \\ \hline
$0$        & 1987.786        & $17.8\%$   & 4  days  &    Shock& $\sigma/2$ & 1990.699 & $19.8\%$ & 12 days & Anti-Bubble\\ \hline
$0$        & 1990.622$\star$ & $15.6\%$   & 6 days   &Anti-Bubble& $\sigma/2$ & 1987.786 & $17.8\%$ & 4  days & Shock     \\ \hline
$0$        & 1990.723$\star$ & $15.0\%$   & 5 days   &Anti-Bubble& $\sigma/2$ & 1990.622 & $15.6\%$ & 6 days  & Anti-Bubble\\ \hline
\end{tabular}
\caption{\label{outlnik} Same as table \ref{outlusdm} for the Nikkei stock 
market index. The events
with a $\star$ have already been studied in previous publications. See text for
an explanation of the class name ``Anti-Bubble.''}  
\end{center}
\end{table}

\subsection{Hong-Kong stock exchange}

The analyzed Hong-Kong stock exchange time series goes from 24/11 1969  to 
13/07 2000. The Hong-Kong stock market is the most volatile of the markets 
considered here and, for example, is twice as volatile as the NASDAQ index in 
the considered historical period. Using the analysis of the distribution of 
$\epsilon$-drawdowns, we identify 8 drawdown outliers, which are listed in 
table \ref{outlhk}. 

In chronological order, the 1973.2 (March and April 1973) events 
have a time series preceding them that is very well fitted
by formula (\ref{lpeq}) (see \cite{SJ2001}). This qualifies these two events 
as endogenous and part of the same ``crash.''

The 1973.75 event is not preceded by an accelerating log-periodic power law 
which puts it in the exogenous class. Indeed, it seems that it can be 
attributed
to the Arab-Israeli  ``Yom Kippur'' War and the subsequent Arab Oil Embargo 
which both occurred in Oct. 1973. 

The 1974.6 event identified as an $\epsilon$-drawdown outlier with 
$\epsilon = \sigma/2$ is not preceded by an accelerating log-periodic power 
law which puts it in the exogenous class. However, no historical event has
been found which this event could casually be linked to. However, as with the
event just below, it could be due to the Crown colony's special situation, 
{\it e.g}, its complete dependence on oil import as well as its sensitivity
to political events in mainland China.

The 1982.7 event is not preceded either by an accelerating log-periodic power 
law and this suggests also that this event is exogenous and resulted
from an external shock. Indeed, we attribute it to the failure of the 
negotiations between British prime minister Margaret Thatcher and 
Deng Xiao-Peng 
and Zhao Zhiyang in Beijing, in Sept. 1982. Deng rejected Thatcher's proposal 
for continued British administration in 
Hong-Kong after 1997. The strong reaction of the market to this decision may be
attributed to the large sensitivity of the Hong-Kong market
to the activities of China proper, before the return of the
colony to China. Since this political event coincided very much in time with 
this very large $\epsilon$-drawdown, we identify this political shock with the 
external source.

The 1987.8 event is preceded by a very nice accelerating log-periodic power 
law time series \\(see \cite{emergent}), qualifying it as an endogenous event.

The 1989.4 event is also preceded by an accelerating log-periodic power law 
time series \\(see \cite{SJ2001}), qualifying it as an endogenous event.

The 1997.8 and 1997.99 events are preceded by a very neat accelerating 
log-periodic power law time series (see \cite{emergent}), qualifying them as 
an endogenous events. 

Compared with our previously published analysis, we recover
the endogenous events preceded by speculative bubbles of 1973.2, 1987.8, 
1989.4 
and 1997. However, the methodology using $\epsilon$-drawdown outliers misses 
the speculative bubbles ending in 1971.7, 1978.7, 1980.9 and 1994 (the Asian 
crisis) that we have previously identified based on the quality of the fits 
with formula (\ref{lpeq}). The reason can be seen from an examination of the 
price time series after these four events (see \cite{SJ2001} and 
\cite{emergent}): the accelerating log-periodic power laws found to fit
the time series indeed signal the end of speculative bubbles, which do not 
lead 
to real crashes but rather to large corrections or to a transition to a 
bearish 
regime lasting many months to years. Such a behavior will not qualify as a 
large
$\epsilon$-drawdown outlier. Nevertheless, these events (1971.7, 1978.7, 
1980.9 
and 1994) are significant in that they signal a strong shift of regime in the 
Hong-Kong stock market. These examples show the limits of our analysis which 
misses important change of regimes where the amplitude is more to be found in 
the duration of the shift rather than in the fast loss of the market. The 
subtle
definition of strong and durable regime shifts are until now eluded our 
attempts
for a rigorous definition, except for the recognition that such strong regime 
shifts are detected by their log-periodic signatures, as are crashes. This 
is a 
weakness of our methodology that we hope to improve upon in the future.

\begin{table}[]
\begin{center}
\begin{tabular}{|c|c|c|c|c|c|c|c|c|c|}\hline
$\epsilon$ & Date             & Size       & Duration & Class & $\epsilon$ & Date     & Size            & Duration   & Class\\\hline
$0$        & 1987.783$\star$ & $41.7\%$   & 4 days   & Bubble& $\sigma/2$ & 1987.753$\star$ & $43.2\%$& 12 days    &Bubble\\\hline
$0$        & 1973.241$\star$ & $38.6\%$   & 7  days  & Bubble& $\sigma/2$ & 1973.241$\star$ & $38.6\%$   &  7 days &Bubble\\\hline
$0$        & 1973.734         & $37.0\%$   & 1 day    & Shock & $\sigma/2$ & 1973.734         & $37.0\%$   & 1  day  &Shock \\\hline
$0$        & 1973.282$\star$ & $32.2\%$   & 8 days   & Bubble& $\sigma/2$ & 1973.282$\star$ & $32.2\%$   & 8 days  &Bubble\\\hline
$0$        & 1989.413$\star$ & $26.4\%$   & 5 days   & Bubble& $\sigma/2$ & 1974.603         & $29.7\%$   & 17 days &Shock \\\hline
$0$        & 1982.732        & $25.5\%$   & 6 days   & Shock & $\sigma/2$ & 1982.721         & $27.0\%$   & 10 days &Shock \\\hline
$0$        & 1997.999$\star$ & $24.5\%$   & 8 days   & Bubble& $\sigma/2$ & 1974.830         & $26.7\%$   & 15 days &Shock \\\hline
$0$        & 1997.796$\star$ & $23.3\%$   & 4 days   & Bubble& $\sigma/2$ & 1989.413$\star$ & $26.4\%$   & 5 days  &Bubble\\\hline
\end{tabular}
\caption{\label{outlhk} Same as table \ref{outlusdm} for the Heng-Seng stock 
market index. The events with a $\star$ have already been studied in 
previous publications.}  
\end{center}
\end{table}

\section{$\epsilon$-drawdown outliers of Bond markets} \label{bm}

\subsection{The U.S. T-bond market}

\begin{figure}
\begin{center}
\parbox[l]{8.5cm}{
\epsfig{file=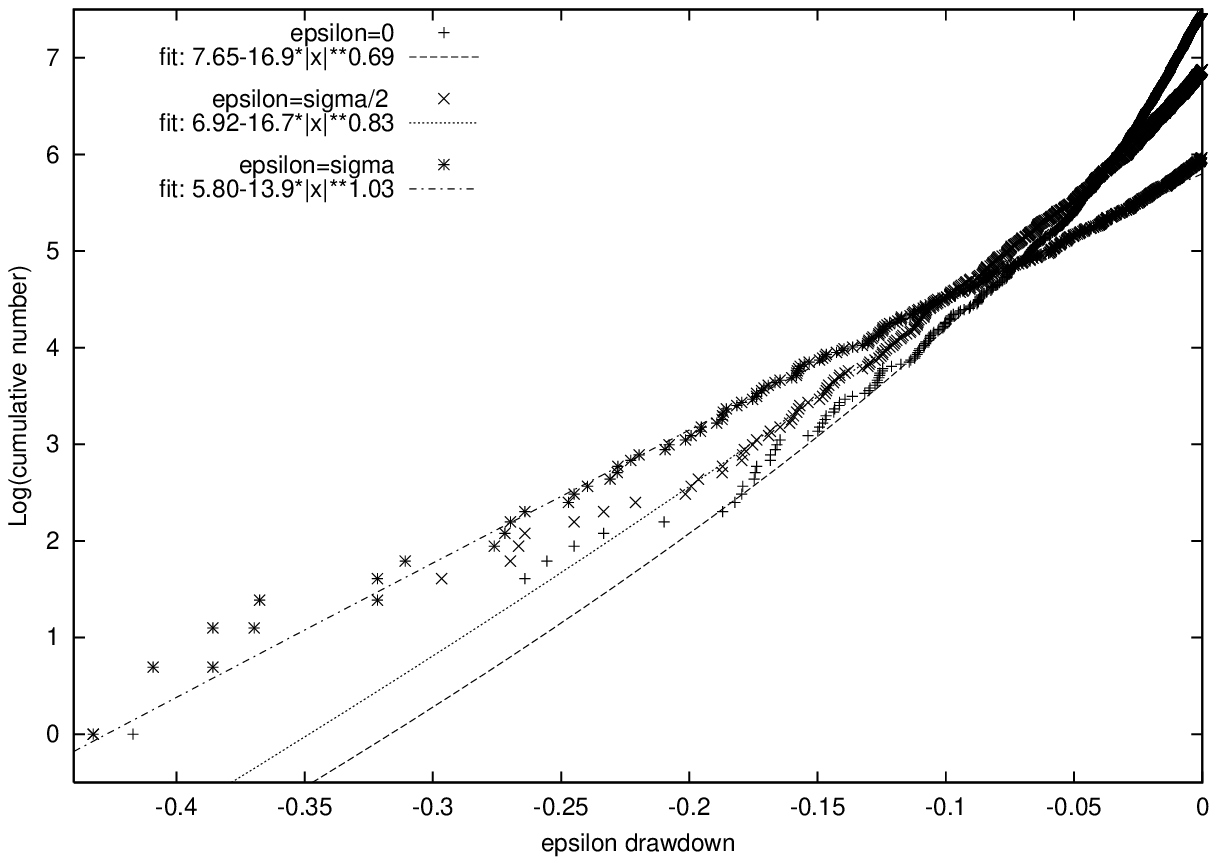,height=8cm,width=8.5cm}
\caption{\label{hkepsdd} Logarithm of the cumulative distribution of 
$\epsilon$-drawdowns in the Hang-Seng using an $\epsilon$ of $0$, $\sigma /2$ 
and $\sigma $, where $\sigma=0.021$ has been obtained from the data.}}
\hspace{5mm}
\parbox[r]{8.5cm}{
\epsfig{file=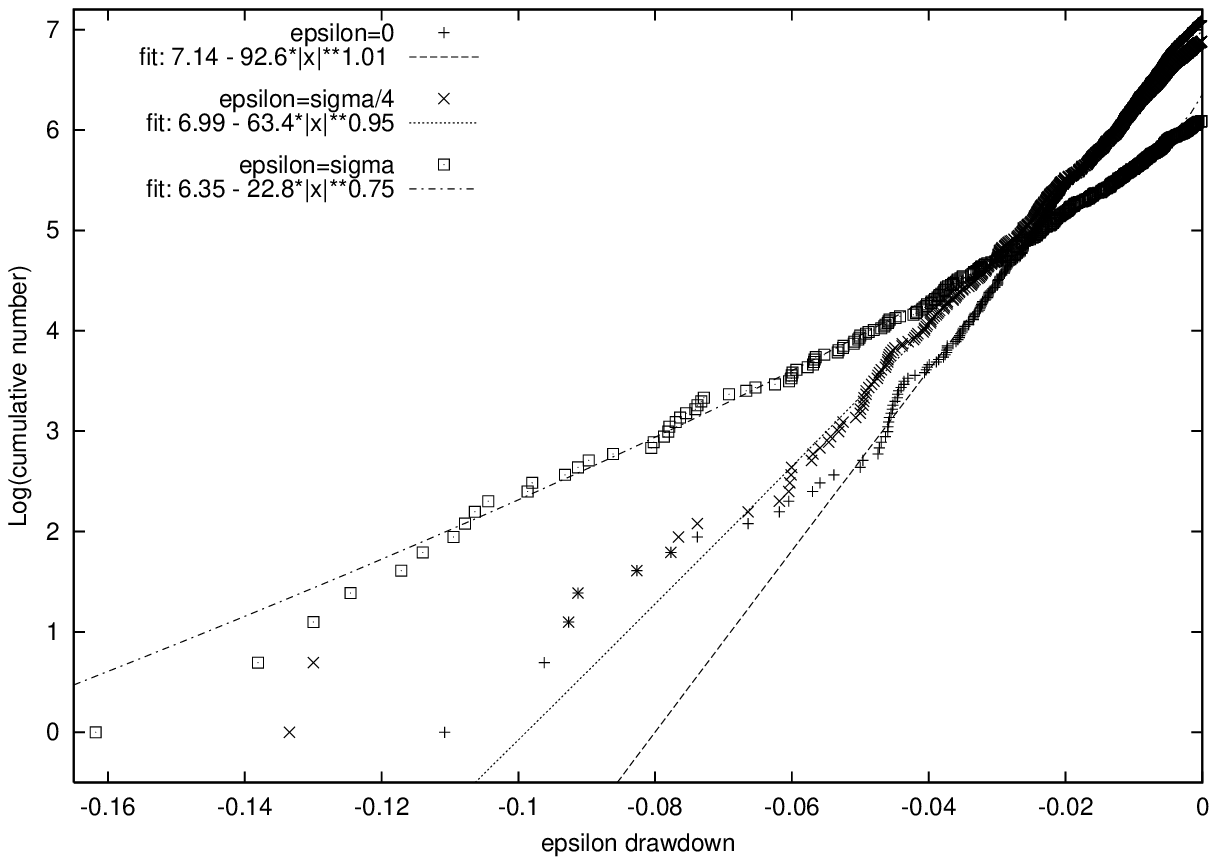,height=8cm,width=8.5cm}
\caption{\label{tbondepsdd}. Logarithm of the cumulative distribution of 
$\epsilon$-drawdowns in the price of the T-bond using an $\epsilon$ of $0$, 
$\sigma /4$ and $\sigma$,where $\sigma=0.012$ has been obtained from the data.
}}
\vspace{1.5cm}
\parbox[l]{8.5cm}{
\epsfig{file=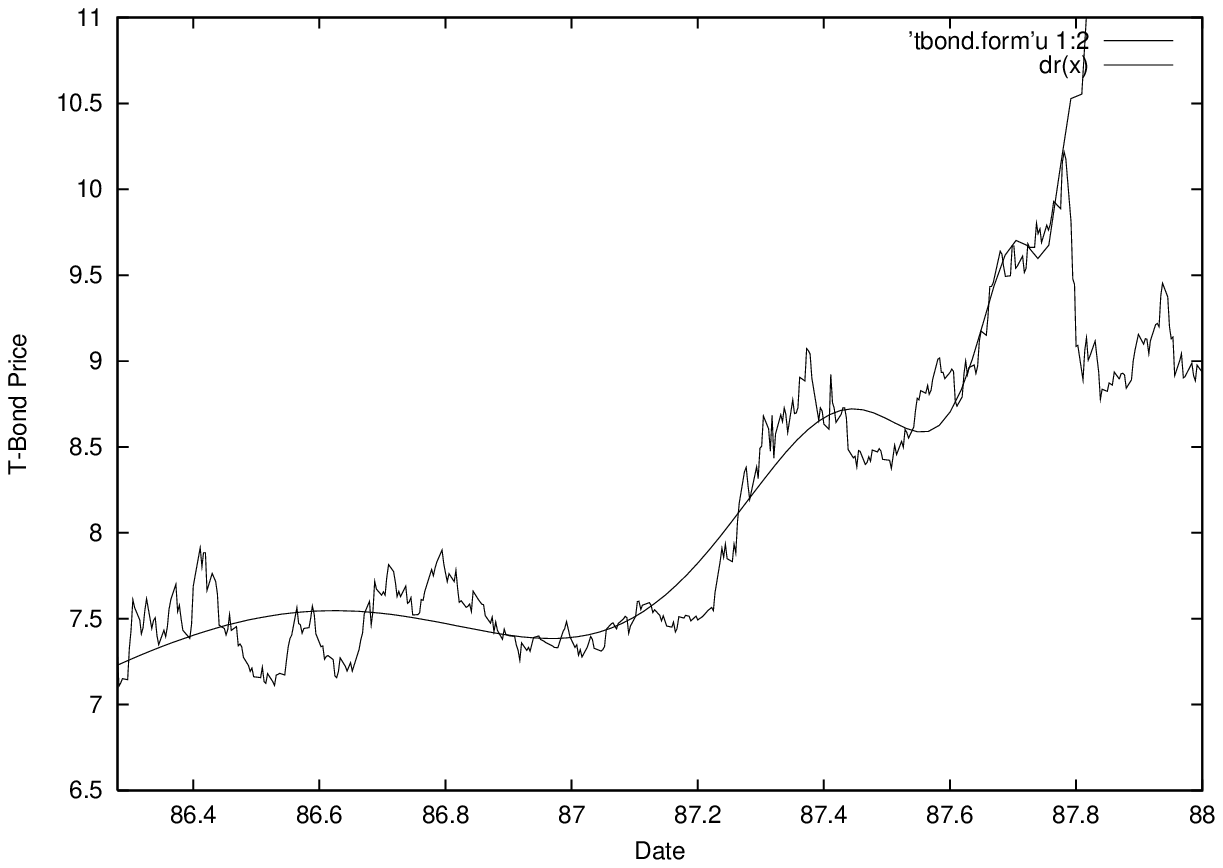,height=8cm,width=8.5cm}
\caption{\label{lptbond87} The T-bond bubble ending in 1987. The fit is eq. 
(\ref{lpeq}) where  $A \approx  14.8  $, $ B\approx  -7.3 $, $C \approx -0.34$,
$z \approx 0.16 $, $ t_c \approx   87.83  $, $ \phi \approx   -1.5 $, $ \omega 
\approx 5.5$.
}}
\hspace{5mm}
\parbox[r]{8.5cm}{
\epsfig{file=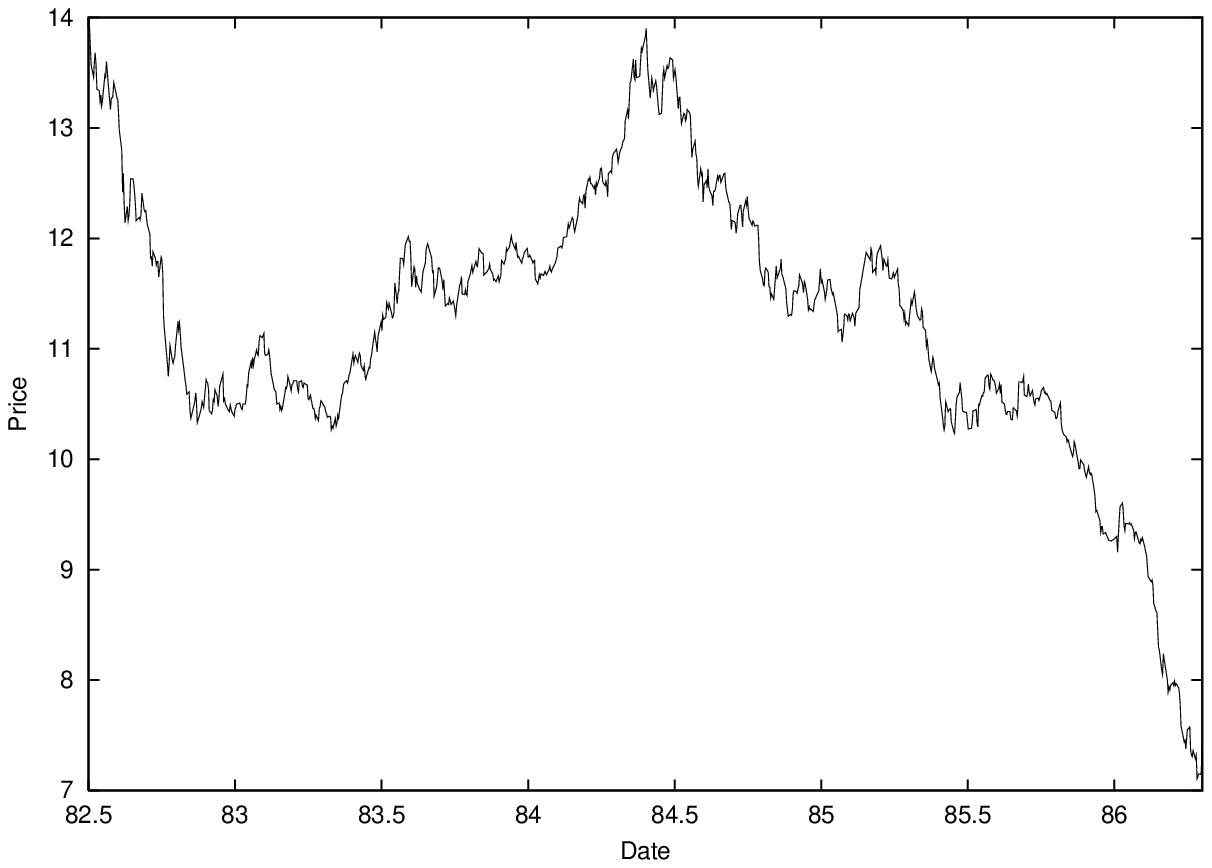,height=8cm,width=8.5cm}
\caption{\label{tbond8286} The price of the T-bond in the period from 1982.5 
to 1986.3. The large drawdowns of Oct. 1982 and  Feb. 1986 are clearly visible.
}}
\end{center}
\end{figure}

The analyzed time series of the U.S. T-bond market extends from 27/11 1980 to 
21/09 1999. Besides the date of the Oct. 1987 crash, the dates of the largest 
drawdowns in the market for the U.S. T-bond do not correlate with those of
the largest drawdowns in the stock markets, see section \ref{usstock}, nor to 
those in the FX markets, see section \ref{fx}. Figure \ref{lptbond87} shows 
that, prior to the T-bond crash of 1987, the bond market also experienced a 
log-periodic bubble similar to that of the stock market, qualifying it as 
endogenous. With respect to the $\epsilon$-drawdown of 1980.9, we do not have 
enough data to decide whether this crash was caused by a bubble or not. 
However, it is tempting to relate this event with the strong decline in the 
Gold price after its crash in early 1980, see table \ref{outlgold}. 
The large drawdowns of 
1982.7 and 1986.1 are part of a very large slow decline, as can be seen in 
figure \ref{tbond8286} and should qualify as exogenous. However, we have not 
been able to identify a ``smoking gun'' for the external sources except that
the 1982.7 event could again be linked to the decline in the Gold price.
Last, we show an example of another log-periodic bubble in the T-bond market,
see figure \ref{lptbond84}, which did not end in a large drawdown but rather 
in a large slow price-slide starting around 1984.4. The case of the T-bond 
market is thus the least convincing of all the markets examined until now in 
terms of the existence of endogenous speculative bubbles: only one case 
(1987.8)
is convincingly associated with an $\epsilon$-drawdown outlier and another one 
(1984.4) also exhibits the accelerated log-periodic signature but is
associated with a change of regime rather than a larger drawdown outlier.

\begin{table}[]
\begin{center}
\begin{tabular}{|c|c|c|c|c|c|c|c|c|c|}\hline
$\epsilon$ & Date             &  Size   & Duration &  Class & $\epsilon$ & Date     & Size            & Duration  &  Class\\ \hline
$0$        & 1987.781$\dagger$& $11.1\%$& 5 days   &  Bubble& $\sigma/4$ & 1986.096 & $13.3\%$        & 16 days   &  shock    \\ \hline
$0$        & 1986.129 & $9.6\%$    & 9  days       &  shock     & $\sigma/4$ & 1987.781$\dagger$& $13.0\%$&  7 days   &Bubble \\ \hline
$0$        & 1980.948 & $9.3\%$    & 4 days        &  ?     & $\sigma/4$ & 1980.948 & $9.33\%$        &  4 days   &  ?    \\ \hline
$0$        & 1982.748 & $9.1\%$    & 6 days        &  shock     & $\sigma/4$ & 1982.748 & $9.1\%$         &  6  days  &  shock    \\ \hline
\end{tabular}
\caption{\label{outltbond} Same as table \ref{outlusdm} for the U.S. Government
T-bond. Bubbles identified by the outlier analysis presented here, {\it i.e.}, 
outliers which have prior log-periodic bubbles not identified in previous 
works are marked by $\dagger$.}  
\end{center}
\end{table}

\subsection{The Japanese Bond market}

\begin{figure}
\begin{center}
\parbox[l]{8.5cm}{
\epsfig{file=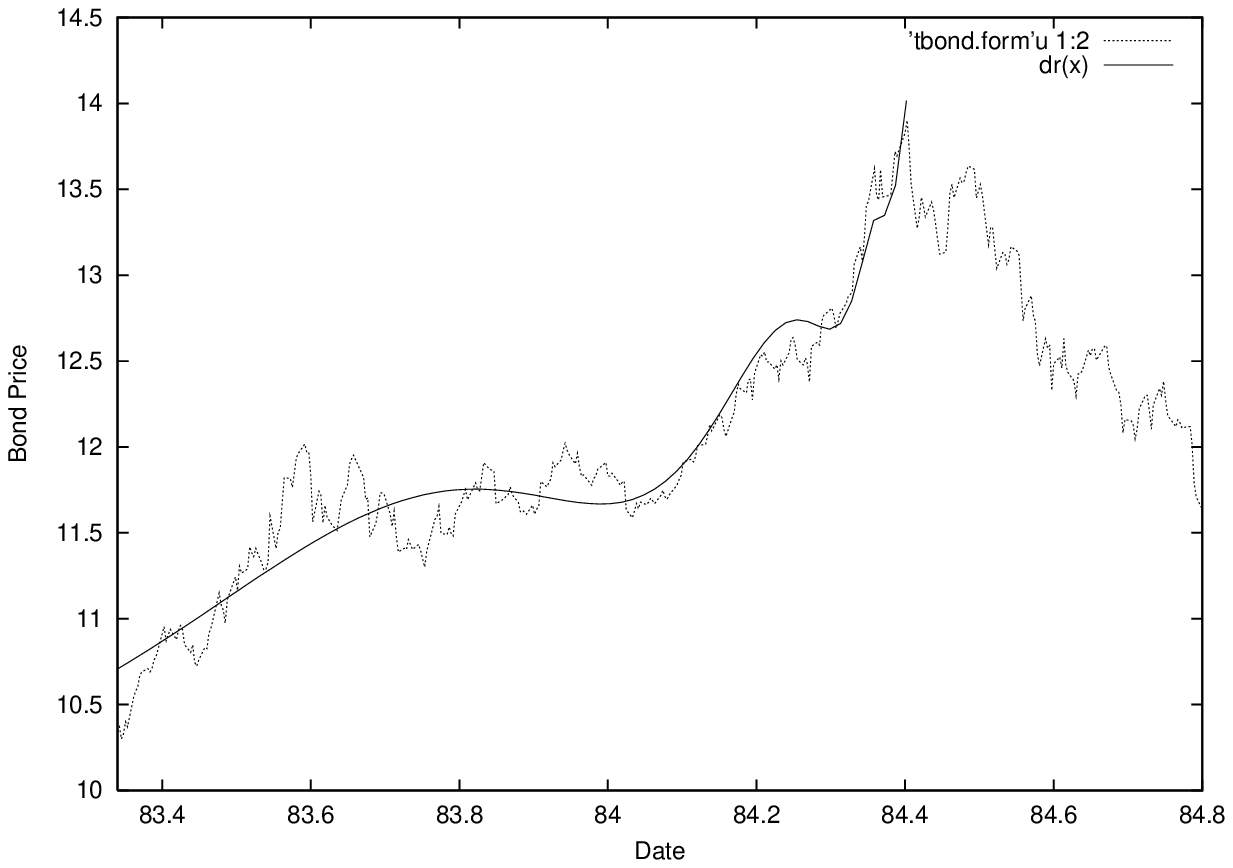,height=8cm,width=8.5cm}
\caption{\label{lptbond84} The T-bond bubble ending with ``a soft landing'' 
in 1984. The fit is eq. (\ref{lpeq}) where  $A \approx  14.5  $, $ B\approx  
-3.5 $, $C \approx   0.36 $, $z \approx 0.33 $, $ t_c \approx   84.4  $, 
$\phi \approx   1.4 $, $ \omega \approx 4.6$}}
\hspace{5mm}
\parbox[r]{8.5cm}{
\epsfig{file=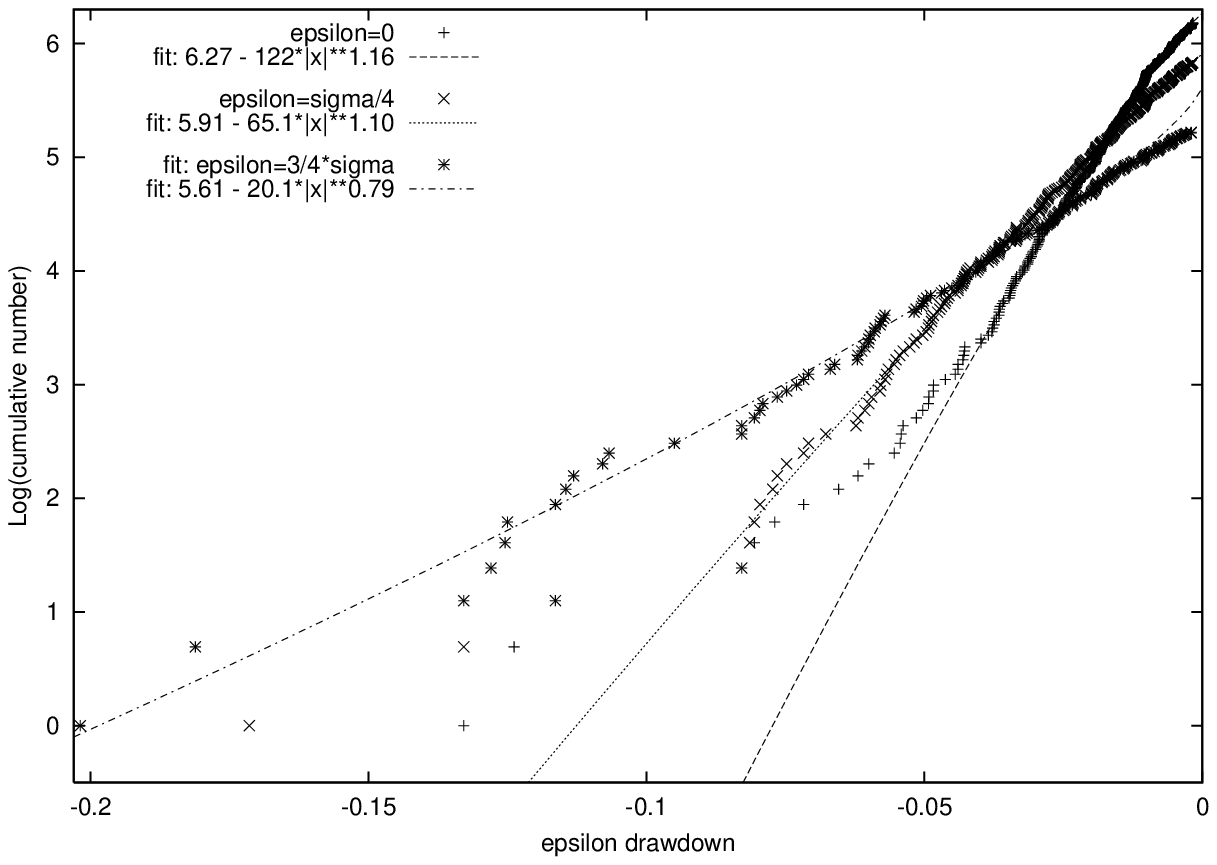,height=8cm,width=8.5cm}
\caption{\label{jgbepsdd}. Logarithm of the cumulative distribution of 
$\epsilon$-drawdowns in the price of the Japanese Government Bond using an 
$\epsilon$ of $0$, $\sigma /4$ and $\sigma $,where $\sigma=0.013$ has been 
obtained from the data.}}
\vspace{1.5cm}
\parbox[l]{8.5cm}{
\epsfig{file=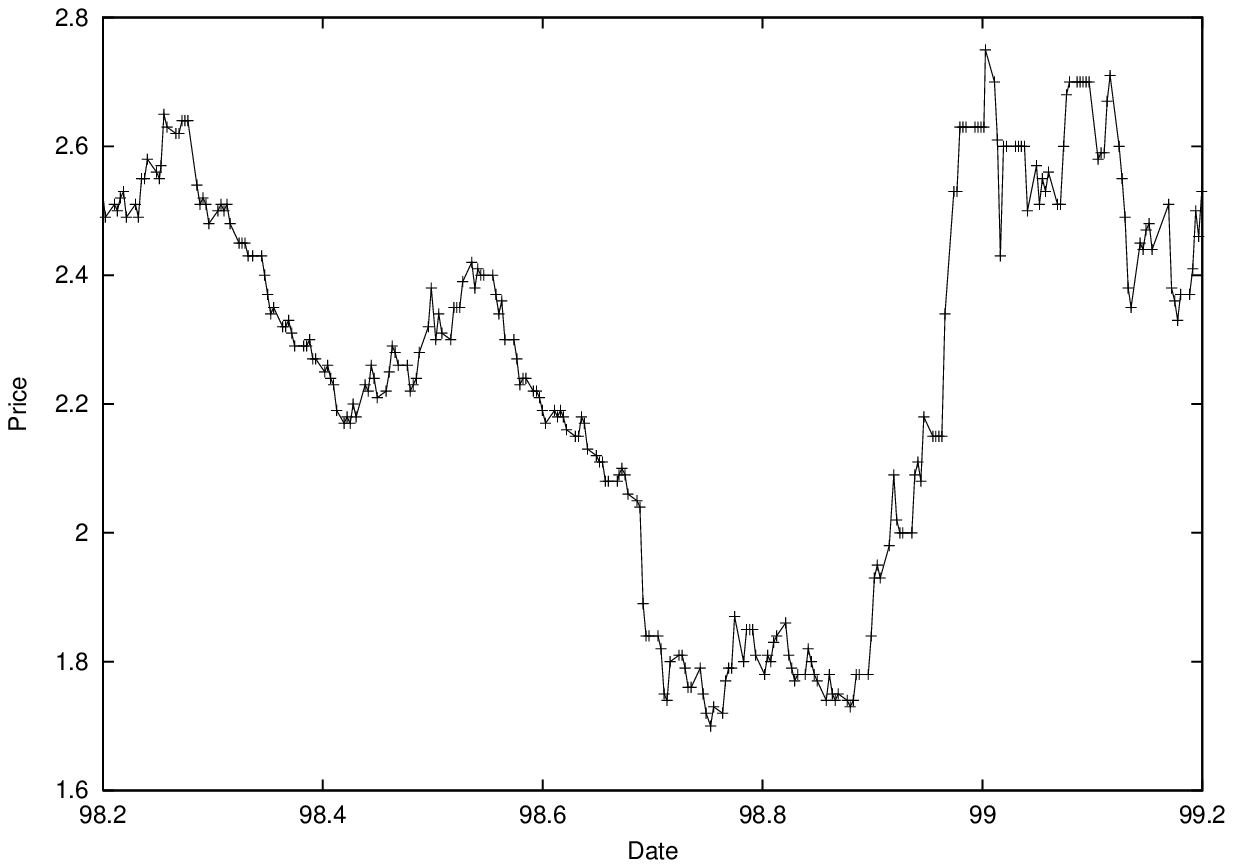,height=8cm,width=8.5cm}
\caption{\label{jgb1999} The price of the  Japanese Government Bond in the 
period from 1998.2 to 1999.2. The large drawdowns of early 1999 are clearly 
visible.}} 
\hspace{5mm}
\parbox[r]{8.5cm}{
\epsfig{file=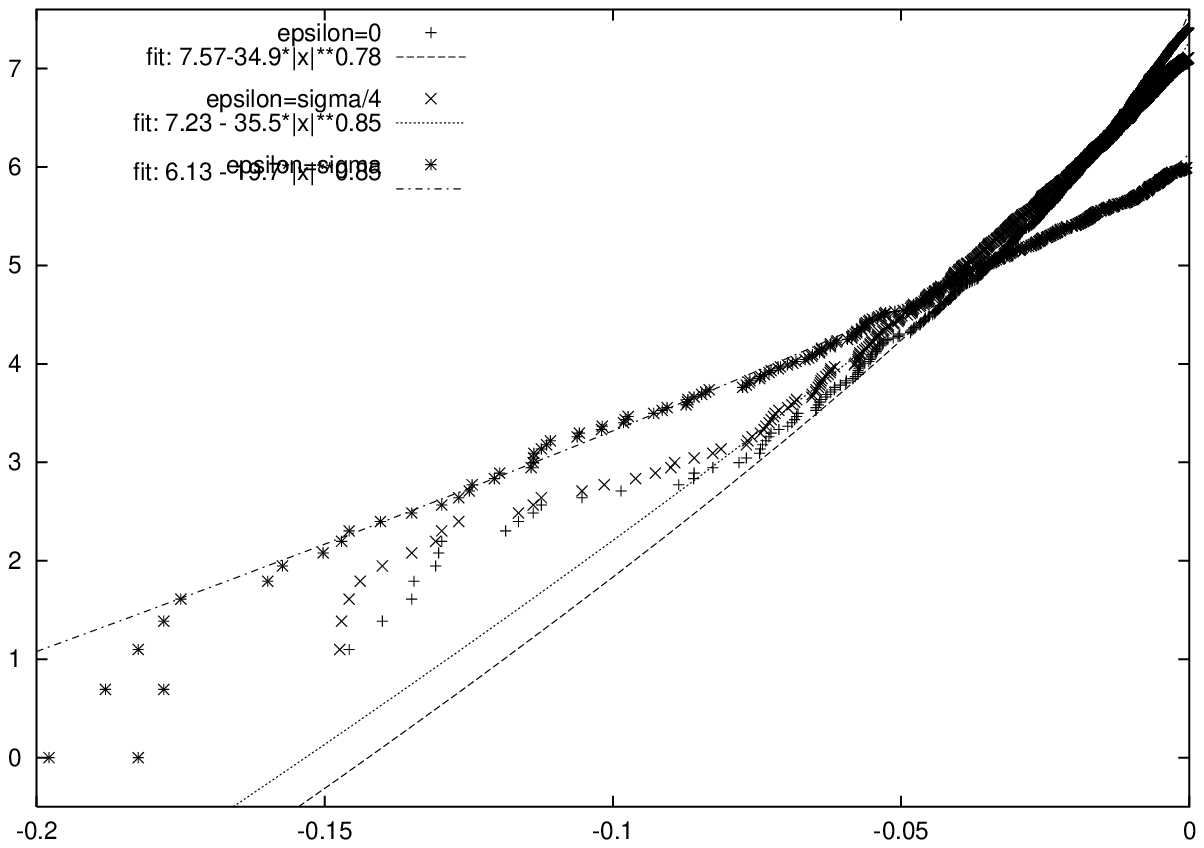,height=8cm,width=8.5cm}
\caption{\label{goldepsdd}. Logarithm of the cumulative distribution of 
$\epsilon$-drawdowns in the price of Gold using an $\epsilon$ of $0$, 
$\sigma /4$ and $\sigma$, where $\sigma=0.013$ has been obtained from the data.
}}
\end{center}
\end{figure}

The analyzed time series of the Japanese bond market extends
from 01/01 1992 to 22/03 1999. We find that one of the three large
drawdown outliers in the Japanese bond-market is correlated with a 
similar event in the stock market, namely the crash
of Aug. 1998. Figure \ref{jgb1999} shows the price of the Japanese Government 
Bonds in the period from 1998.2 to 1999.2. All three drawdowns listed in table 
\ref{outljgb} are clearly visible. In the absence of any reasonable fit with 
formula (\ref{lpeq}), we qualify these three events as exogenous. It is 
probable
that the driving source of the 1998.67 event was the synchronous crash on 
the stock market itself being associated with the Russian crisis.
We also observe the occurrence of a large drawup in figure \ref{jgb1999} 
preceding the two large drawdowns. Such behavior is observed in synthetic 
time series of the GARCH(1,1) model calibrated to this time series.

\begin{table}[]
\begin{center}
\begin{tabular}{|c||c|c|c|c|c|c|c|c|c|c|}\hline
$\epsilon$ & Date     & Size       & Duration& Class      & $\epsilon$ & Date     & Size       & Duration & Class      \\ \hline
$0$        & 1999.116 & $13.3\%$   & 5 days  & ?          & $\sigma/4$ & 1998.672 & $17.1\%$   & 11 days & Shock\\ \hline
$0$        & 1998.672 & $12.4\%$   & 6 days  &Shock& $\sigma/4$ & 1999.116 & $13.3\%$   &  5 days & ?           \\ \hline
$0$        & 1999.003 & $11.6\%$   & 3 days  & ?          &$\sigma/4$ & 1999.002 & $11.6\%$    &  3 days & ?           \\ \hline
\end{tabular}
\caption{\label{outljgb} Same as table \ref{outlusdm} for the Japanese 
Government Bonds.}  
\end{center}
\end{table}

\section{$\epsilon$-drawdown outliers of the Gold Market} \label{gold}

The analyzed time series of the Gold market goes from 2/1 1975 to 24/7 1998.
The number of large drawdowns not following the fit with eq.~(\ref{stretched}) 
is surprisingly larger for Gold than for all previous markets, with up to 20 
events qualifying as ``outliers'', see figure \ref{goldepsdd}. Table 
\ref{outlgold} lists only the four largest ones.

\begin{table}[]
\begin{center}
\begin{tabular}{|c|c|c|c|c|c|c|c|c|c|}\hline
$\epsilon$ & Date     & Size               & Duration & Class     &$\epsilon$ & Date     & Size        & Duration& Class\\ \hline
$0$        & 1980.057$\star$ & $18.2\%$   &2 days& Bubble    &$\sigma/4$ & 1980.057$\star$ & $18.2\%$& 2 days  &Bubble \\ \hline
$0$        & 1980.199$\star$ & $17.8\%$   &4 days& Bubble    &$\sigma/4$ & 1980.199$\star$ & $17.8\%$& 4 days  &Bubble \\ \hline
$0$        & 1981.057 & $14.6\%$           &6 days&Anti-bubble&$\sigma/4$ & 1981.480 & $14.7\%$        &11 days  &Anti-bubble\\\hline
$0$        & 1983.149 & $14.0\%$           &2 days&  ?  &$\sigma/4$ & 1980.113$\star$ & $14.7\%$& 7 days  &Bubble   \\ \hline
\end{tabular}
\caption{\label{outlgold} Same as table \ref{outlusdm} for the Gold market. 
The events
with a $\star$ have already been studied in previous publications.}  
\end{center}
\end{table} 

\begin{figure}
\begin{center}
\parbox[l]{8.5cm}{
\epsfig{file=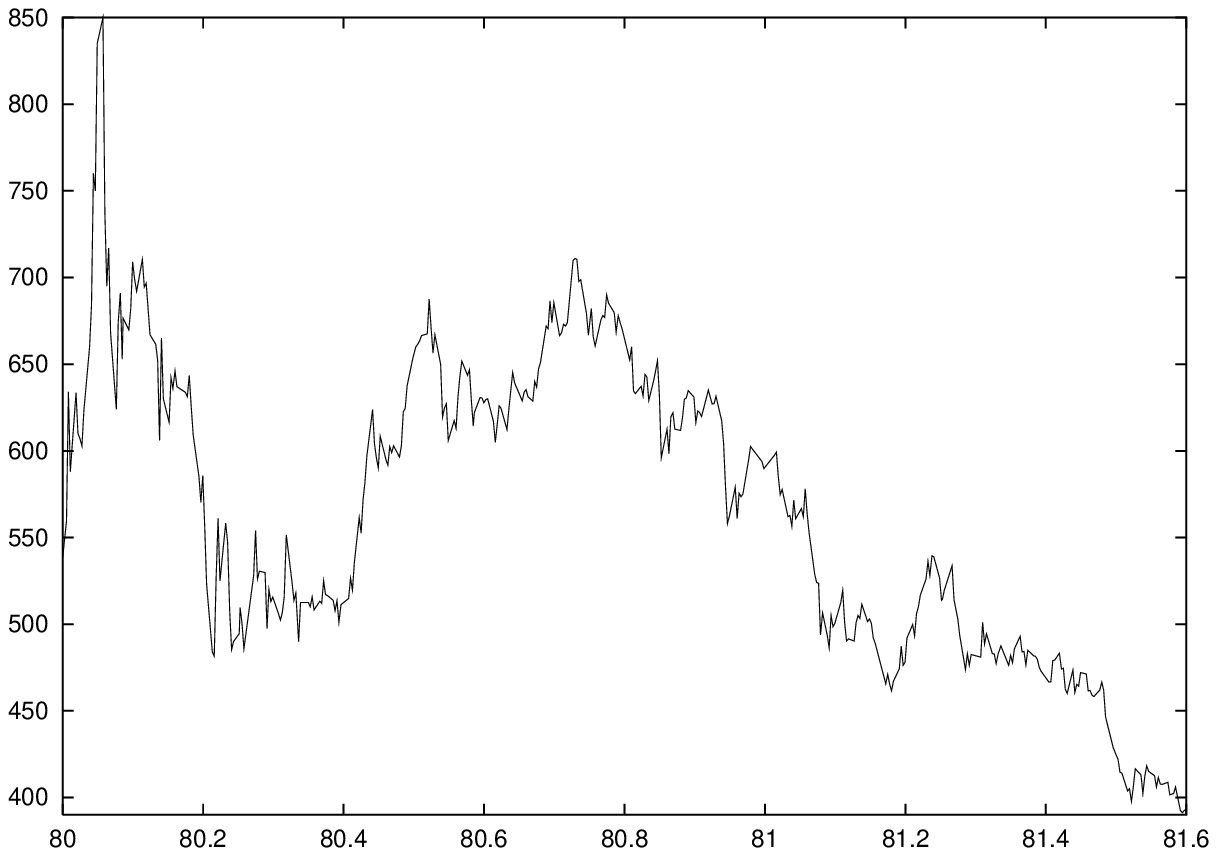,height=8cm,width=8.5cm}
\caption{\label{au81} The price of Gold in the period from 1980.0 to 1981.6. 
The large drawdowns of early 1980 as well as early and mid- 1981 are 
clearly visible.
}}
\hspace{5mm}
\parbox[r]{8.5cm}{
\epsfig{file=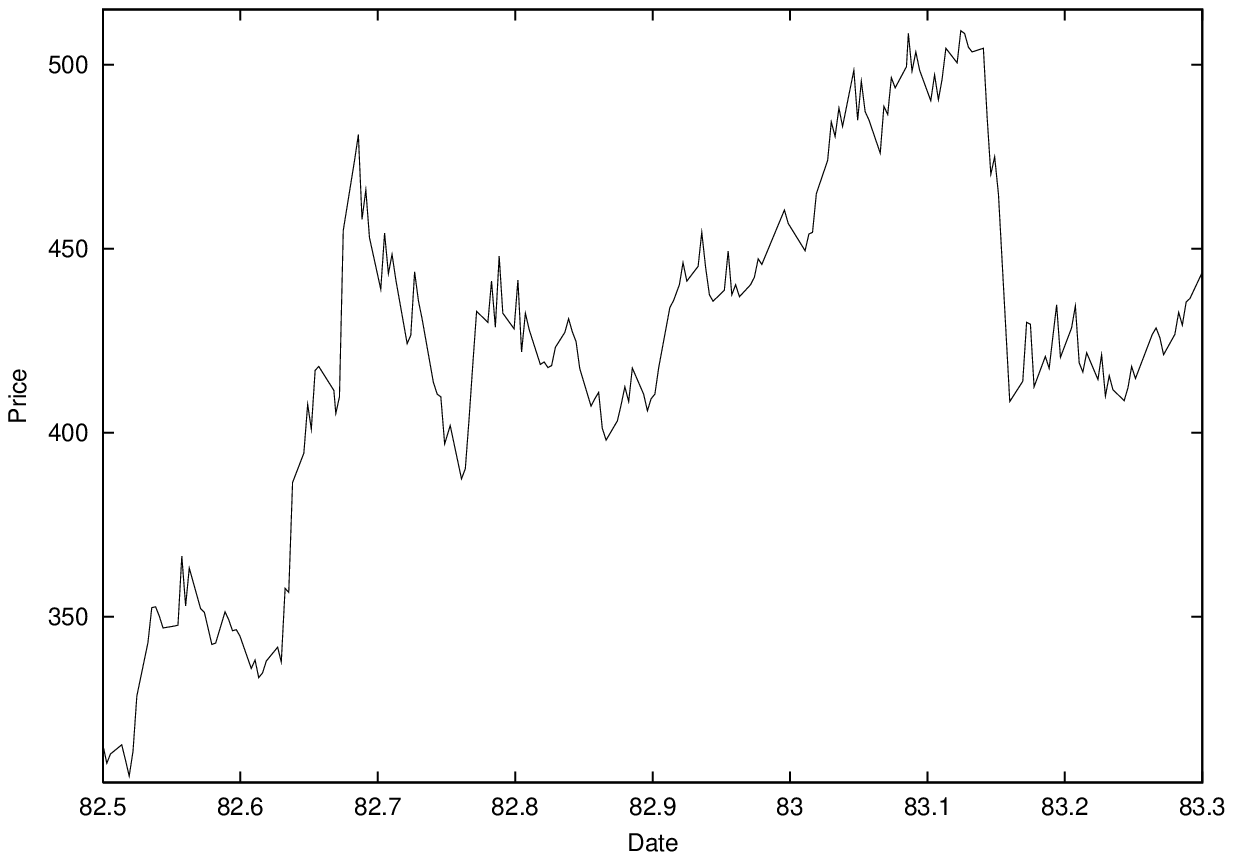,height=8cm,width=8.5cm}
\caption{\label{au83} The price of Gold in the period from 1982.5 to 1983.3. 
The large drawdown of early 1983 is clearly visible.}}
\end{center}
\end{figure}

The two largest events are the crash of the Gold price in 1980 and a related 
aftershock already analyzed in \cite{antibub1}. A second aftershock turns up 
for $\epsilon = \sigma/4$. Two events, 1981.1 for $\epsilon = 0$ and 1981.5 
for $\epsilon = \sigma/4$, are related to the anti-bubble following the crash 
of 1980 previously published in \cite{antibub1}, see also figure \ref{au81}. 
With respect to the event of 1983.1 for $\epsilon = 0$, figure \ref{au83} 
shows that its preceding time series cannot be fitted 
convincingly with formula (\ref{lpeq}), which qualifies it as exogenous.

\section{Conclusion}

By combining the statistical evidence of two methods, that is, the analysis
of financial data from the point of view of detecting (i) anomalous events 
seen as outliers to the parameterization (\ref{stretched}) and (ii) 
accelerating log-periodic power laws quantified by eq.(\ref{lpeq}) with
values of $\omega$ and $z$ satisfying (\ref{zomegaval}) qualifying endogenous 
speculative bubble, we have provided an objective test of the hypothesis that 
the largest negative markets moves are special and form two distinct 
populations. The exogenous crashes can be attributed to extraordinary important
external perturbations and/or news impacting the market. The endogenous crashes
can be seen as the natural deaths of self-organized self-reinforcing 
speculative bubbles giving rise to specific precursory signatures in the form 
of log-periodic power laws accelerating super-exponentially.

The generalization of the analysis of drawdowns to drawdowns coarse-grained in 
amplitude ($\epsilon$-drawdowns) has strengthened the evidence for the presence
of anomalously large drops/corrections/crashes in the financial markets. 
A particularly remarkable result is that a large majority of the crashes and 
associated bubbles previously studied for the presence of strong LPPS have 
been identified with an outlier or extremely large drawdown. Furthermore, 
several new cases of LPPS have been identified on the basis of our generalised
outlier analysis, most notably for the FTSE where {\it all} outliers were
found to have LPPS. This confirm previous tests \cite{JLS2000} that showed that
LPPS is a strong discriminator of bubbles preceding strong corrections. It also
strengthen the conclusion that LPPS (which is by construction 
transient) is almost uniquely associated with a speculative phase announcing a 
strong change of regime. The analysis reported here can thus be seen as 
refuting possible criticism concerning data picking of a few cases. Here, we 
have developed a systematic approach which compares the occurrence of the 
largest drawdowns with the existence of log-periodicity and super-exponential 
growth (LPPS). Doing this, we have identified
novel occurrences of log-periodicities which were unnoticed before.

Globally over all the markets analyzed, we identify 49 drawdown outliers, of 
which 25 are classified as endogenous, 22 as exogeneous and 2 as associated
with the Japanese anti-bubble. Restricting to the world market indices, 
we find 31 outliers, of which 19 are endogenous, 10 are exogenous and 
2 are associated with the Japanese anti-bubble. For the FX market, 
we identify 7 outliers (3 endogeneous with LPPS and 4 exogeneous shocks);
for the US markets, we find 12 outliers (7 endogenous with LPPS and
5 exogenous shocks); in the London stock market, we identify 4
outliers, all of them qualified endogeneous with LPPS; In the German market,
we find 4 outliers (1 endogenous with LPPS and 3 exogenous shocks);
for the Japanese market, we find 3 outliers (1 endogeneous with LPPS,
and 2 shocks associated with the anti-bubble phase
starting in Jan. 1990); in the Hong-Kong market, we find 8 outliers
(6 endogeneous LPPS and 2 exogenous shocks); In the US T-Bond market, 
we find 4 outliers (1 endogeneous with LPPS and 3 exogenous shocks); in
the Japanese bond market, we identify 3 outliers, all of them being
exogeneous shocks; finally in the gold market, we find 4 outliers
(2 endogenous with LPPS and 2 exogeneous shocks).

Two cases with exogenous shocks stand out, in the sense
that these external shock were not due to political or economic news but
rather to the impact of a crash in the US market and in most
of the other stock markets in the 
world \cite{krach87}. These two drawdown outliers are
the 1987 drawdown outlier on the German DAX index 
and the 1987 drawdown outlier on the Japanese Nikkei index. They have
been qualified as due to exogenous shocks because of the abscence of any LPPS
in the time series preceding them, implying that these markets did not crash
by an instability associated with a developing bubble. These two
phenomena are reminiscent of the contagion literature 
(see \cite{conta1,conta2,contaMS} for reviews), which 
refer to manifestations of propagating
crises resulting from an increase in
the correlation (or linkage) across markets during turmoil periods.
Our present analysis suggest new avenues for research to establish the role of the
presence and abscence of instabilities ripening in local markets
in the propagation and amplitude of contagions.

Notwithstanding the overall positive results presented
in this work, there are still cases that are unclear.
For instance, as shown by the example of the T-bond bubble ending in 1984, 
shown in fig. \ref{lptbond84}, we
can observe a strong log-periodic signal with super-exponential acceleration 
of the price without
the occurrence of a crash ending this spell. True, there is a noticeable 
change of regime but
it occurs rather smoothly over a long time scale. This calls for more work 
to characterize this
change of regime or regime switch, possibly by combining techniques of 
Markov-switching models \cite{Kim} 
with those developed here.

\section*{Acknowledgments}

The authors are grateful to Ingve Simonsen for providing the data used in
constructing figures \ref{wave} and \ref{djiawave}. 
Cited papers by the authors are available from 
http://www.nbi.dk/\~\/johansen/pub.html and \\
http://www.ess.ucla.edu/faculty/sornette/

\newpage

\section*{Appendix: Data} \label{data}

The markets and corresponding time periods analyzed in this paper are:
\begin{itemize}
\item The foreign exchange market (FX) in terms of the exchange between the 
US\$ and the German Mark (DM) and the exchange between the US\$ and the 
Japanese Yen. The data are from 04/01 1971 to 19/05 1999 for both exchange 
rates.

\item The U.S. stock market in terms of the Dow Jones Industrial Average 
(DJIA), the SP500 and the NASDAQ. The data are from 01/01 1900 to 17/07 2000 
for the DJIA, 29/11 1940 to 17/07 2000 for SP500 and 05/02 1971 to 17/07 2000 
for the NASDAQ.

\item The London stock market index FTSE. The data are from 02/03 1984 to 
13/07 2000.

\item The Frankfurt stock market index DAX. The data are from 02/01 1970 to 
13/07 2000.

\item The Tokyo stock market index Nikkei. The data are from 28/03 1986 to 
13/07 2000.

\item The Hong-Kong stock market index Hang-Seng. The data are from 24/11 1969 
to 13/07 2000.

\item T-bonds and Japanese Government Bonds (JGB).  The data are from 27/11
1980 to 21/09 1999 for the T-bonds and 01/01 1992 to 22/03 1999 for the JGB.

\item The Gold market. The data are from 2/1 1975 to 21/08 1998.
\end{itemize}
Many other markets, especially the emergent markets of Asia and Latin America,
as well as individual stocks have been analysed previously using pure drawdowns
\cite{outl2}. The main reason for not analyzing German Government Bonds is the 
special historical situation that Germany has been in since the unification on 
3. Oct. 1990.

\newpage

\clearpage

\clearpage

\clearpage

\begin{figure}
\begin{center}
\parbox[l]{8.5cm}{
\epsfig{file=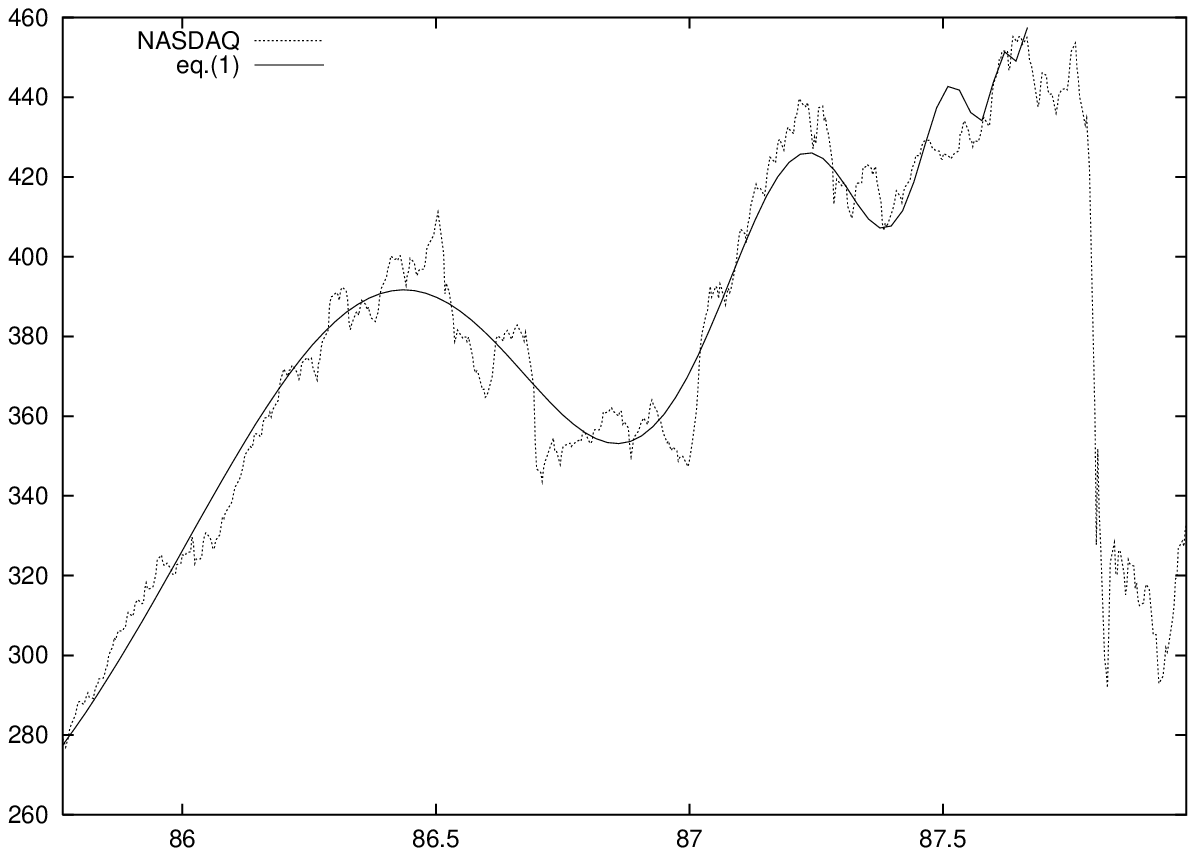,height=8cm,width=8.5cm}
\caption{\label{nas1987} The NASDAQ bubble ending in 1987. The fit is eq. 
(\ref{lpeq}) where  $A \approx  460  $, $ B\approx  -92 $, $C \approx   -34 $, 
$z \approx 0.68 $, $ t_c \approx   87.67  $, $ \phi \approx   -1.6 $, $ \omega 
\approx  6.1$. } }
\hspace{5mm}
\parbox[r]{8.5cm}{
\epsfig{file=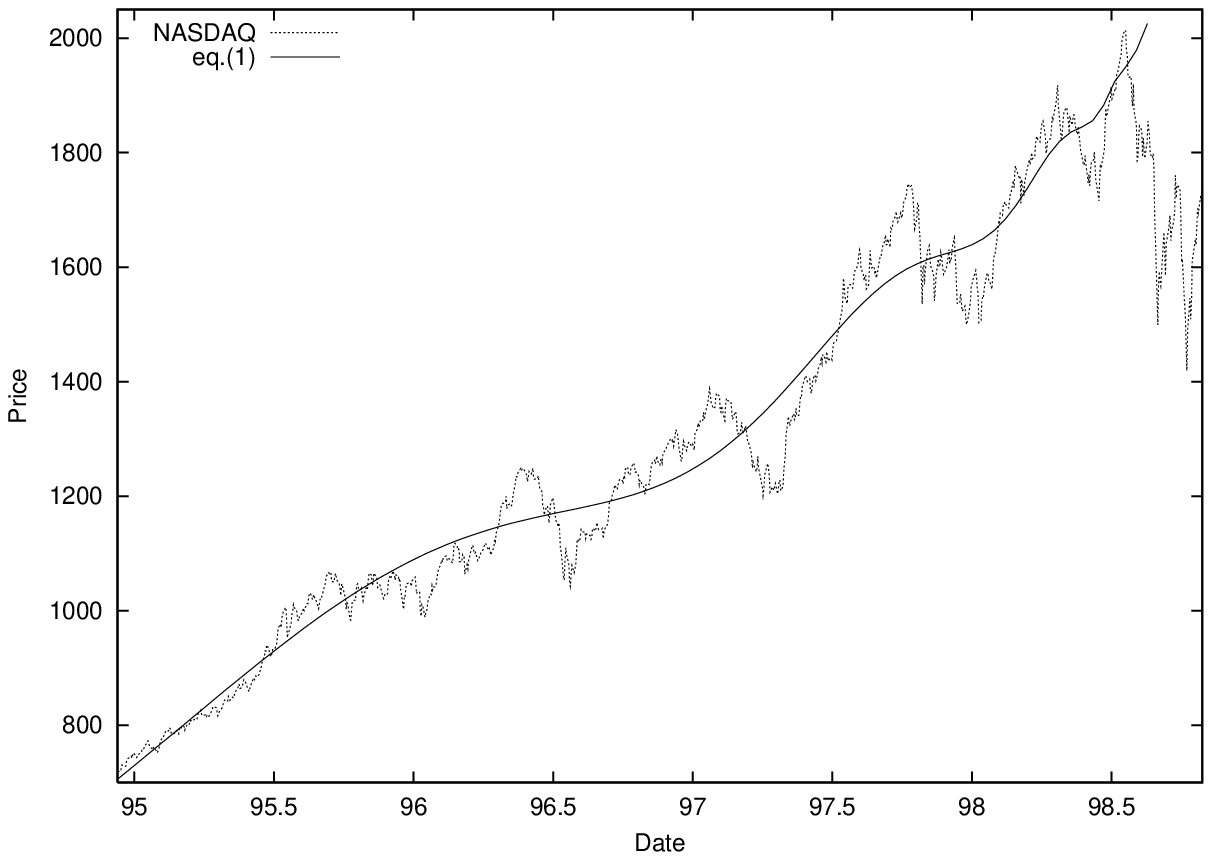,height=8cm,width=8.5cm}
\caption{\label{nas1998} The NASDAQ bubble ending in 1998. The fit is eq. 
(\ref{lpeq}) where  $A \approx  2051  $, $ B\approx  -535 $, $C \approx   39 $,
$z \approx 0.68 $, $ t_c \approx   98.64  $, $ \phi \approx   -0.4 $, $ \omega 
\approx  6.0$. } }
\end{center}
\end{figure}

\clearpage

\clearpage

\clearpage

\newpage

\begin{thebibliography}{}

\bibitem[Barro et al., 1989]{krach87} 
Barro, R.J., E.F. Fama, D.R. Fischel, A.H. Meltzer, R. Roll
and L.G. Telser (1989) Black monday and the future of financial markets, edited by
R.W. Kamphuis, Jr., R.C. Kormendi and J.W.H. Watson (Mid American Institute for
Public Policy Research, Inc. and Dow Jones-Irwin, Inc.).

\bibitem[Checkhlov et al., 2000]{Checkhlov}
Checkhlov, A., Uryasev, S. and M. Zabarankin (2000), Portfolio 
optimization with drawdown constraints, Discussion paper, ISE Dept., 
Univ. of Florida.

\bibitem[Claessens et al. (2001)]{conta1} Claessen, S., R.W. Dornbush and Y.C. Park,
(2001) Contagion: Why crises spread and how this can be stopped, in S.
Cleassens and K.J. Forbes eds, International Financial Contagion (Kluwer
Academic Press).

\bibitem[Coe, 2002]{Coe} Coe, P.J., 2002,
Financial crisis and the great depression:  A regime switching approach,
Journal of Money, Credit, \& Banking 34 (1), 76-93.

\bibitem[Cutler et al. (1989)]{Cutler} Cutler, D., J. Poterba and L. Summers,
What Moves Stock Prices? 1989, Journal of Portfolio Management, Spring, 4-12.

\bibitem[Cvitanic and Karatzas, 1995]{Cvitanic} 
Cvitanic, J. and I. Karatzas, 1995, On Portfolio Optimization Under "Drawdown"
Constraints, IMA Lecture Notes in Mathematics \& Applications 65, 77-88.

\bibitem[NIST]{handbook} Engineering Statistical Handbook, National Institute 
of Standards and Technology. 
See http://www.itl.nist.gov/div898/handbook/prc/section1/prc16.htm

\bibitem[Fama, 1991]{Fama} Fama, E.F. , 1991,
Efficient Capital Markets: II, Journal of Finance 46, 1575-1617.

\bibitem[Fisher, 1998]{Fisher} Fisher, M.E., 1998,
Renormalization group theory: Its basis and formulation in statistical
physics, Review of Modern Physics 70, 653-681.

\bibitem[Forbes and Rigobon (2002)]{conta2} Forbes, K.J. and R. Rigobon (2002) 
No contagion, only interdependence:
measuring stock market co-movements, forthcoming Journal of Finance.

\bibitem[Grossman and Zhou, 1993]{Grossman} Grossman, S. J. and Z. Zhou., 1993,
Optimal Investment Strategies for Controlling
Drawdowns, Mathematical Finance, 3 (3), 241-276.

\bibitem[Johansen, 1997]{thesis} Johansen, A., 1997,
Discrete scale invariance and other cooperative phenomena in spatially extended
systems with threshold dynamics  Ph.D. Thesis, Niels Bohr Inst. 1997. 

\bibitem[Johansen, 2002]{crashcom} Johansen, A., 2002 
Comment on "Are financial crashes predictable?" 
Eur. Phys. Lett. (in press)

\bibitem[Johansen, 2002]{bali} Johansen, A, 2002,
Characterization of large price variations in financial markets,
Proceedings of Econophysics conference 28-31 Aug. 2002. To be published in
Physica A.

\bibitem[Johansen {\it et al.}, 1999]{JLS2000}
Johansen, A., O. Ledoit and D. Sornette, 2000,
Crashes as critical points,  International Journal of Theoretical and Applied 
Finance 3 (2),  219-255.

\bibitem[Johansen and Sornette, 1998]{outl1}
Johansen, A. and D. Sornette, 1998, Stock market crashes are outliers, 
European Physical Journal B 1, 141-143.

\bibitem[Johansen and Sornette, 1999]{JS1999}
Johansen, A. and D. Sornette, 1999, Critical Crashes, RISK 12 (1), 91-94.

\bibitem[Johansen and Sornette, 2000]{JS2000}
Johansen, A. and D. Sornette, 2000, 
The Nasdaq crash of April 2000: Yet another example of
log-periodicity in a speculative bubble ending in a crash,
Eur. Phys J. B 17 pp. 319-328.

\bibitem[Johansen and Sornette, 2001b]{emergent} Johansen, A. and D. Sornette, 
2001,
Bubbles and anti-bubbles in
Latin-American, Asian and Western stock markets: An empirical study, 
International Journal of Theoretical and
Applied Finance 4 (6), 853-920.

\bibitem[Johansen and Sornette, 2001b]{outl2}
Johansen, A. and D. Sornette, Winter 2001/02, 
Large Stock Market Price Drawdowns Are Outliers,
Journal of Risk, 4(2), 69-110.

\bibitem[Johansen {\it et al.}, 1999]{JSL1999}
Johansen, A., D. Sornette and O. Ledoit, 1999,
Predicting Financial Crashes using discrete scale invariance,
Journal of Risk 1 (4), 5-32.

\bibitem[Johansen and Sornette, 1999]{antibub1}
Johansen, A. and D. Sornette, 1999,
Financial ``Anti-Bubbles'': Log-Periodicity in Gold and Nikkei collapses 
Int. J. Mod. Phys. 10, 563-575.

\bibitem[Johansen and Sornette, 2000]{antibub2}
Johansen, A. and D. Sornette, 2000, 
Evaluation of the quantitative prediction of a trend reversal on the Japanese 
stock market in 1999, Int. J. Mod. Phys. C 11 no. 2,  359-364.

\bibitem[Johansen, Simonsen and Sornette, 2002]{JSS} 
Johansen, A., I. Simonsen and D. Sornette, 2002, 
Time Scales of Outliers. In preparation.

\bibitem[Johansen, 2002]{tauhori}
Johansen, A, 2002,
Outlier analysis using ``elastic time''. In preparation.

\bibitem[Lillo and Mantegna, 2000]{LilloMantegna} 
Lillo, F., and Mantegna, R.N., 2000,
Symmetry alteration of ensemble return distribution in crash and rally
days of financial markets, European Physical Journal B 15, 603-606.

\bibitem[Malevergne and Sornette, 2002]{contaMS} Malevergne, M. and D. Sornette (2002)
Investigating Extreme Dependences: Conditioning Effect Versus
Contagion in Latin-American Crises, 
(e-print at http://arXiv.org/abs/cond-mat/0203166)

\bibitem[Mansilla, 2001]{Mansilla} 
R. Mansilla, Algorithmic Complexity in Real Financial Markets, cond-mat/0104472

\bibitem[Kim and Nelson, 1999]{Kim} Kim, C.-J. and C.R. Nelson, 1999,
State-space models with regime switching: classical and Gibbs-sampling
approaches with applications (Cambridge, Mass. : MIT Press).

\bibitem[Laherr\`ere and Sornette, 1998]{Lahe} Laherr\`ere, J. and D. Sornette,
1998, Stretched exponential distributions in Nature and Economy: ``Fat tails''
with characteristic scales, European Physical Journal B 2, 525-539.

\bibitem[Press {\it et al.}, 1992]{Book:NR-1992} 
Press, W.H., Teukolsky, S.A., Vetterling, W.T. \& Flannery, B.P. 
Numerical Recipes in Fortran, 2nd ed. (Cambridge University Press, New
York, 1992).

\bibitem[Simonsen {\it et al.}, 2002]{Simonsen-EPJB-2002}
 I.\ Simonsen, M.\ H.\ Jensen and A.\ Johansen, Eur. Phys. J. B {\bf 27}  583 
(2002).

\bibitem[Sornette, 1998]{Sordsi} Sornette, D., 1998,
Discrete scale invariance and complex dimensions, Physics Reports 297, 239-270.

\bibitem[Sornette and Helmstetter, 2002]{Sorhelm} Sornette, D. and A. 
Helmstetter, 2002, Endogeneous Versus Exogeneous Shocks in Systems with Memory,
in press in Physica A (http://arXiv.org/abs/cond-mat/0206047)

\bibitem[Sornette and Johansen, 1997]{SJ1997} Sornette, D. and A. Johansen, 
1997, Large financial crashes, Physica A 245, 411-422.

\bibitem[Sornette and Johansen, 1998]{SJ1998} Sornette, D. and A. Johansen, 
1998, A Hierarchical Model of Financial Crashes, Physica A 261 (Nos. 3-4), 

\bibitem[Sornette and Johansen, 2001]{SJ2001}
Sornette, D. and A. Johansen, 2001
Significance of log-periodic precursors to financial crashes,   
Quantitative Finance 1, 452-471.

\bibitem[Sornette et al., 1996]{SJB} Sornette, D., A. Johansen and 
J.-P. Bouchaud, 1996,
Stock market crashes, Precursors and Replicas, J.Phys.I France 6, 167-175.

\bibitem[Sornette et al, 2002]{SorMRW} Sornette, D., Y. Malevergne and 
J.F. Muzy, 2002,
Volatility fingerprints of large shocks: Endogenous versus exogenous,
submitted to Risk Magazine
(working paper at http://arXiv.org/abs/cond-mat/0204626)

\bibitem[Sornette and Zhou, 2002]{predsz} Sornette, D. and W.-X. Zhou, 2002,
 The US 2000-2002 Market Descent: How Much Longer and Deeper?
in press in Quantitative Finance
(preprint at http://arXiv.org/abs/cond-mat/0209065)

\bibitem[White (1996)]{white} White E.N., 1996, Stock market crashes and
speculative manias. In The international library of macroeconomic and
financial history, 13 (An Elgar Reference Collection, Cheltenham, UK;
Brookfield, US).

\bibitem[Zhou and Sornette, 2002]{zhouturb1} Zhou, W.-X. and D. Sornette, 2002,
Evidence of Intermittent Cascades from Discrete Hierarchical Dissipation
in Turbulence, Physica D 165, 94-125.

\bibitem[Zhou et al., 2002]{zhouturb2} Zhou, W.-X., D. Sornette and V.F. 
Pisarenko, 2002,
New Evidence of Discrete Scale Invariance in the Energy Dissipation of
Three-Dimensional Turbulence: Correlation Approach and Direct
Spectral Detection, submitted to Physical Review E
(preprint at http://arXiv.org/abs/cond-mat/0208347)




\end{thebibliography}
\end{document}